\documentclass[review,11pt,3p]{elsarticle}

\usepackage{amsmath}
\usepackage{bm}
\usepackage{amsfonts}
\usepackage{dsfont}
\usepackage{fancyhdr}
\usepackage{euscript}
\usepackage{graphicx}
\usepackage{mathrsfs}
\usepackage{color}
\usepackage{footnote}
\usepackage[pdftex,colorlinks,bookmarksopen,bookmarksnumbered,citecolor=red,urlcolor=red]{hyperref}
\usepackage{cancel}
\usepackage{textcomp}
\usepackage{algorithm} 
\usepackage{epstopdf}
\usepackage{subfigure}
\usepackage{siunitx}
\usepackage[colorinlistoftodos]{todonotes} 
\usepackage{bm}

\newcommand{\x}{\boldsymbol{x}}

\newcommand{\eq}[1]{Eq.~(\ref{#1})}

\newcommand{\invisible}[1]{}
\graphicspath {
    {./}
    {./figures/}
}

\newcommand{\vx}{\vm{x}}
\newcommand{\bsym}[1]{\boldsymbol{#1}}
\newcommand{\vm}[1]{\mathbf{#1}}

\newcommand{\jim}[1]{\ensuremath{\mathrm{#1}}} 

\usepackage{calc}

\usepackage[colorlinks]{hyperref}

\usepackage{xcolor}

\newenvironment{MyColorPar}[1]{%
	\leavevmode\color{black}\ignorespaces%
}{%
}%

\newcommand{\red}[1]{\textcolor{black}{#1}}

\makeatletter
\def\ps@pprintTitle{%
	\let\@oddhead\@empty
	\let\@evenhead\@empty
	\def\@oddfoot{}%
	\let\@evenfoot\@oddfoot}
\makeatother

\begin{document}

\begin{frontmatter}



\title{High-order maximum-entropy collocation methods \footnote {\textcopyright 2020. This manuscript version is made available under the CC-BY-NC-ND 4.0 license http://creativecommons.org/licenses/by-nc-nd/4.0/. \newline Published version available at https://doi.org/10.1016/j.cma.2020.113115. }    }


\author[label1]{F.~Greco\corref{cor1}}
\author[label1]{M.~Arroyo}

\address[label1]{Laboratori de C\`alcul Num\`eric, Universitat Polit\`ecnica de Catalunya-BarcelonaTech, Jordi Girona 1, E-08034 Barcelona, Spain.}

\cortext[cor1]{Correspondence to: francesco.greco@upc.edu}

\begin{abstract}
This paper considers the approximation of partial differential equations with a point collocation framework  based on high-order local maximum-entropy schemes (HOLMES). 
In this approach, smooth basis functions are computed through an optimization procedure and the strong form of the problem is directly imposed at the collocation points, reducing significantly the computational times with respect to the Galerkin formulation. Furthermore, such a method is truly meshfree, since no background integration  grid  is necessary.
The validity of HOLMES collocation is verified with supportive numerical examples, where the expected convergence rates are obtained. This includes the approximation of PDEs on domains bounded by implicit and explicit (NURBS) curves, illustrating a direct integration between   geometric modeling and   numerical analysis.
 \end{abstract}


\end{frontmatter}

\section{Introduction}
\label{section:introduction}

The numerical approximation of partial differential equations (PDEs) is a crucial task in many fields of science and engineering.
Over the last years, the most widely used methodology  has been the finite element method (FEM), based on the Galerkin method and the weak form of the PDEs, which is compatible with $C^0$ basis functions.
However, if an approximation of the solution is expressed as a linear combination of smoother basis functions, then collocation approaches are also possible, which consist in the imposition of the strong form of the governing equations  in a set of isolated  points called collocation points. Since no numerical integration is involved, this can introduce several computational and implementation advantages.

A number of approximation methods rely on smooth basis functions and are then compatible with collocation. A notable example is isogeometric analysis (IGA) \cite{cottrell2009isogeometric} whose key idea is to employ spline functions both for the geometry representation and for the numerical analysis. These functions possess a higher inter-element continuity than FEM, which makes them suitable for collocation methods \cite{Auricchio2010,Schillinger2013b,Reali2014}.
IGA collocation has indeed been applied to a variety of fields \cite{Kruse2015,Gomez2014,Marino2016,Kiendl2017,Maurin2018f} 
and ongoing research on the topic  is very promising \cite{Gomez2016a,Montardini2017}.
However, the tensor product structure of B-splines and NURBS (non-uniform rational B-splines) basis functions introduces an intrinsic limitation in volume meshing and local refinement, which can be only partially avoided with alternative spline functions \cite{scott_thesis,Wei2016}.

Another family of approximation schemes with smooth basis functions are meshfree (or meshless) methods  \cite{Belytschko96,Huerta04,li2002meshfree,Chen2015}. 
As the name suggests, these schemes  do not rely on a mesh or a structured grid to construct the numerical approximation, which makes them particularly suited for problems involving large deformations.
Thanks to the high continuity of their  basis functions, most meshfree methods have been used both in Galerkin and collocation approaches. Nodally integrated formulations have also been  developed as an alternative integration strategy \cite{Chen2001}.

In the context of meshfree collocation, many methodologies have been successfully applied to different types of problems and  an exhaustive list  is beyond the scope of this paper.
The Smooth Particle Hydrodynamics \cite{Benz1990} can be considered as the first  meshfree method and is based on collocation. Other schemes that have been used with collocation are Multiquadrics least square methods          \cite{Kansa1990}, the Reproducing Kernel Particle Method   \cite{Aluru2000,Hu2011a}, the Radial Basis Functions  \cite{Yao2007,Zhang2000b} and the Finite Point Method     \cite{Onate1996a}. 
The interested reader is referred to \cite{Garg2018} and to \cite{Hillman2018a} for a more detailed classification.

This paper focuses on the development of a collocation framework based on \textit{maximum-entropy} (max-ent) basis functions. Max-ent methods were firstly introduced in computational mechanics to compute polygonal interpolants in          \cite{Sukumar04} and then they were extended to meshfree approximation schemes in     \cite{Arroyo2006}.
The underlying idea behind these methods is to consider an equivalence between basis functions and probability distributions and  to maximize an entropy functional in order to compute the most unbiased basis functions, subject to reproducibility constraints.
Obtained through an optimization procedure, these functions are able to  reproduce constant and linear fields.

Since these pioneering works, it was noted that in order to impose higher order consistency conditions, the reproducibility constraints need  to be carefully designed for feasibility. This allowed the definition of quadratic maximum entropy schemes with non-negative basis functions    \cite{Cyron,rosolen2013second} and also the definition of high-order local maximum entropy schemes (HOLMES) with signed functions \cite{bompadre2012convergent}. 
In these schemes, although the positivity of the basis functions is lost, their $C^\infty$ continuity is preserved and an excellent accuracy is obtained in different applications, such as Helmholtz and shell problems \cite{bompadre2012convergent}.
\red{Remarkably, even if the evaluation of the basis functions requires the iterative solution of a nonlinear optimization problem, the computational times are comparable to those required for other meshfree methods. 
	In addition, 
	 signed basis functions are needed for high-order reproducibility conditions also for other methods, although the negative part can be reduced to some extent \cite{Wang2014a}.}

Max-ent methods have been successfully applied to a variety of fields using the standard Galerkin formulation with a background integration mesh \cite{ortiz2010maximum,amiri2014phase,Millan10,cyron2010stable,Greco2016} but only a limited amount of research has been done on their use in  a collocation framework \cite{Perazzo2017,Fan2018c,Fan2019}. This previous literature on max-ent collocation is based on the linear consistent schemes developed in \cite{Arroyo2006}. 
However, examination of the literature on meshfree collocation methods suggests that at least quadratic consistency of the basis functions  \cite{Wang2018a,Hu2009a} is required for the method to converge in second-order partial differential equations. If $n$ is the order of the basis functions, this gives an order of convergence of $n$ for even orders   and $n-1$ for odd orders, which also follows the analogous IGA results     \cite  {Auricchio2010}.
Starting from this consideration,   this work focuses on the development of a collocation framework based on HOLMES approximants, which provide arbitrary orders of consistency, are smooth and locally refineable.

In addition, this work explores the application of  the collocation approach  to   PDEs on domains whose boundary has a smooth geometric definition, given either by an implicit description or by a NURBS curve.
This second case is particularly relevant because it allows a tight integration between CAD information and numerical analysis.
Furthermore, in contrast to IGA collocation, arbitrary shapes can be easily discretized with the proposed meshfree collocation method --- see
  \cite{Mirfatah2019a} for related work based on exponential basis functions. 
 
The outline of the paper is the following:
in Section \ref{section:collocaiton} the point collocation framework is described and its advantages with respect to the Galerkin formulation are discussed;
in Section \ref{section:holmes} the HOLMES approximants are introduced and their properties are briefly discussed;
then the accuracy of the proposed methodology is proved with supportive numerical examples in Section \ref{section:numerical_examples}; finally, in Section \ref{section:conclusions},  concluding remarks are given and possible future developments are discussed.

\section{Formulation of the point collocation framework}
\label{section:collocaiton}

Given a computational domain $\Omega$ with boundary $\Gamma$, the strong form of a boundary value problem can be expressed as
\begin{equation}
\label{pde}
\begin{cases}
\mathcal{L} \textbf{u} + \textbf{b}  = \textbf{0} & \text{in } \Omega \\ 
\mathcal{B}^g \textbf{u} = \textbf{g} & \text{on } \Gamma^g \\ 
\mathcal{B}^t \textbf{u} = \textbf{t} & \text{on } \Gamma^t,
\end{cases}
\end{equation}
where $\textbf{u}=\textbf{u}(\vx)$ is the unknown field, $\mathcal{L}$ is  a differential operator in $\Omega$, $\mathcal{B}^g$ is an operator on the Dirichlet part of the boundary $\Gamma^g$, $\mathcal{B}^t$ is a differential operator on the Neumann part of the boundary $\Gamma^t$, $ \textbf{b}$ is a source term and $ \textbf{g}$ and  $ \textbf{t}$ are the imposed boundary conditions.

In contrast to the Galerkin formulation, in the point collocation framework the   strong form of the problem is directly enforced in a set of points, called collocation points, and the solution $\textbf{u}(\vx)$ is approximated as a linear combination of basis functions.
Different strategies are possible for the choice of the collocation points and have been considered in the literature of both isogeometric and meshfree collocation methods.
In the former case, it was shown that using suitable coordinates for the collocation points and using over-determined systems of equations leads to increased orders of convergence, equal to those of the Galerkin formulation \cite{Gomez2016a,Montardini2017}.
Given the tensor product structure of IGA basis functions, these suitable coordinates can be easily identified in any dimension. The situation is different for meshfree formulations applied to unstructured points distributions, where the most natural choice for the collocation points are the nodes, which are adopted in this work. We do not pursue here the identification of optimal or privileged sets of collocations points because of the unstructured nature of the discretization that characterizes meshfree methods.

For a vectorial field $\textbf{u}(\vx): \Omega \rightarrow \Re^d$, the numerical
  approximation is expressed as
\begin{equation}\label{eq:uh}
\textbf{u}^h(\vx)= \sum_{a=1}^m \phi_a(\vx) \textbf{u}_a,
\end{equation}
where $\phi_a(\vx)$ is the basis function associated with the node $\vx_a$ \red{($a = 1, 2, \ldots, m$)}  and $\textbf{u}_a$ is a vector of coefficients. 
Using this approach and enforcing the strong form equation at the nodal coordinates, a linear system is obtained:
\begin{equation}
\textbf{K} \textbf{d}    = \textbf{f},
\end{equation}
where  $\textbf{K}$, $\textbf{d}$ and  $\textbf{f}$ are often called stiffness matrix, displacement and force vector, although they have another meaning with respect to the Galerkin formulation. Their expression is given by:
\begin{equation*}
\textbf{K}   =\left[
\begin{array}{c}
\textbf{K}^i \\ 
\textbf{K}^g \\ 
\textbf{K}^t \\ 
\end{array}\right],
\;\;\;
\textbf{f}   =\left[
\begin{array}{c}
\textbf{f}^i \\ 
\textbf{f}^g \\ 
\textbf{f}^t \\ 
\end{array}\right]
\end{equation*}
 
\begin{equation}
\begin{cases}
\textbf{K}^i_{ab}=\mathcal{L} [\phi_b(\vx_a)], &\textbf{f}^i_{a}=\textbf{b}(\vx_a) \\ 
\textbf{K}^g_{ab}=\mathcal{B}^g [\phi_b(\vx_a)], &\textbf{f}^g_{a}=\textbf{g}(\vx_a) \\ 
\textbf{K}^t_{ab}=\mathcal{B}^t [\phi_b(\vx_a)], &\textbf{f}^t_{a}=\textbf{t}(\vx_a) 
\end{cases}
\end{equation}
and the numerical implementation is simple and straightforward.

\begin{MyColorPar}{red}

\subsection{Computational advantages with respect to the Galerkin formulation}
Given the large support of max-ent basis functions, the point collocation framework introduces   significant computational advantages with respect to the Galerkin formulation, which can be summarized by the two following points:
\begin{itemize}
\item the basis functions are evaluated in a much smaller number of points;
\item the stiffness matrix filling procedure is straightforward with no memory bottlenecks;
\end{itemize}

The first point results from the fact that the number of quadrature points in the Galerkin formulation is in general higher than the number of nodes. Table~\ref{tab2d} summarizes the expected ratios of quadrature points to nodes for the integration rules of 1\textsuperscript{st} to 5\textsuperscript{th} order given in \cite{zhang2009set}, calculated for infinite regular triangular and tetrahedral  meshes.  
Although the optimal number of quadrature points required for the weak form integration depends on the specific application and has not been studied in detail yet, it can be assumed that higher order basis functions require integration rules of higher polynomial order.
This may lead for instance to a ratio of 84 quadrature points per node when using a 4\textsuperscript{th} order rule in three dimensions.
In addition, the Galerkin formulation requires in each   point the multiplication of the basis functions derivatives of each pair of neighbors to the point, whose number is in the order of several hundred in three dimensions.

While the evaluation of the contribution of each integration point is by itself very computationally demanding, in general, this part is  not  the bottleneck in Galerkin implementations. In fact, when dealing with large scale problems and sparse matrix implementations, the matrix filling procedure becomes particularly challenging due to memory bottlenecks \cite{Peco2015}. Instead, in  collocation methods the basis functions are evaluated at each collocation point  and the matrix is directly filled row by row.
Based on these considerations, a quantitative comparison of the computational times through  numerical examples should strongly depend on the details of the  Galerkin implementation  and is beyond the scope of this work.
 
We finally note that, as it has been argued in the context of IGA, the bandwidth of the collocation matrix is smaller than that of the Galerkin matrix \cite{Auricchio2010,Schillinger2013b}.
 

 \begin{table}[h] 
 	\centering
 	\caption{Number of quadrature points for Galerkin implementations of max-ent methods integrated over triangles (left) and tetrahedra (right).}
 	\label{tab2d}
 	
 \begin{tabular}{ cc } 

 	2D & 3D \\

  	\begin{tabular}{|c|c|c|c|c|c|}
 	\hline
 	Order  	&  1 & 2 & 3 & 4 & 5  \\ \hline
 	Points per triangle		&  1 & 3 & 6 & 6 & 7 \\ \hline
 	Points per node	&  2 & 6 & 12 & 12 & 14 \\ \hline
 \end{tabular}
&
 	\begin{tabular}{|c|c|c|c|c|c|}
	\hline
	Order  	&  1 & 2 & 3 & 4 & 5  \\ \hline
	Points per tetrahedron		&  1 & 4 & 8 & 14 & 14 \\ \hline
	Points per node	& 6 & 24 & 48 & 84 & 84 \\ \hline
\end{tabular}

 \end{tabular}

 \end{table}

%
%
%
%
%

\end{MyColorPar}

\section{The HOLMES approximants}
\label{section:holmes}

Maximum entropy basis functions were firstly developed by Sukumar \cite{Sukumar04}, who was inspired by the max-ent formalism introduced by Jaynes  in statistical mechanics \cite{jaynes1957information}.
In this approach, an entropy functional that depends on a discrete probability measure $\{p_a\}_{a=1}^m$ is maximized, subject to linear constraints on $p_a$. 
On noting the correspondence between basis functions $\{\phi_a\}_{a=1}^m$ and discrete probability measures $\{p_a\}_{a=1}^m$, the max-ent formalism was adopted for the construction of polygonal interpolants \cite{Sukumar04}. 
An extension to meshfree approximation schemes was introduced with the local maximum-entropy (LME) basis functions, developed by Arroyo and Ortiz in
\cite{Arroyo2006}.
The main idea of LME schemes is to  consider a Pareto optimum between the maximization of the entropy functional  and the   locality of the basis functions. 
Given a set of distinct nodes in $\Re^d$, located at $\vx_a$,
and being  $\Omega = \textrm{con}(\vx_1, \dots, \vx_m) \subset \Re^d$  the convex hull of the nodal set, 
the associated formulation is to
find $\vx \mapsto \bsym{\phi}(\vx): \Omega \to \Re_+^m$  as the
solution of the following optimization problem:
\begin{equation}
\label{LME}
\min_{\bsym{\phi} \in \Re_+^m} 
\sum_{a=1}^m  \phi_a(\vx) \ln \phi_a(\vx)  +\beta \sum_{a=1}^m   \phi_a(\vx) |\vx-\vx_a|^2,
\end{equation}
subject to the linear reproducing conditions:
\begin{equation}
	\label{rep_constr}
	\begin{aligned}
	\sum_{a=1}^m \phi_a(\vx) &= 1, \\
	\sum_{a=1}^m \phi_a(\vx) ( \vx_a - \vx) &= \vm{0},
	\end{aligned}
\end{equation}
where $\Re_+^m$ is the non-negative orthant. In this formulation, the shape and the locality of the basis functions is controlled by the Pareto parameter $\beta$, which is normally replaced by a dimensionless parameter $\gamma=\beta h^2$, being $h$ a measure of the node spacing.
In practice,  LME schemes correspond to the polygonal interpolants of  \cite{Sukumar04} for $\gamma=0$, while it can be proved that they recover  linear Delaunay FEM when  $\gamma \to  \infty$.

The computation of this type of approximants with high-order consistency conditions is not straightforward, since adding the corresponding constraints to the ones in Eq.~\eqref{rep_constr} may lead to an unfeasible problem, even for   quadratic constraints \cite{Arroyo2006}.
This motivated several attempts to develop quadratic max-ent schemes, that include enrichment \cite{Greco2013} following the partition-of-unity framework \cite{Babuska1996_partition}, combination of the max-ent formalism with de Boor's algorithm \cite{Gonzalez2010} and relaxation of the constraints.
In this latter framework,   \cite{Cyron} and   \cite{rosolen2013second} propose feasible quadratic constraints whereas in \cite{bompadre2012convergent}   the positivity constraints are released to formulate the HOLMES approximants, following an idea previously suggested in \cite{Sukumar2007}.

In the HOLMES formalism, the basis functions $\phi_a(\vx)$ are expressed as the difference between two non-negative contributions 
\begin{equation}
\phi_a(\vx)=\phi_a^+(\vx)-\phi_a^-(\vx).
\end{equation}
The entropy of $\phi_a$ cannot be computed since the logarithm is only defined on  $\Re^+$. However, a modified entropy functional is defined as:
\begin{equation}
\label{ent_mod}
H(\bsym{\phi^+},\bsym{\phi^-})=- \sum_{a=1}^m  \phi_a^+(\vx) \ln \phi_a^+(\vx) - \sum_{a=1}^m  \phi_a^-(\vx) \ln \phi_a^-(\vx),
\end{equation}
which is no longer the entropy of a probability distribution, like the one used in \cite{Sukumar04}.
However, on using such a functional, problem \eqref{LME} is generalized to arbitrary orders  of consistency $n$ in the following way:
\begin{equation}
\begin{aligned}
\label{HOLMES}
\min_{\bsym{\phi}^+,\bsym{\phi}^- \in \Re_+^m} \;\;\;
\sum_{a=1}^m  \phi_a^+(\vx) \ln \phi_a^+(\vx) + \sum_{a=1}^m  \phi_a^-(\vx) \ln \phi_a^-(\vx)+\\
+\dfrac{\gamma}{h^p} \sum_{a=1}^m   \left( \phi_a^+(\vx)-\phi_a^-(\vx)\right)  |\vx-\vx_a|_p^p
\end{aligned}
\end{equation}
subject to
\begin{equation}
	\label{n_constr}
	\begin{aligned}
	\sum_{a=1}^m \left( \phi_a^+(\vx)-\phi_a^-(\vx)\right) &= 1, \\
	\sum_{a=1}^m \left( \phi_a^+(\vx)-\phi_a^-(\vx)\right) ( \vx_a - \vx)^{\bm{\alpha}} &= 0, \;\;\;\forall \bm{\alpha} \in \mathbb{N}^d, \; 1\leq |\bm{\alpha}| \leq n.
	\end{aligned}
\end{equation}
In the equation above, the \textit{multi-index} notation adopted in \cite{bompadre2012convergent} is followed, where  $\bm{\alpha}$ is a vector in $\mathbb{N}^d$ and  $|\bm{\alpha}|=\sum_{i=1}^{d}\alpha_i$ is the order of $\bm{\alpha}$.  At the same time,  for a given point $\tilde{\vx}\in \mathbb{R}^d$, the expression   $\tilde{\vx}^{\bm{\alpha}} $ denotes the product $ \Pi_{i = 1}^d \tilde{x}_i^{\alpha_i}$.  

In the definition of problem \eqref{HOLMES},  a \textit{p}-norm $|\vx|_p=( \sum_{i=1}^{d}|x|_i^p)^{1/p} $ is used, which generalizes the quadratic norm employed in \eq{LME} for LME schemes.
In fact, in the context of Galerkin formulations,  a sufficient condition for the convergence of HOLMES is $p>n$, which ensures  that the basis functions have a proper exponential decay \cite{bompadre2012convergent,bompadre2012convergence}. Hence, this sufficient condition suggests values of $p>2$ for higher-order max-ent schemes.
This aspect is discussed more in detail in Section~\ref{section:numerical_examples}, where the numerical examples show that such a condition is not necessary in the context of collocation.
We note that, although in most of the literature the polynomial order of the basis functions and the orders of convergence are referred to as \textit{p}, here we follow the notation in \cite{bompadre2012convergent}, where \textit{p} refers to the \textit{p}-norm whereas \textit{n} refers to orders of consistency.

\begin{figure}  
	\centering
	\subfigure[$n=2,p=2,\gamma=1.521  \;(\hat{R}=4)   $]
	{\includegraphics[width=0.32\textwidth,clip,keepaspectratio,angle=0]{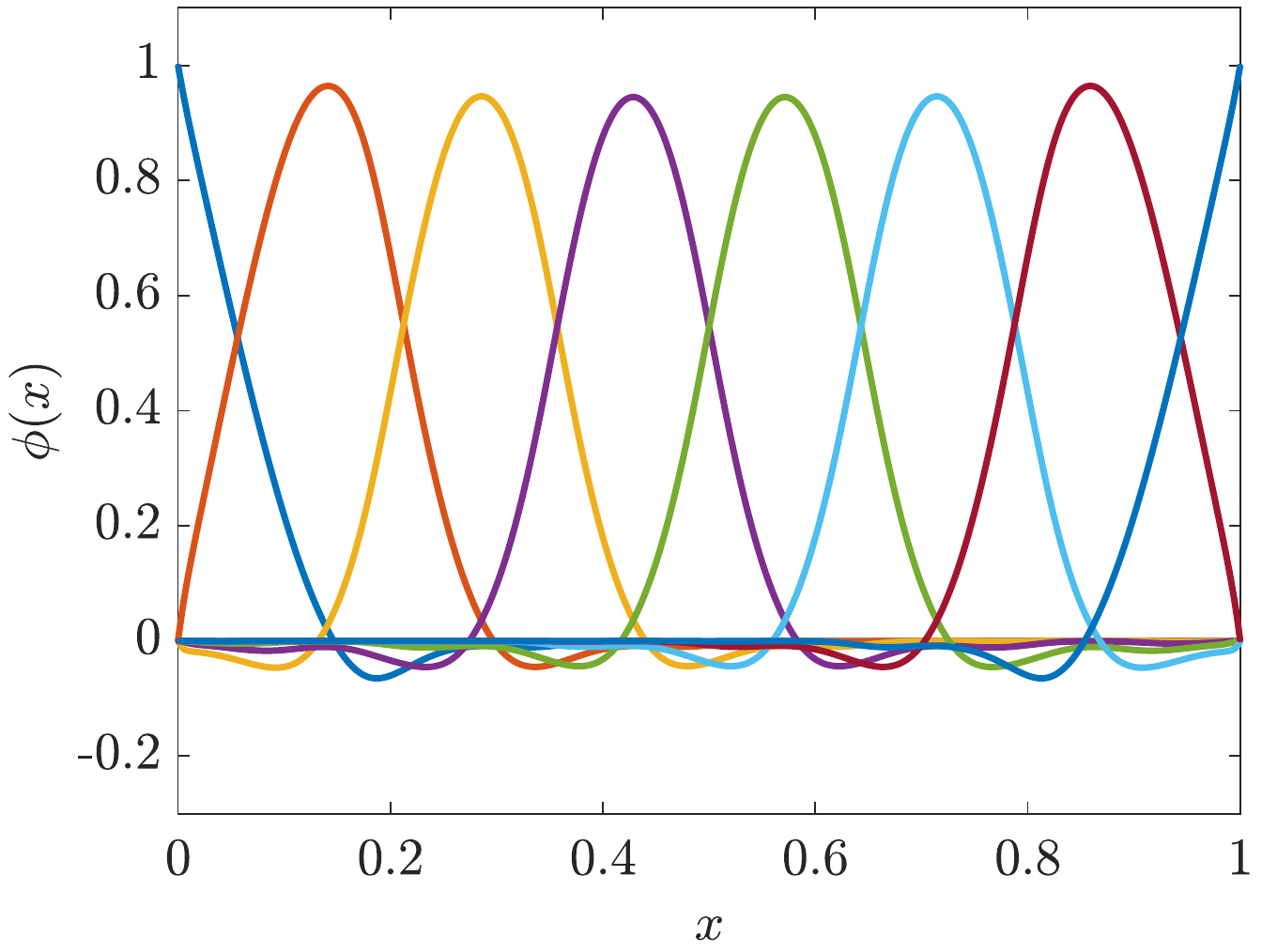}}
	{\includegraphics[width=0.32\textwidth,clip,keepaspectratio,angle=0]{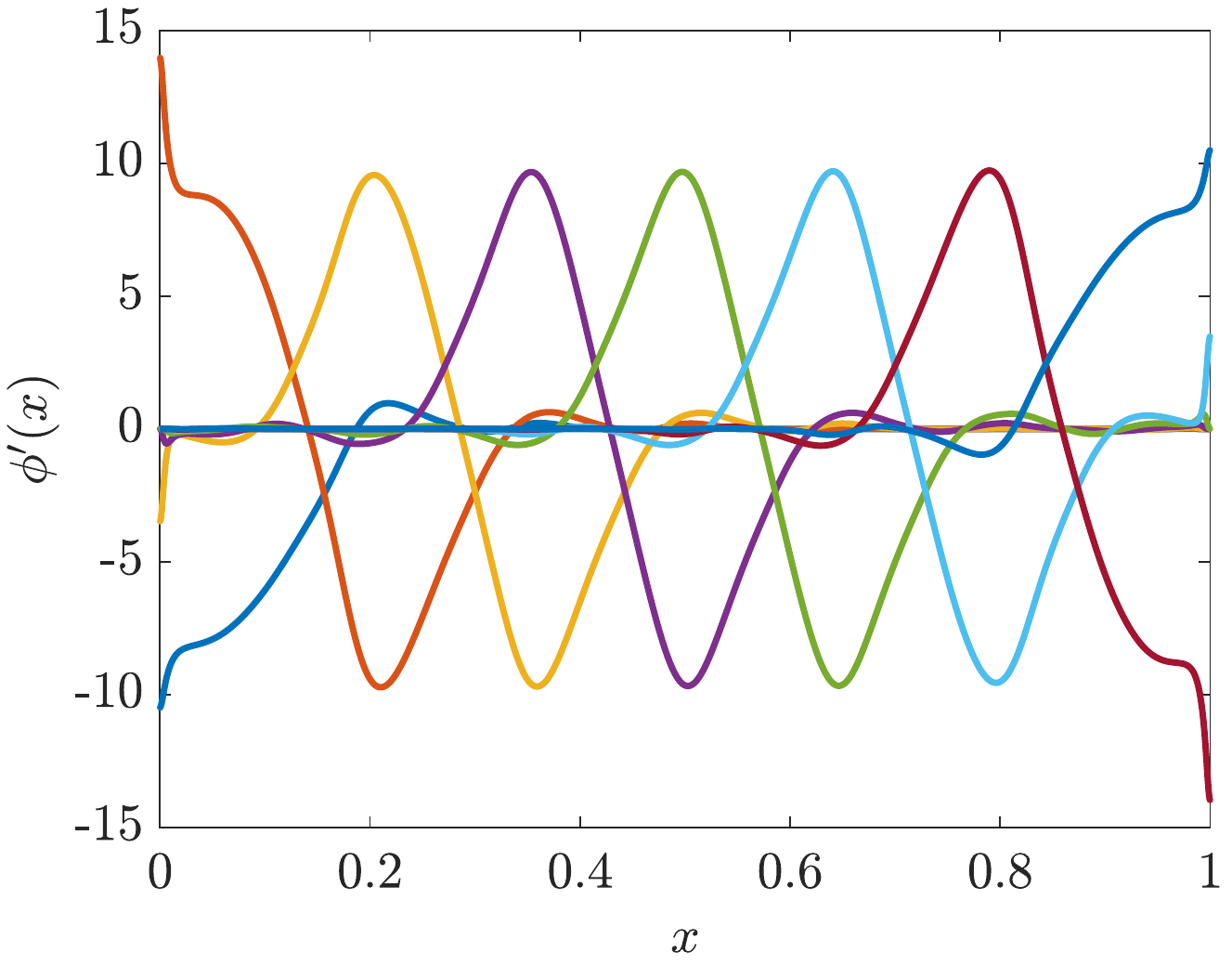}}
	{\includegraphics[width=0.32\textwidth,clip,keepaspectratio,angle=0]{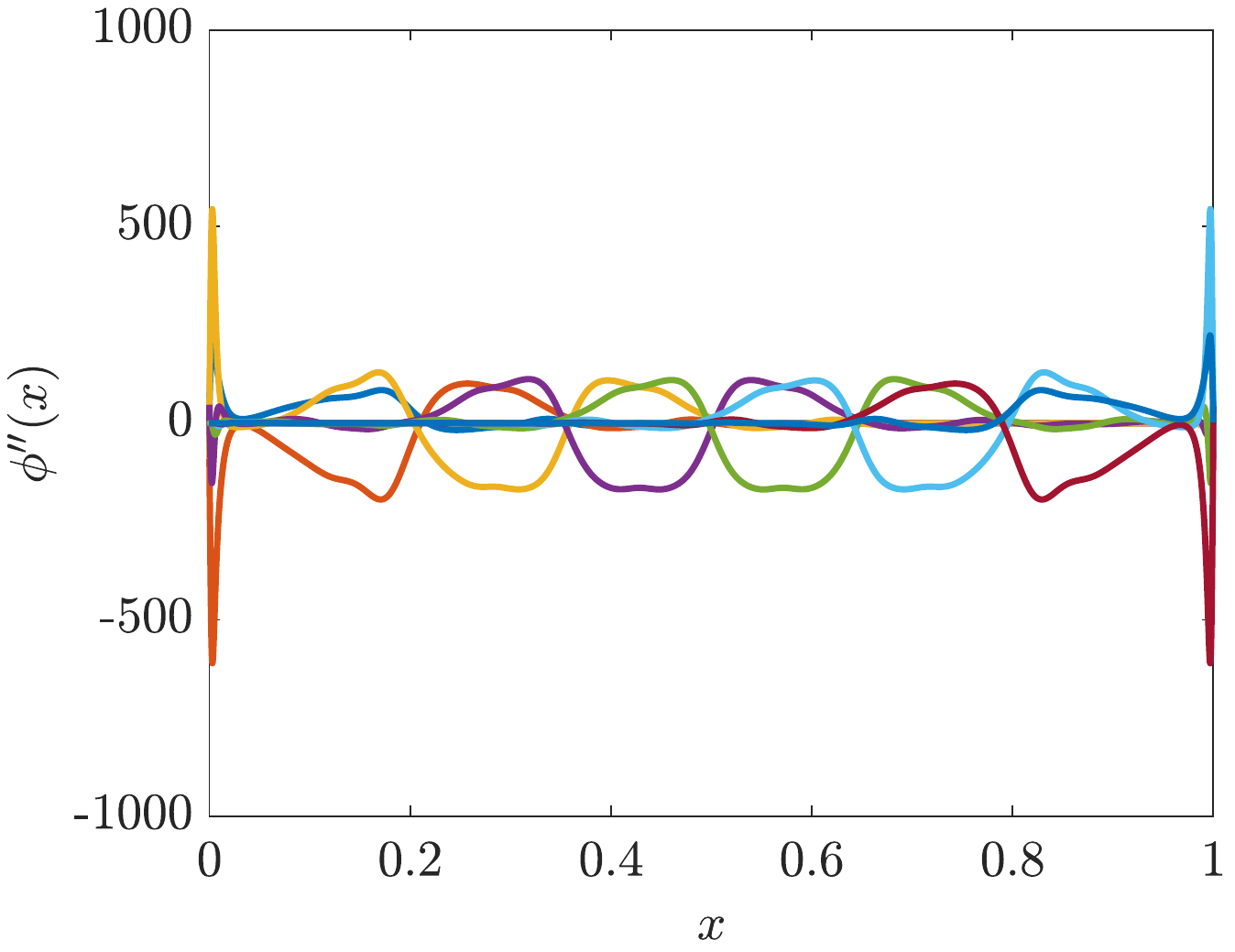}}
	\subfigure[$n=2,p=2,\gamma=0.676  \;(\hat{R}=6)  $]
	{\includegraphics[width=0.32\textwidth,clip,keepaspectratio,angle=0]{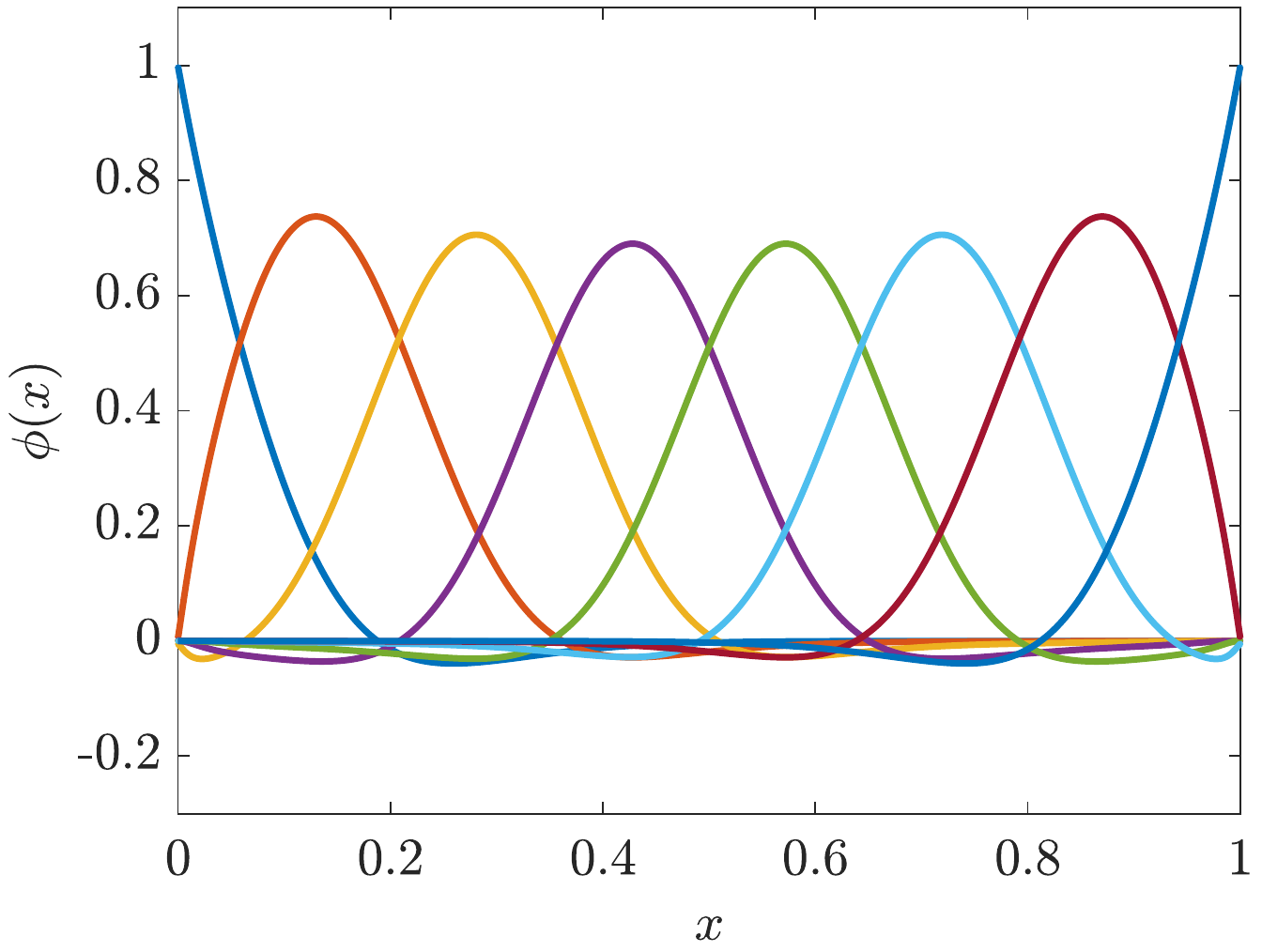}}	
	{\includegraphics[width=0.32\textwidth,clip,keepaspectratio,angle=0]{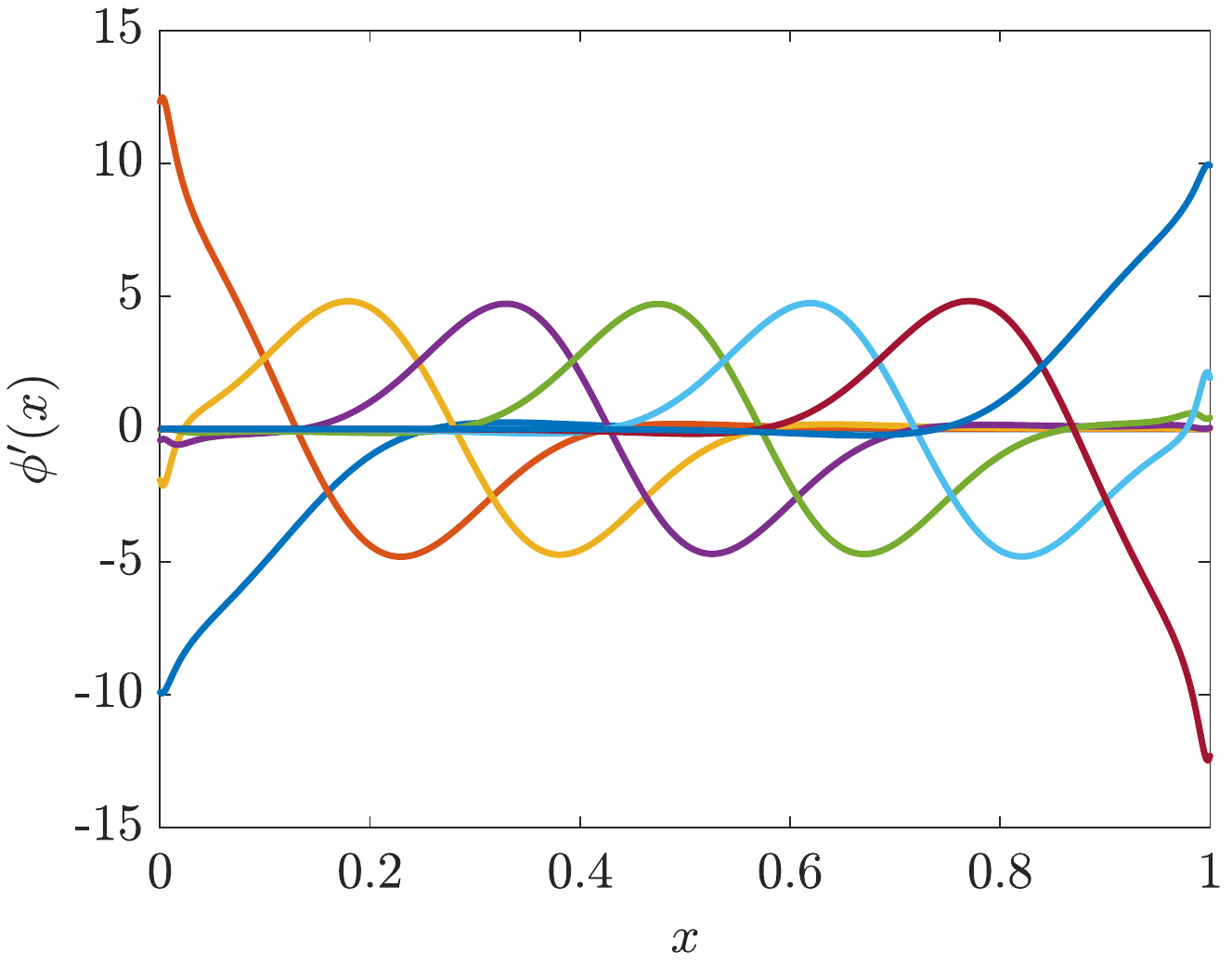}}
	{\includegraphics[width=0.32\textwidth,clip,keepaspectratio,angle=0]{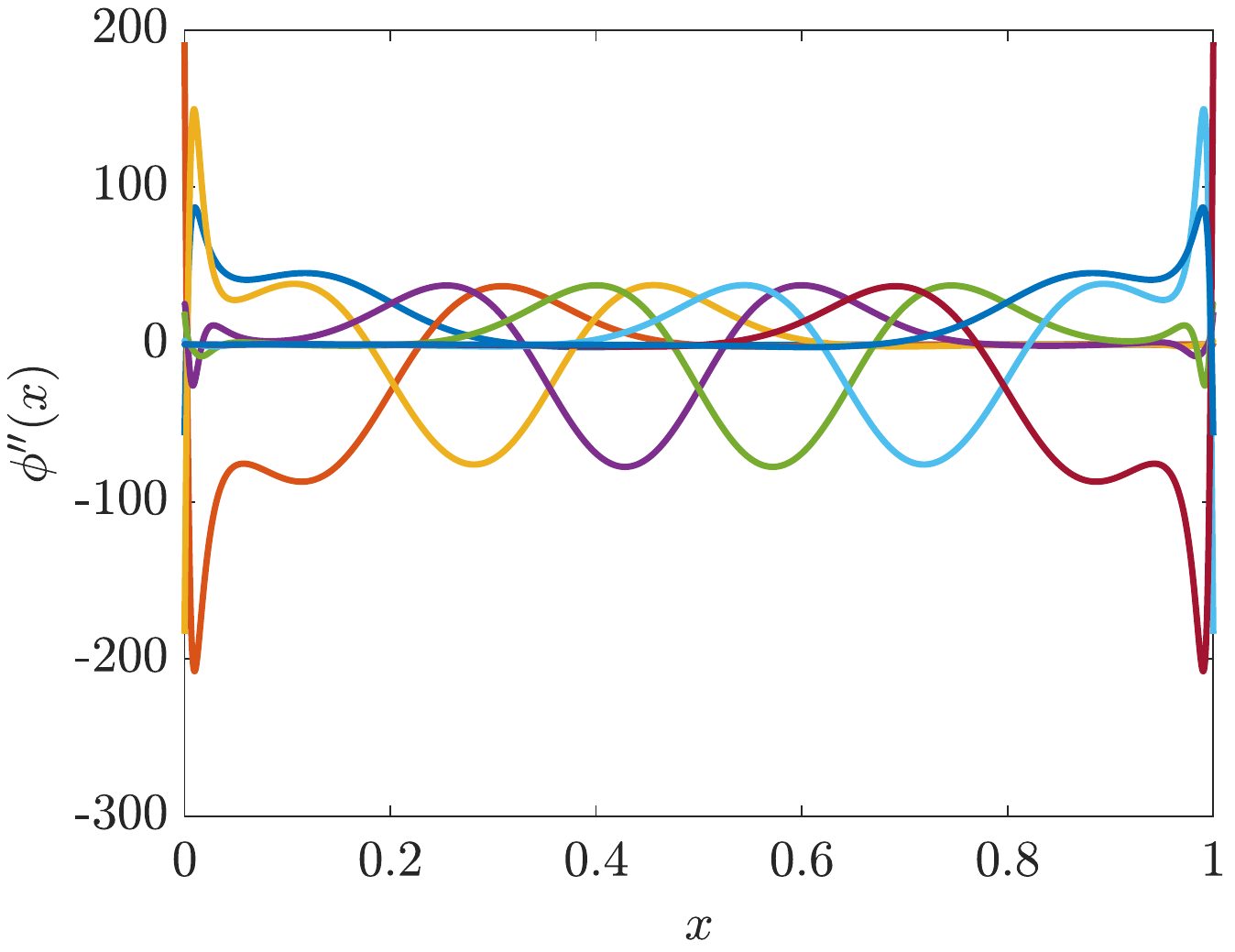}}
	\subfigure[$n=2,p=4,\gamma=0.095  \;(\hat{R}=4)  $]
	{\includegraphics[width=0.32\textwidth,clip,keepaspectratio,angle=0]{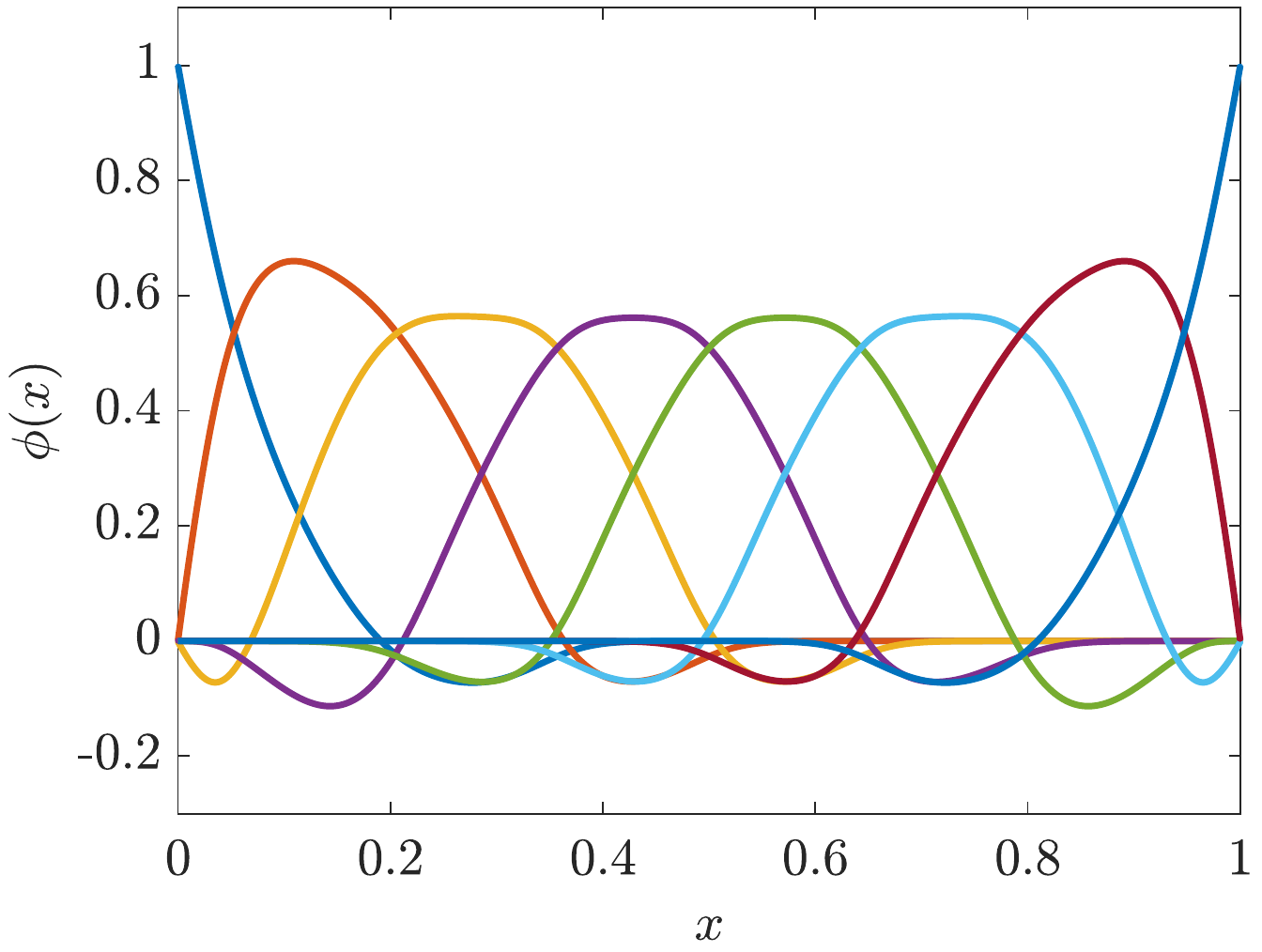}}	
	{\includegraphics[width=0.32\textwidth,clip,keepaspectratio,angle=0]{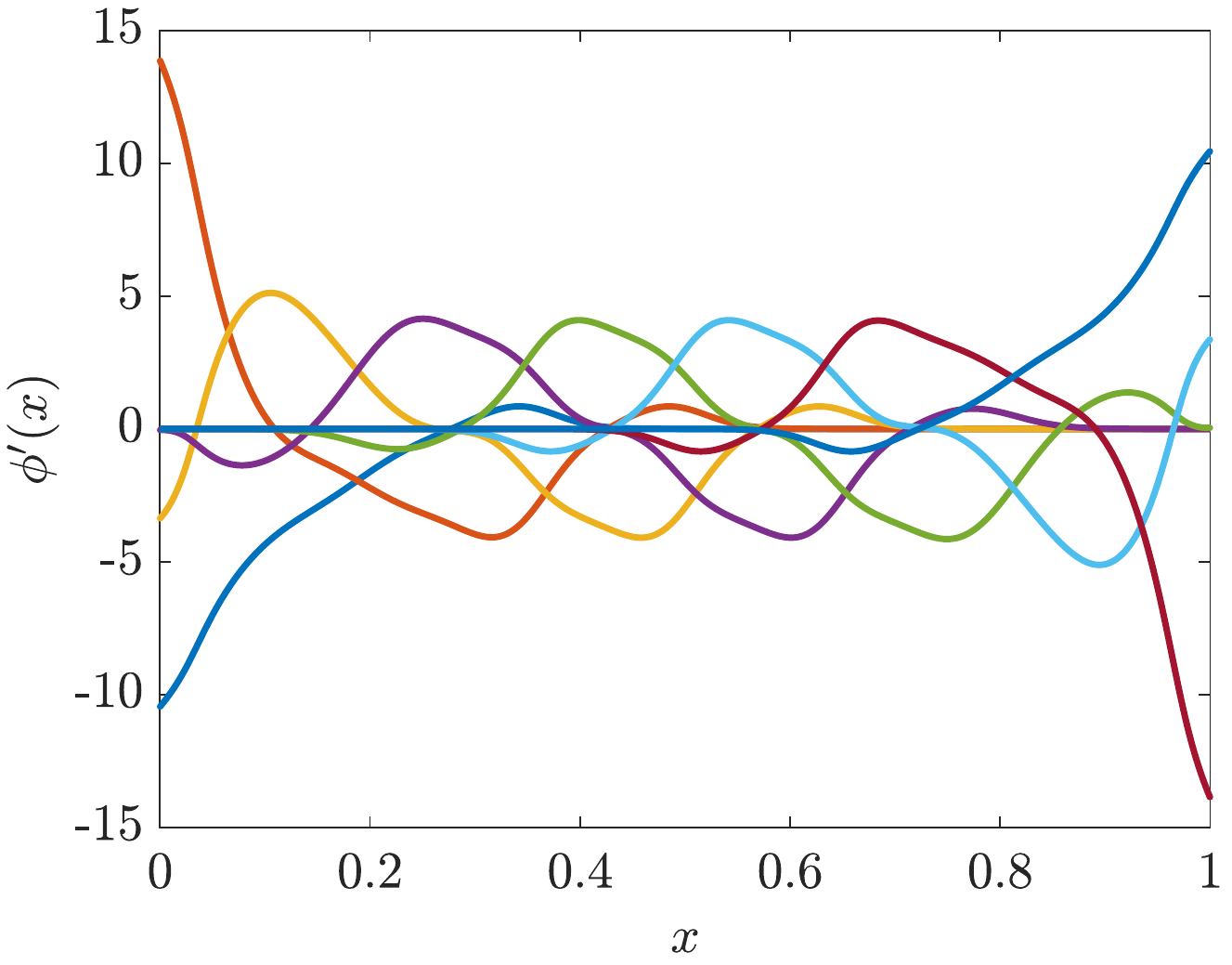}}
	{\includegraphics[width=0.32\textwidth,clip,keepaspectratio,angle=0]{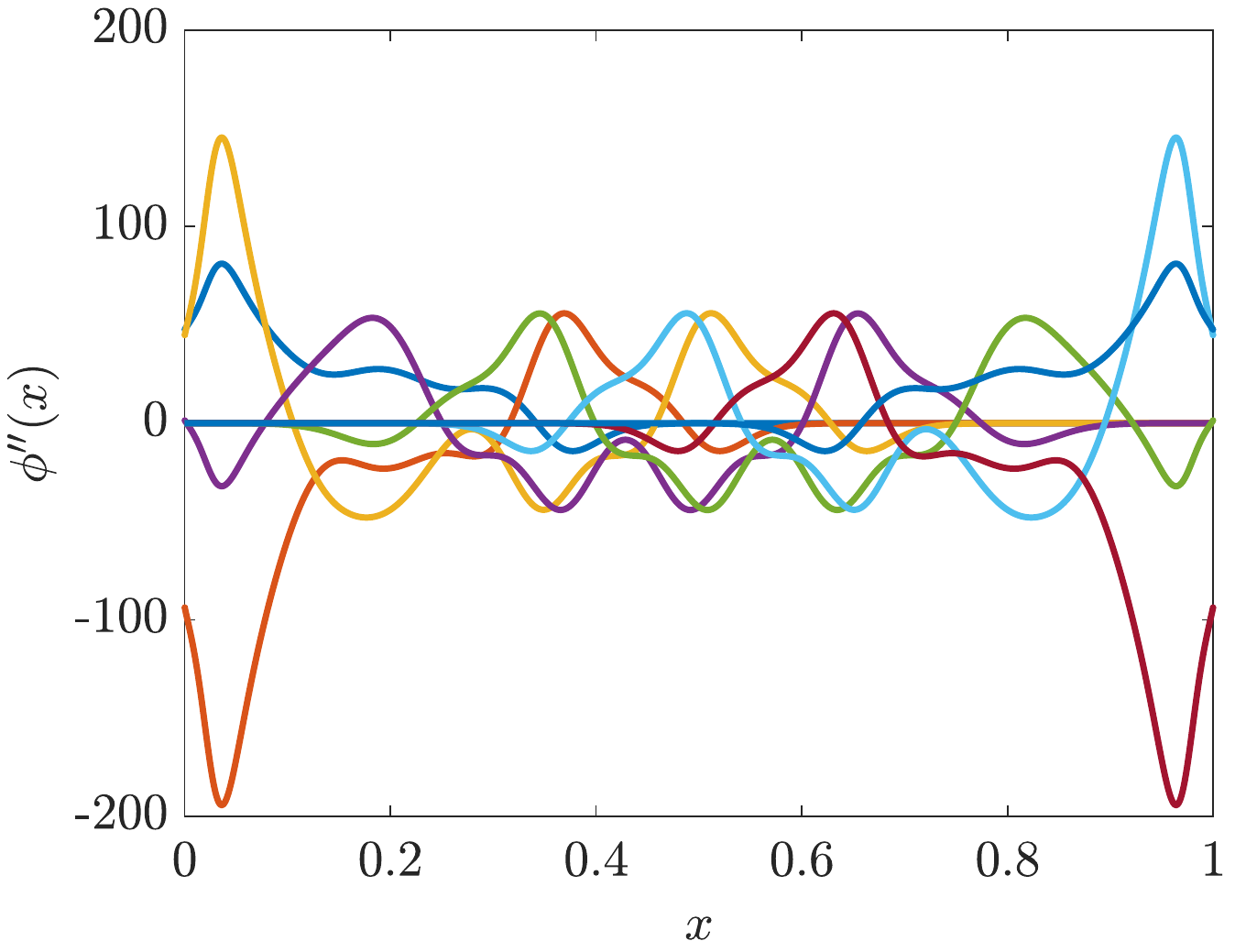}}
	\subfigure[$n=4,p=4,\gamma=0.006  \;(\hat{R}=6)  $]	{\includegraphics[width=0.32\textwidth,clip,keepaspectratio,angle=0]{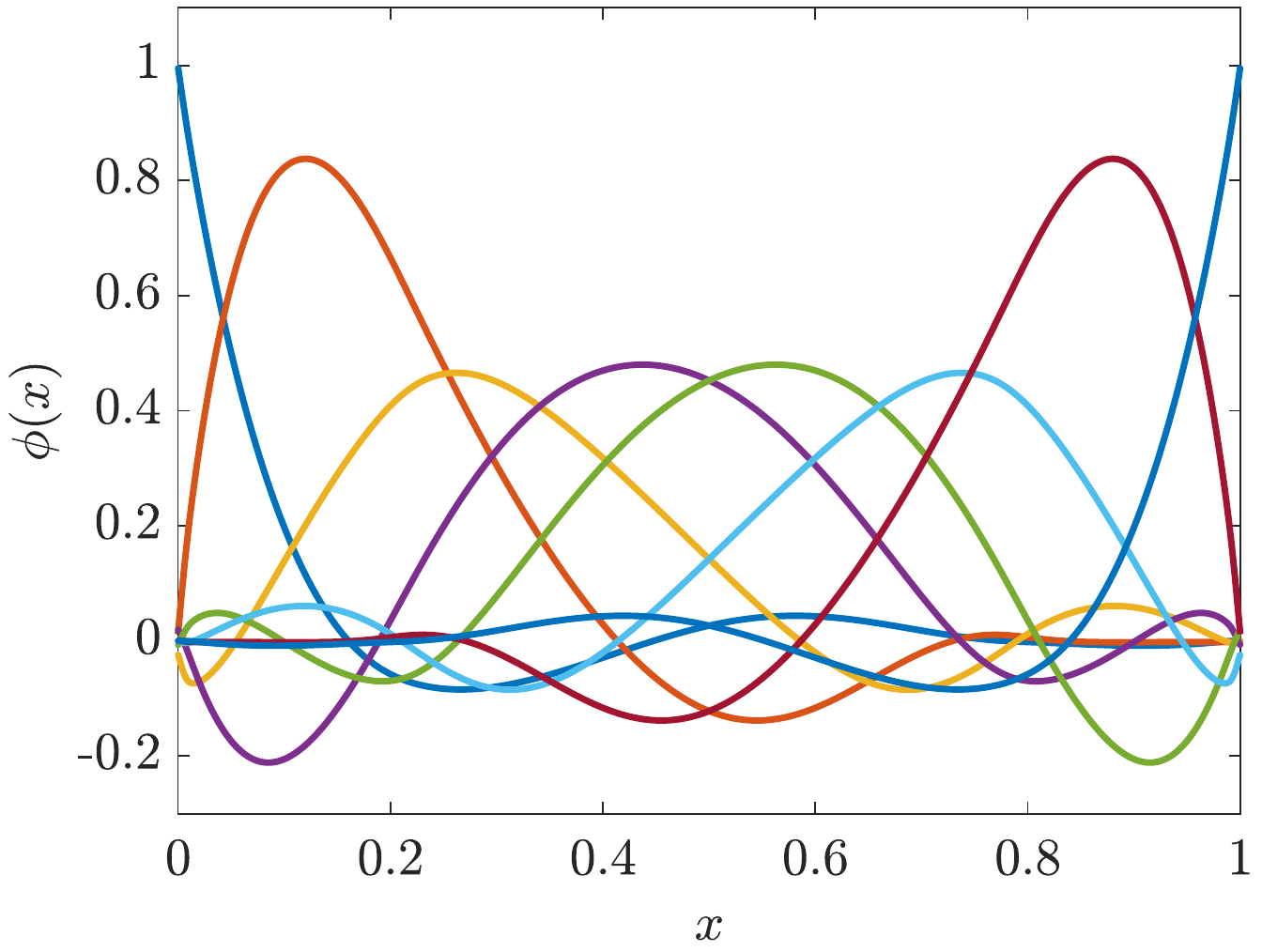}}	
	{\includegraphics[width=0.32\textwidth,clip,keepaspectratio,angle=0]{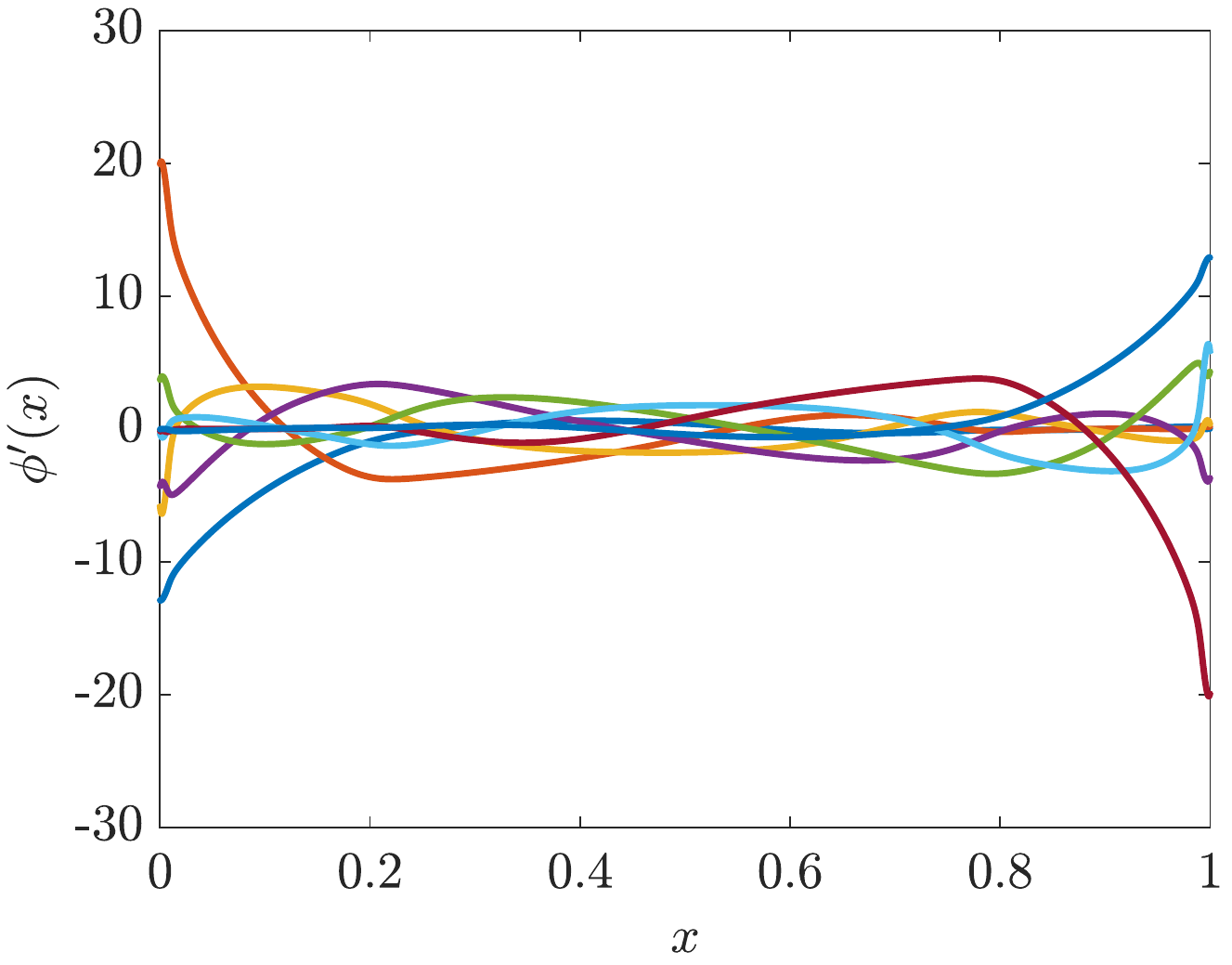}}
	{\includegraphics[width=0.32\textwidth,clip,keepaspectratio,angle=0]{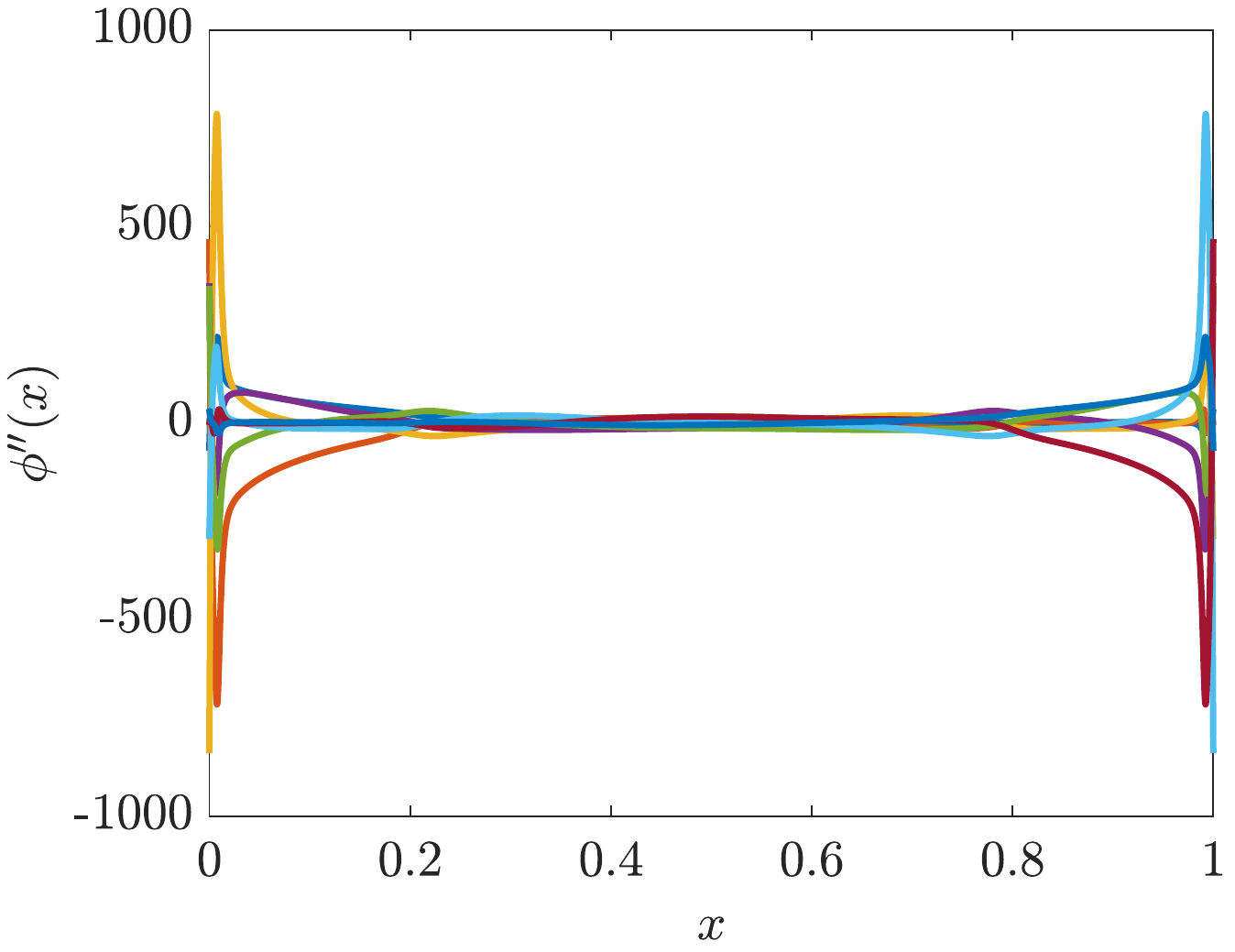}}	
	\caption{HOLMES basis functions in one dimension, computed for different orders $n$, different values of the locality norm $p$ and different estimated support sizes $\hat{R}$, for a truncation tolerance $\epsilon=10^{-11}$.	} 
	\label{fig:holmes_fi}
\end{figure}

\begin{figure}  
	\centering
	\subfigure[$n=2,p=2,\gamma=1.521  \;(\hat{R}=4)   $]
	{\includegraphics[width=0.49\textwidth,clip,keepaspectratio,angle=0]{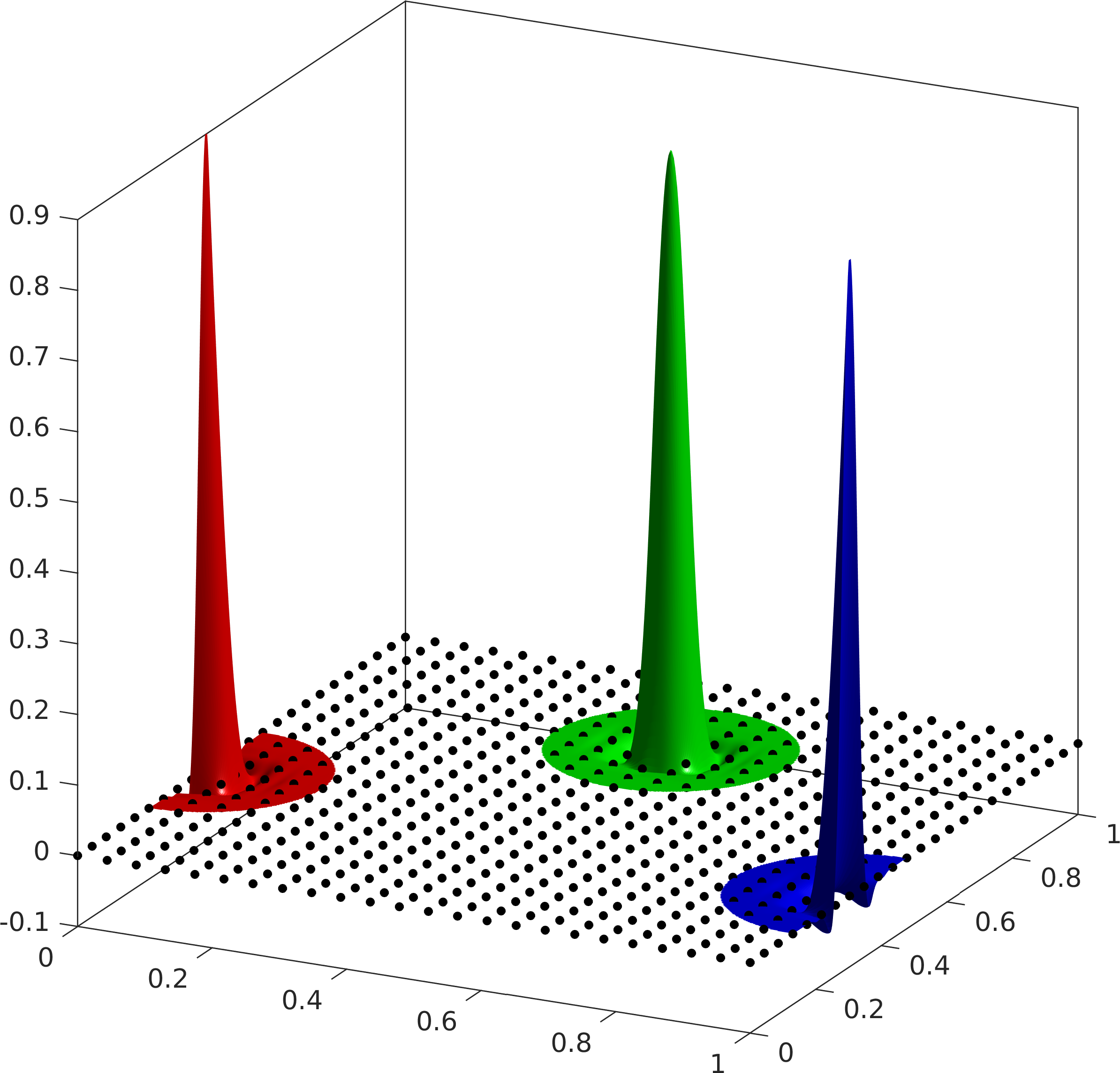}}
	\subfigure[$n=2,p=2,\gamma=0.676  \;(\hat{R}=6)  $]
	{\includegraphics[width=0.49\textwidth,clip,keepaspectratio,angle=0]{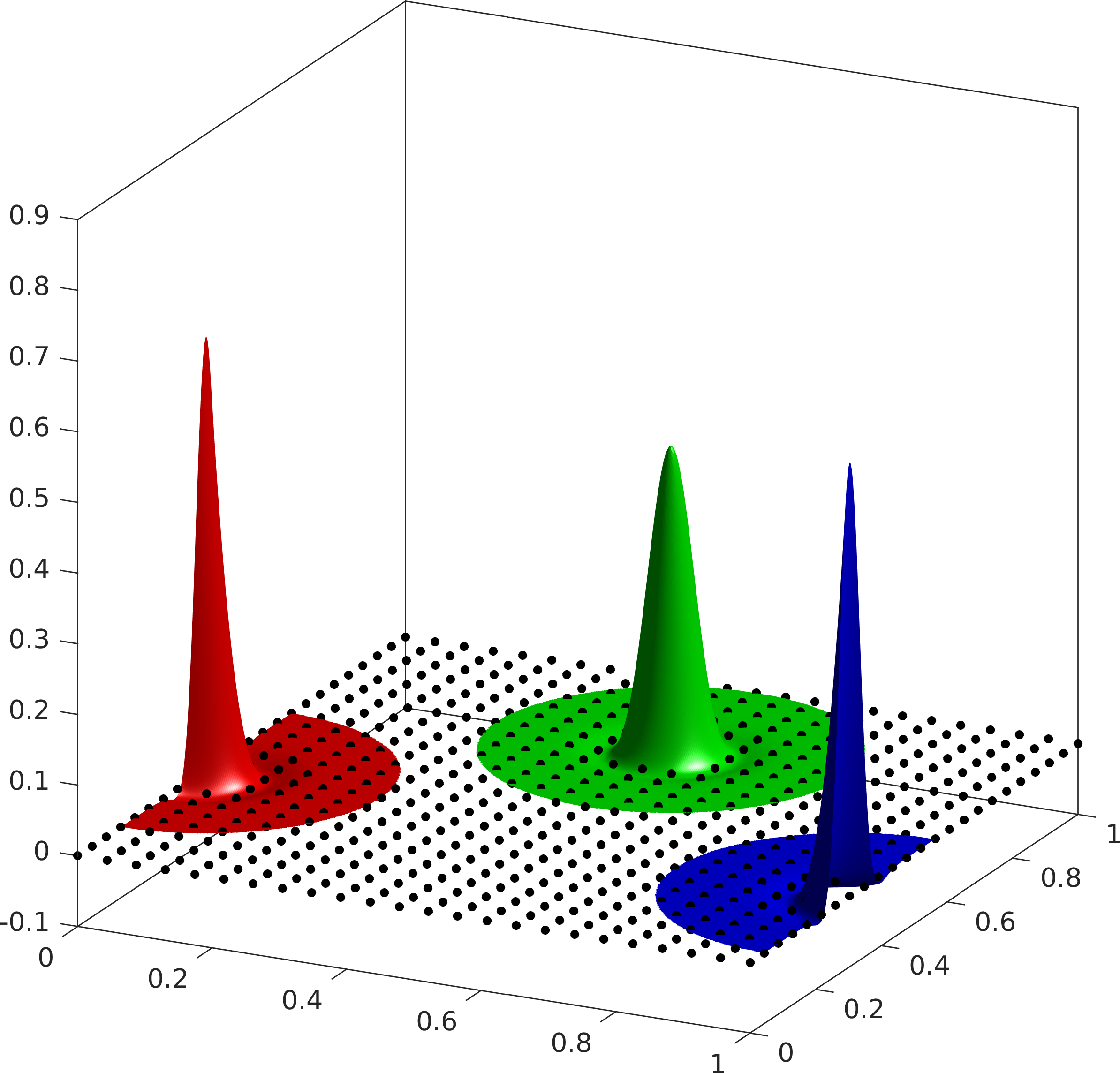}}	
	\subfigure[$n=2,p=4,\gamma=0.095  \;(\hat{R}=4)  $]
	{\includegraphics[width=0.49\textwidth,clip,keepaspectratio,angle=0]{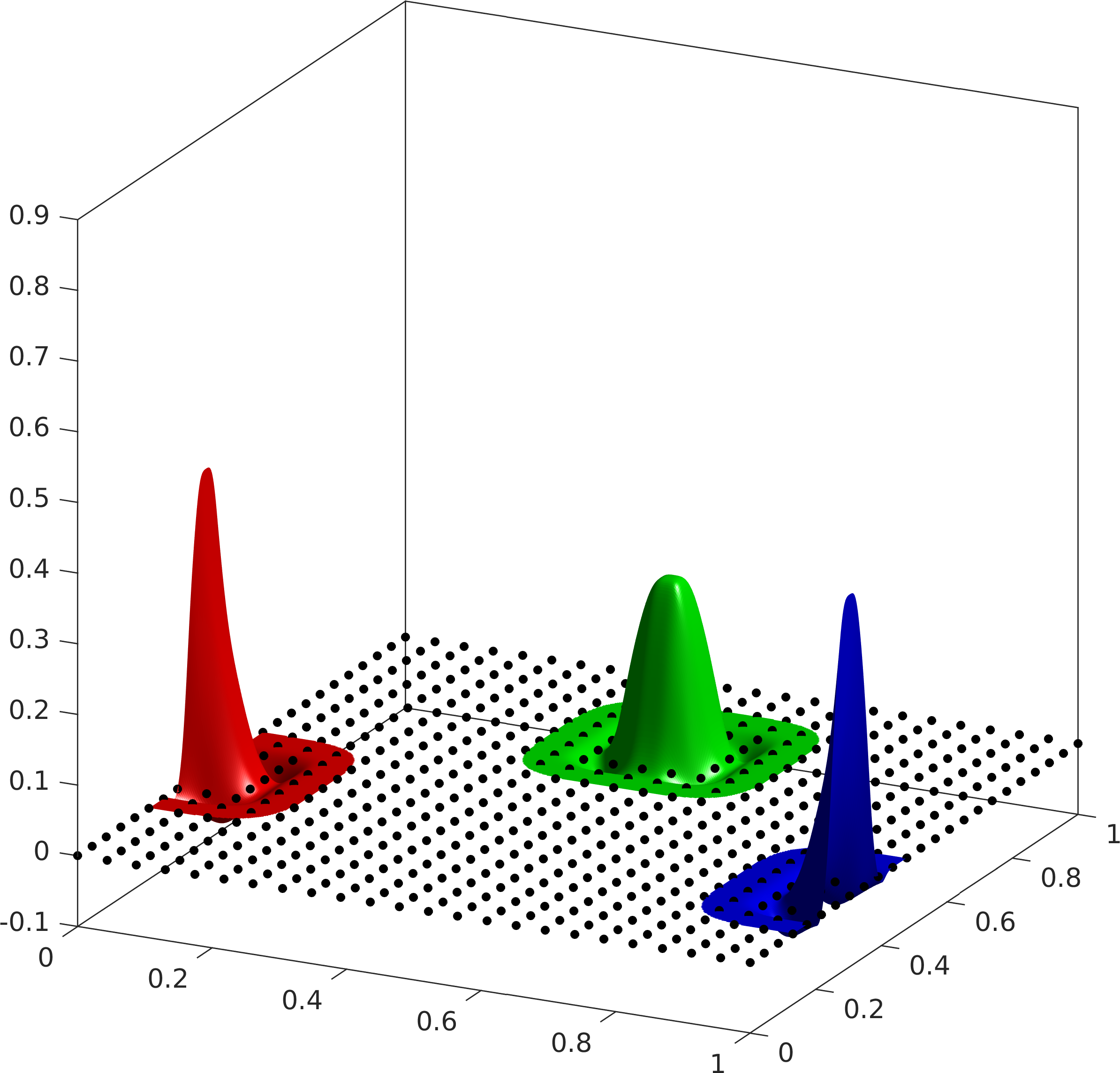}}	
	\subfigure[$n=4,p=4,\gamma=0.006  \;(\hat{R}=6)  $]	
	{\includegraphics[width=0.49\textwidth,clip,keepaspectratio,angle=0]{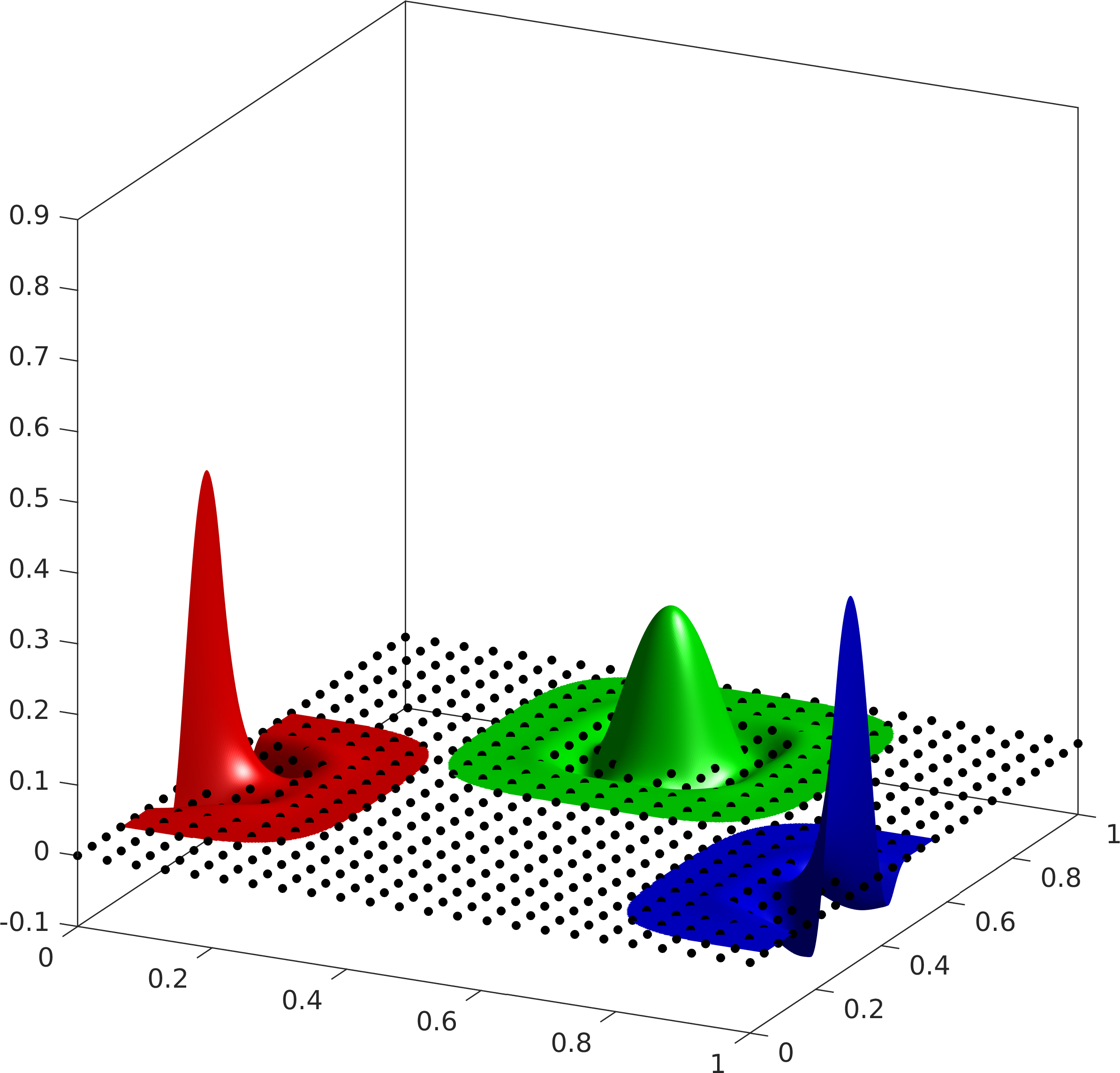}}	
	\caption{HOLMES basis functions in two dimensions, computed for different orders $n$, different values of the locality norm $p$ and different estimated support sizes $\hat{R}$, for a truncation tolerance $\epsilon=10^{-11}$. Note that the truncated support is circular only for $p=2$ and becomes square for $p \to \infty$. 	} 
	\label{fig:holmes_fi2d}
\end{figure}

The   HOLMES problem in \eqref{HOLMES} is solved by mean of duality methods, where  the computation of the Lagrange multipliers is performed in an efficient and robust way using the Newton-Raphson scheme.
The reader is referred to \cite{bompadre2012convergent} for more details about the dual formulation and  the computation of HOLMES first and second derivatives. 
A plot of HOLMES basis functions and their derivatives is given in Fig.~\ref{fig:holmes_fi} and Fig.~\ref{fig:holmes_fi2d}, for the one and two-dimensional case respectively. 
It can be observed how the locality of the basis functions depends both on $\gamma$ and $p$ and, in particular, higher values of either $\gamma$ or $p$ lead to a more local character of the approximation.
The figures also  show the continuity of the basis functions, which are in general $C^\infty$-continuous if $ p $ is even and $C^{p-1}$-continuous if $ p $ is odd.

\begin{figure}
 	\centering
 	\subfigure[$\epsilon=10^{-8}$]
 	{\includegraphics[width=0.32\textwidth,clip,keepaspectratio,angle=0]{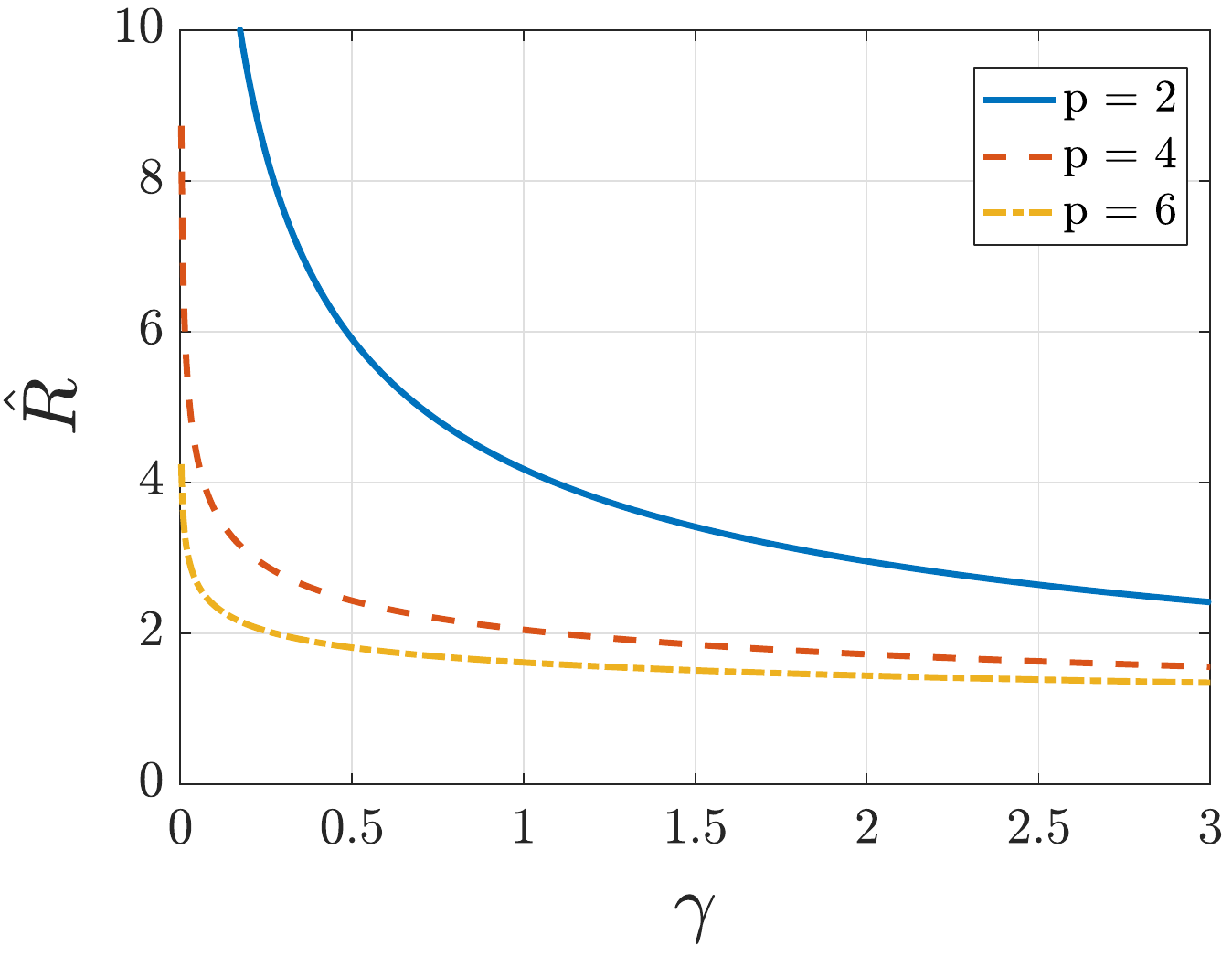}}
 	\subfigure[$\epsilon=10^{-11}$]
 	{\includegraphics[width=0.32\textwidth,clip,keepaspectratio,angle=0]{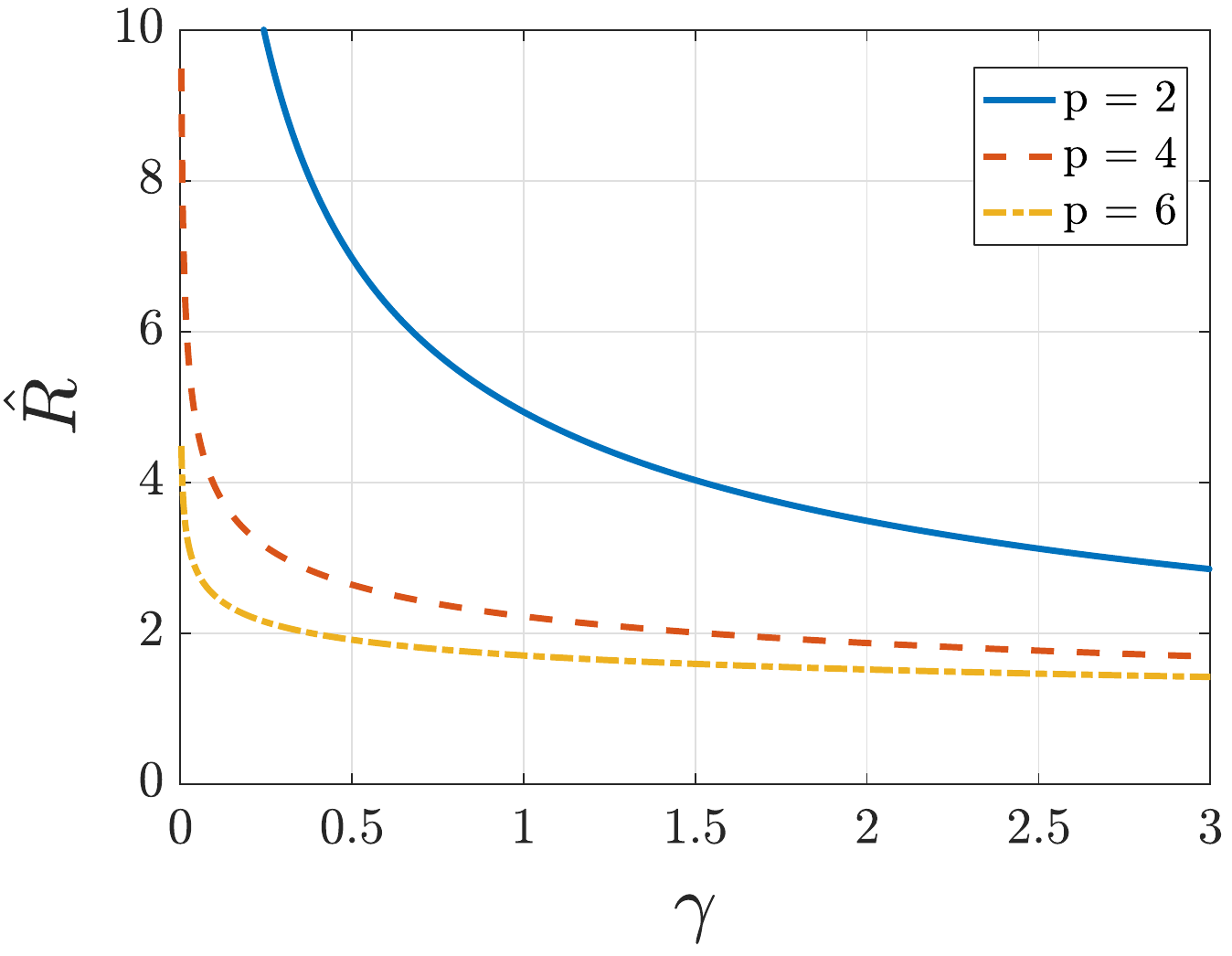}}
  	\subfigure[$\epsilon=10^{-14}$]
  	{\includegraphics[width=0.32\textwidth,clip,keepaspectratio,angle=0]{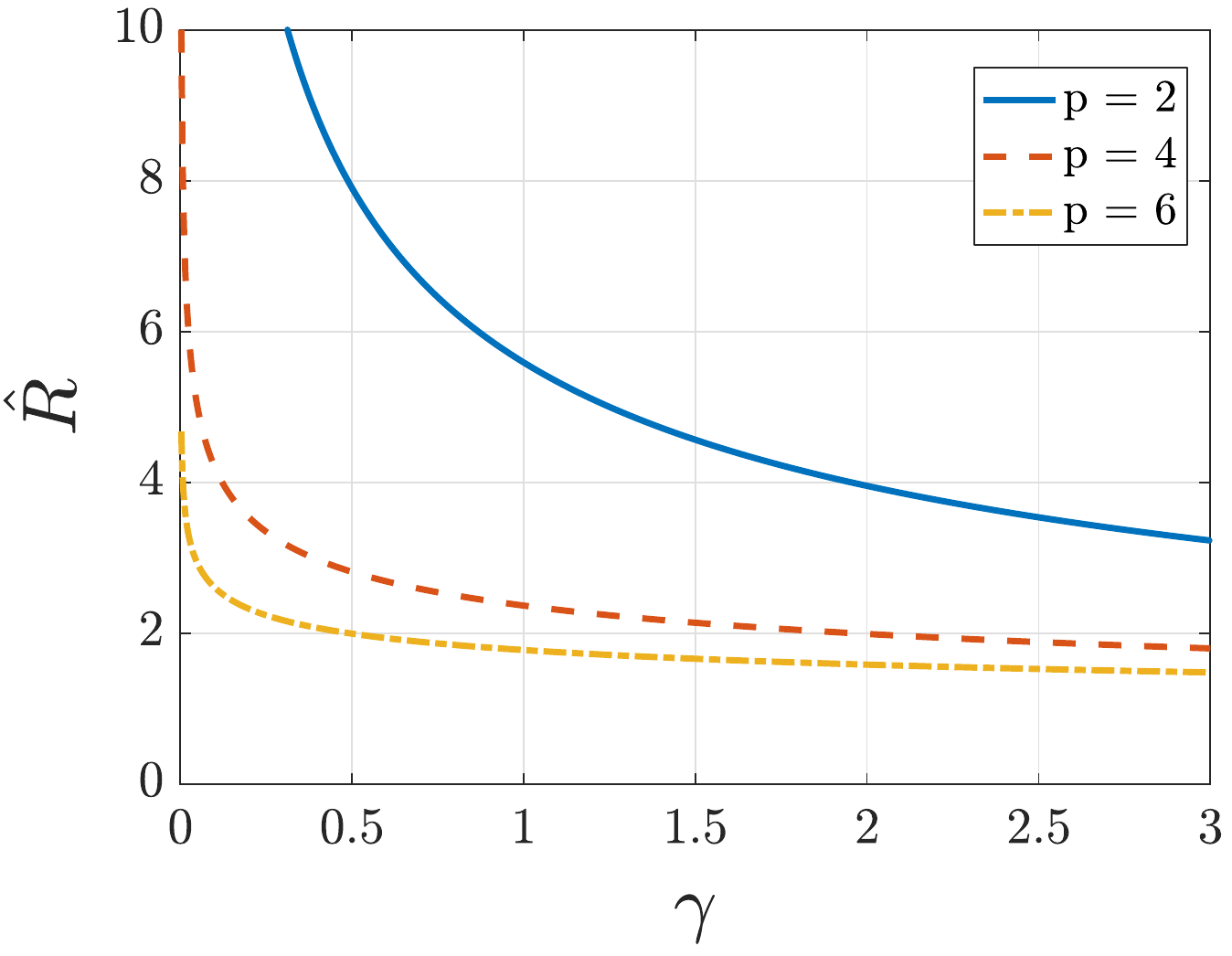}}
 	\caption{
 	Dependence of the normalized   support size of HOLMES basis functions with respect to $\gamma$ and $p$, for different values of the truncation tolerance $\epsilon$.
 	} 
 	\label{fig:support}
 \end{figure}

Strictly speaking, the support of HOLMES basis functions is  the entire computational domain.
However, the locality term included in the formulation confers them with a fast exponential decay and, in the numerical practice, they can be truncated when their value   is below a given threshold $\epsilon$.
This is   crucial   to preserve the sparsity and the conditioning of the stiffness matrix.
To implement the truncation, a common practice  is to define an effective support of $\phi_a$ around $\vx_a$ which, according to \cite{bompadre2012convergent},  can be approximated as
	\begin{equation} 
	r_p = h \left( -\dfrac{\log \epsilon +1}{\gamma}\right) ^{\frac{1}{p}} .
	\end{equation}	
Once this value is estimated, the basis functions are truncated when the $L^p$ distance from the evaluation point to a node is greater than $r_p$. 
Since the value of $r_p$ is proportional to the grid spacing $h$,  a normalized support size $\hat{R}$ can be defined as
	\begin{equation} 
\hat{R}=\frac{r_p }{h}	= \left( -\dfrac{\log \epsilon +1}{\gamma}\right) ^{\frac{1}{p}}  
\label{trunc}
	\end{equation}	
and is used in this work to compare the different types of basis functions.
Since their locality depends both on $\gamma$ and $p$, when $p$ changes  different values of $\gamma$ are required to have the same support size, as it can be observed in Fig.~\ref{fig:support}. This is also noticeable in the basis functions plot of Fig.~\ref{fig:holmes_fi}, where a radius $\hat{R}=4$ can be obtained either with  $ p=2 $ and $\gamma=1.521$ (Fig.~\ref{fig:holmes_fi}a) or  with $ p=4 $ and $\gamma=0.095$ (Fig.~\ref{fig:holmes_fi}c).
Therefore, it is not straightforward to have a direct understanding of the basis functions character by  looking only at the values of $\gamma$ and $ p$.
Here, for a given value of $p$, we fix the desired value of the support size $\hat{R}$ and the relative truncation tolerance $\epsilon$ and, then,   compute the required $\gamma$ as
	\begin{equation} 
	\gamma=  -\frac{\log \epsilon +1}{\hat{R}^p}.
		\end{equation}
The obtained value of  $\gamma$ is then employed to compute the basis functions.

\section{Numerical examples}
\label{section:numerical_examples}
The performance of HOLMES collocation is studied in this section through a selection of numerical examples.
Given the large amount of parameters involved in the computation of HOLMES basis functions, a one-dimensional application is considered first, in order to have a preliminary understanding of their effect on the convergence and accuracy  of the method.
Then,   benchmark applications for acoustic Helmholtz problems and for linear elasticity are considered. 
Finally, the application of HOLMES collocation to   domains defined by a NURBS curve is discussed more in detail.

\begin{MyColorPar}{red}
In all the examples, relative discrete errors are considered for the convergence studies, which for a given vector field $\textbf{v}(\vx)$ and its numerical approximation $\textbf{v}^h(\vx)$, are defined   as
\begin{equation}\label{eq:er}
E_r(\textbf{v},\textbf{v}^h)= \sqrt{\dfrac{\sum_{a=1}^m  (\textbf{v}_a-\textbf{v}_a^h)(\textbf{v}_a-\textbf{v}_a^h)^T}{\sum_{a=1}^m  \textbf{v}_a \textbf{v}_a^T}},
\end{equation}
where $\textbf{v}_a$ and $\textbf{v}_a^h$ are respectively the values of $\textbf{v}$ and $\textbf{v}^h$  at the node $\vx_a$.

 
%
%
%

As far as  theory is concerned, an abstract mathematical framework for the numerical analysis of collocation methods has not been developed yet. 
For meshfree collocation, some preliminary results can be found in \cite{Aluru2000,Hu2011a,Wang2018a,Hu2009a}.
In the case of IGA, a thorough theoretical analysis that includes proofs of stability, convergence and error estimates  is  available for the one-dimensional case in \cite{Auricchio2010}. For higher dimensions, convergence results and error estimates are available only in numerical studies  \cite{Auricchio2010,Schillinger2013b}. 
Based on these references, \textit{a-priori} estimates for the  convergence rate are used   in this work to validate the numerical results.
In particular, for a given order of consistency $n $, the approximate solution is expected to converge with a rate of $n$ for the error in both the $L^2$ norm and the $H^1$ semi-norm if $n $ is even, and with a rate of $ n-1$ for both norms
if $n $ is  odd.

 \end{MyColorPar}
  
%
%
%

\subsection{Preliminary study in one dimension}
\label{section:1d}
\begin{figure}  
 	\centering
 	\subfigure[$n=2,p=2$]
 	{\includegraphics[width=0.32\textwidth,clip,keepaspectratio,angle=0]{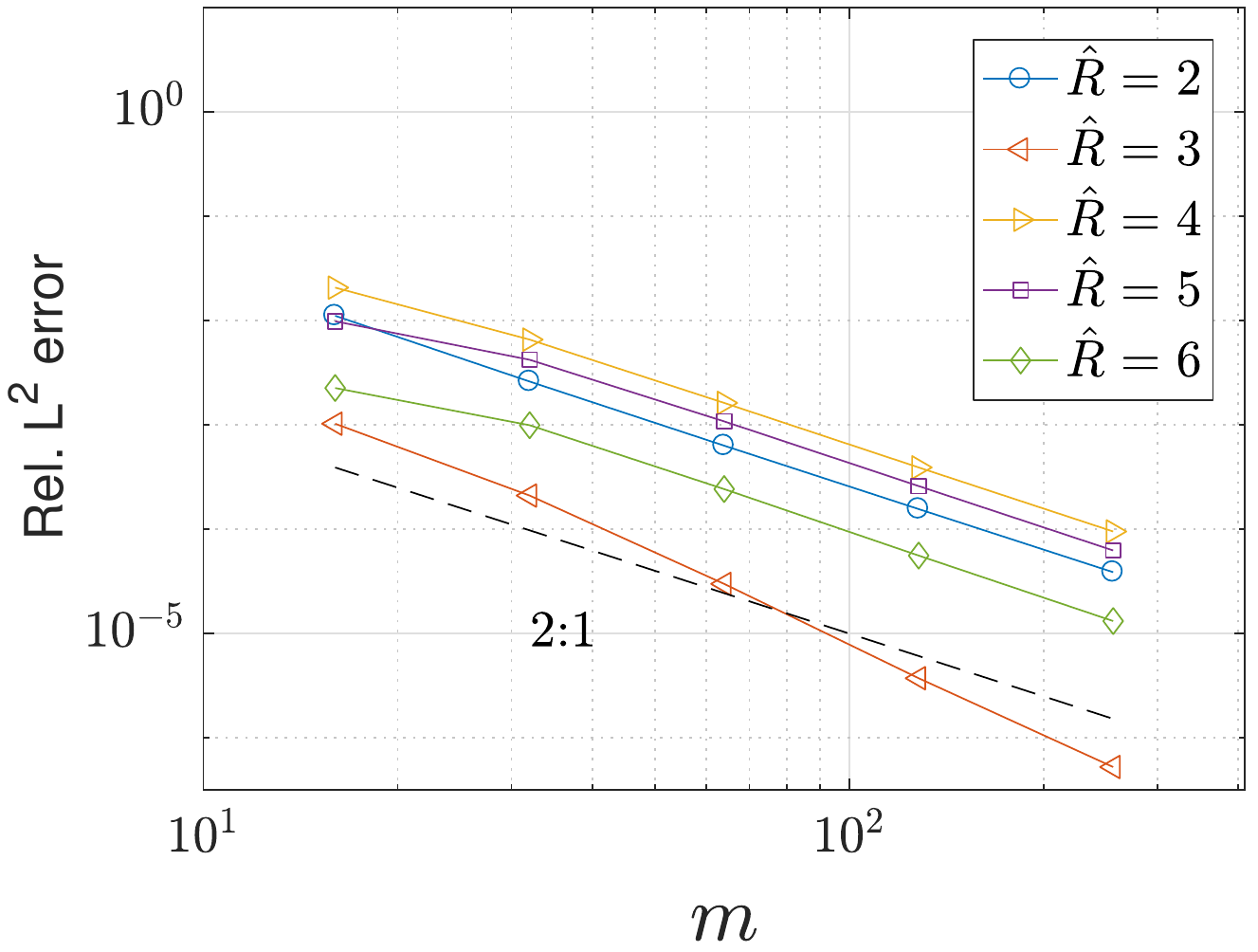}}
 	\subfigure[$n=2,p=4$]
 	{\includegraphics[width=0.32\textwidth,clip,keepaspectratio,angle=0]{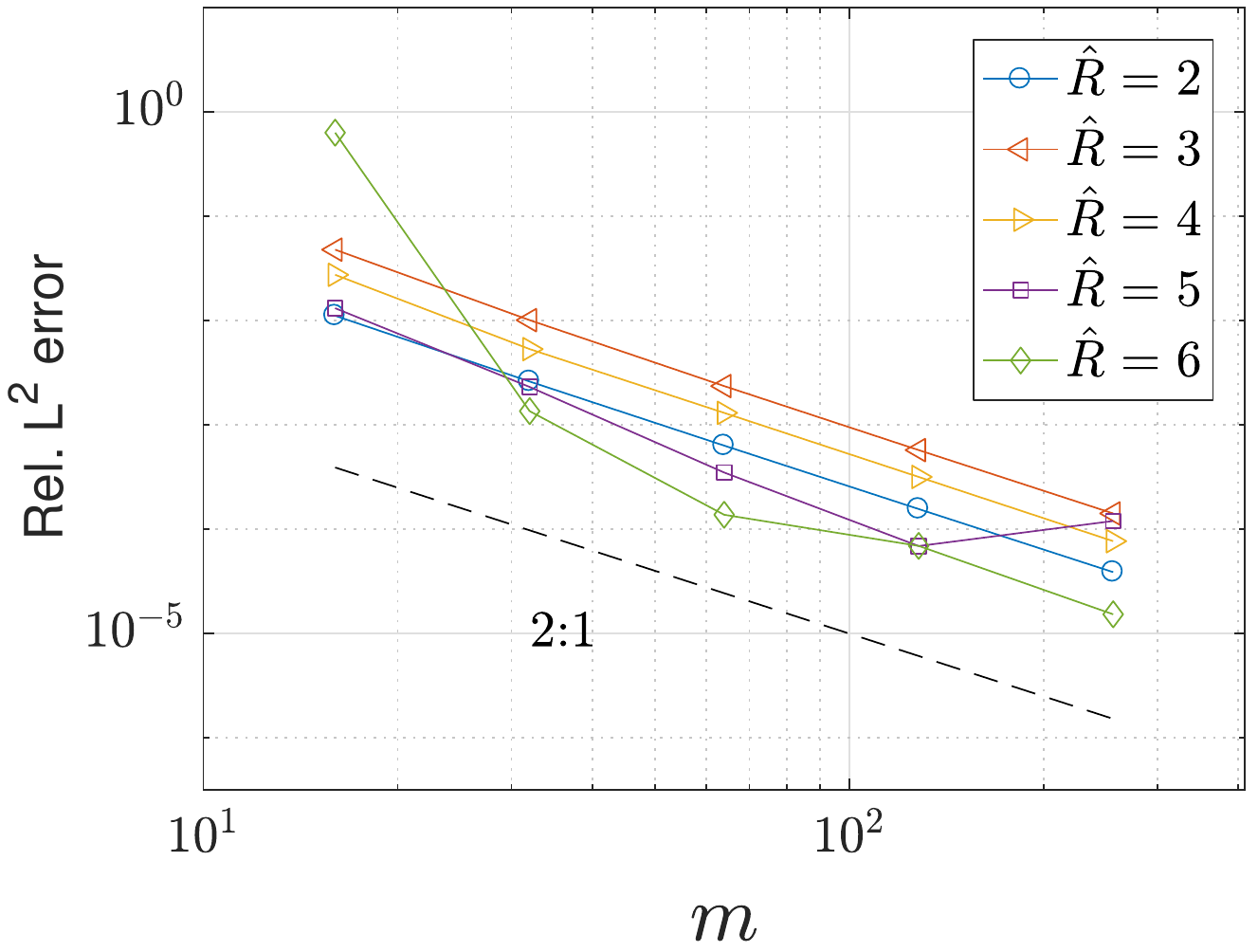}}
 	\subfigure[$n=2,p=6$]
 	{\includegraphics[width=0.32\textwidth,clip,keepaspectratio,angle=0]{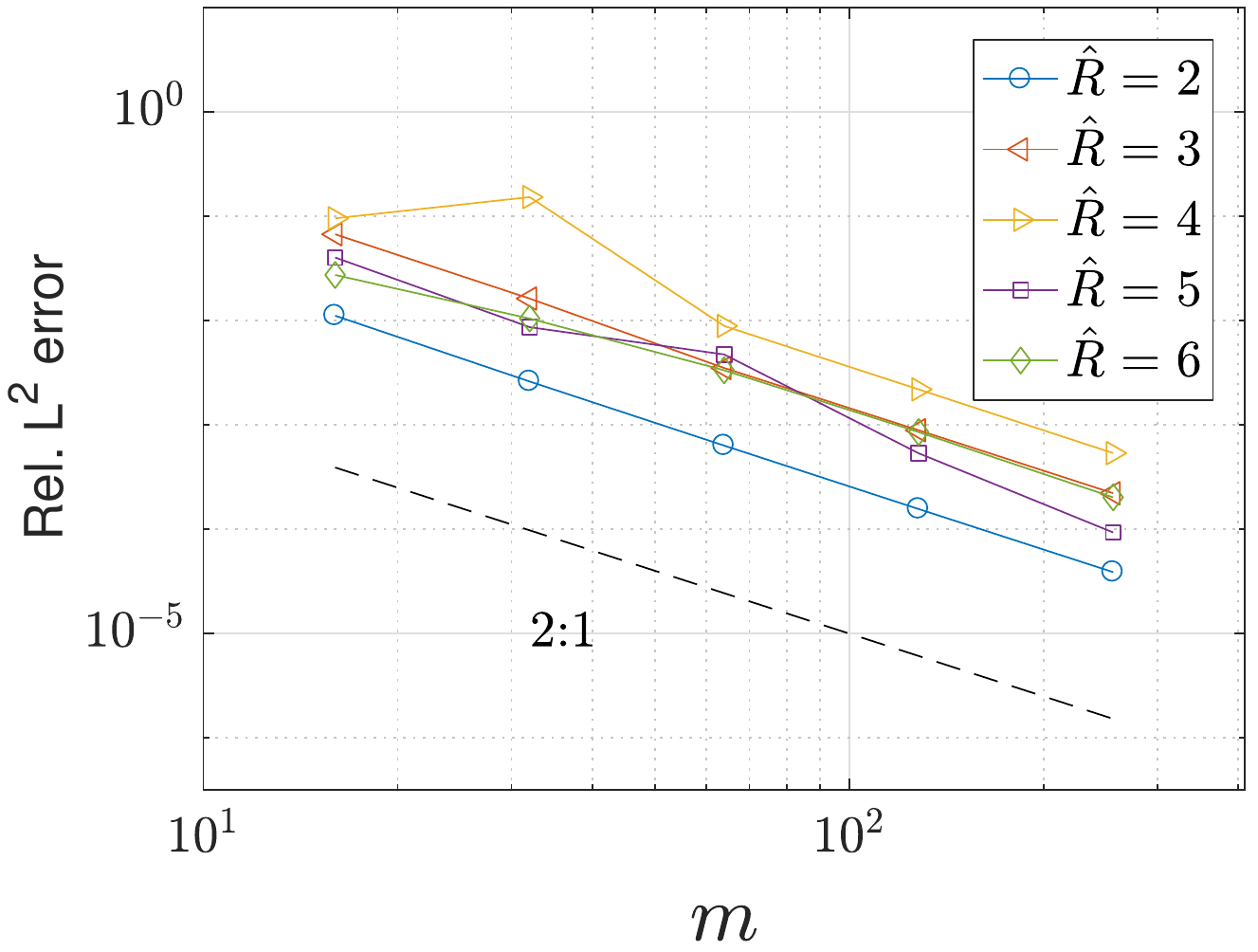}}
 	\subfigure[$n=4,p=2$]
 	{\includegraphics[width=0.32\textwidth,clip,keepaspectratio,angle=0]{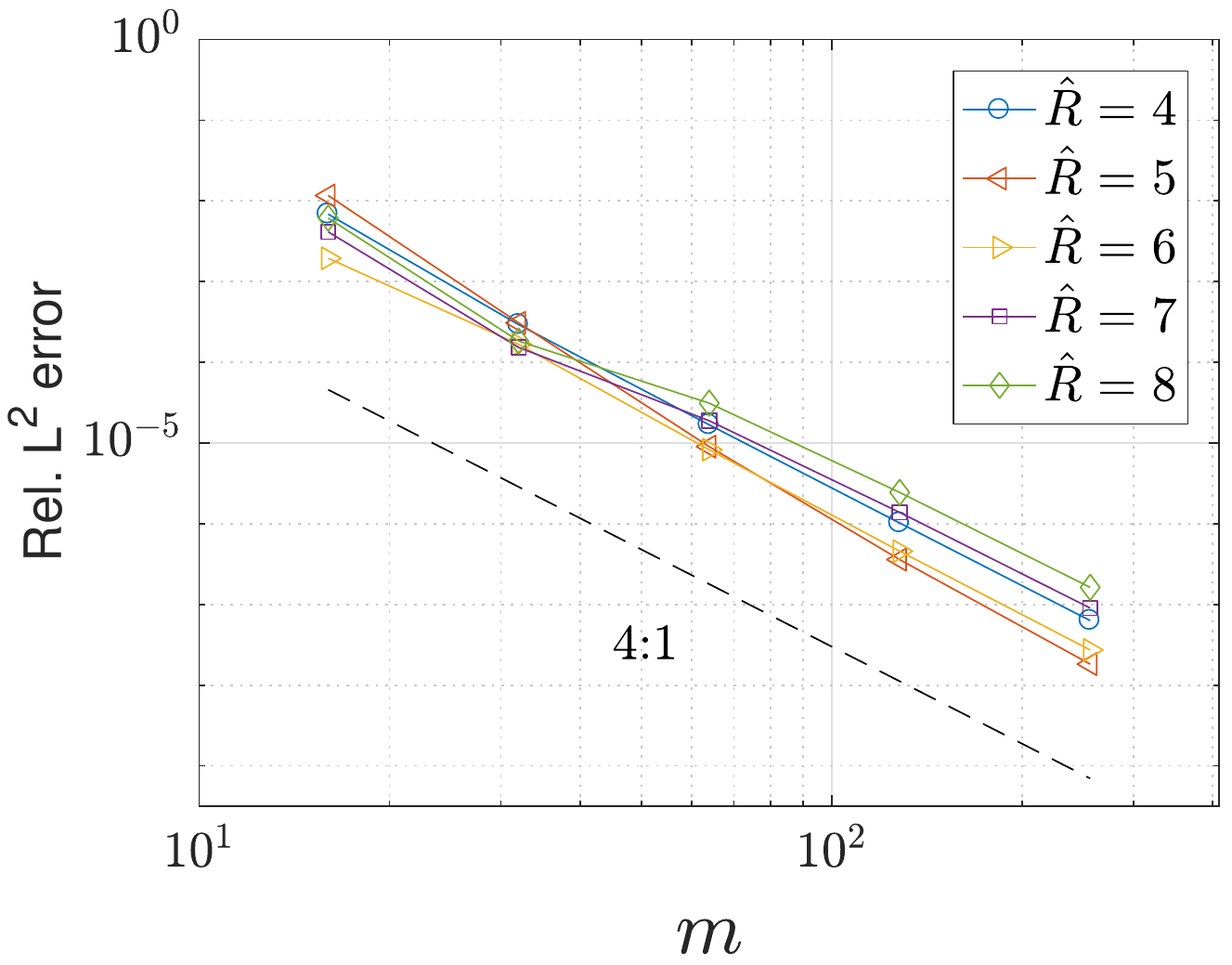}}	
 	\subfigure[$n=4,p=4$]
 	{\includegraphics[width=0.32\textwidth,clip,keepaspectratio,angle=0]{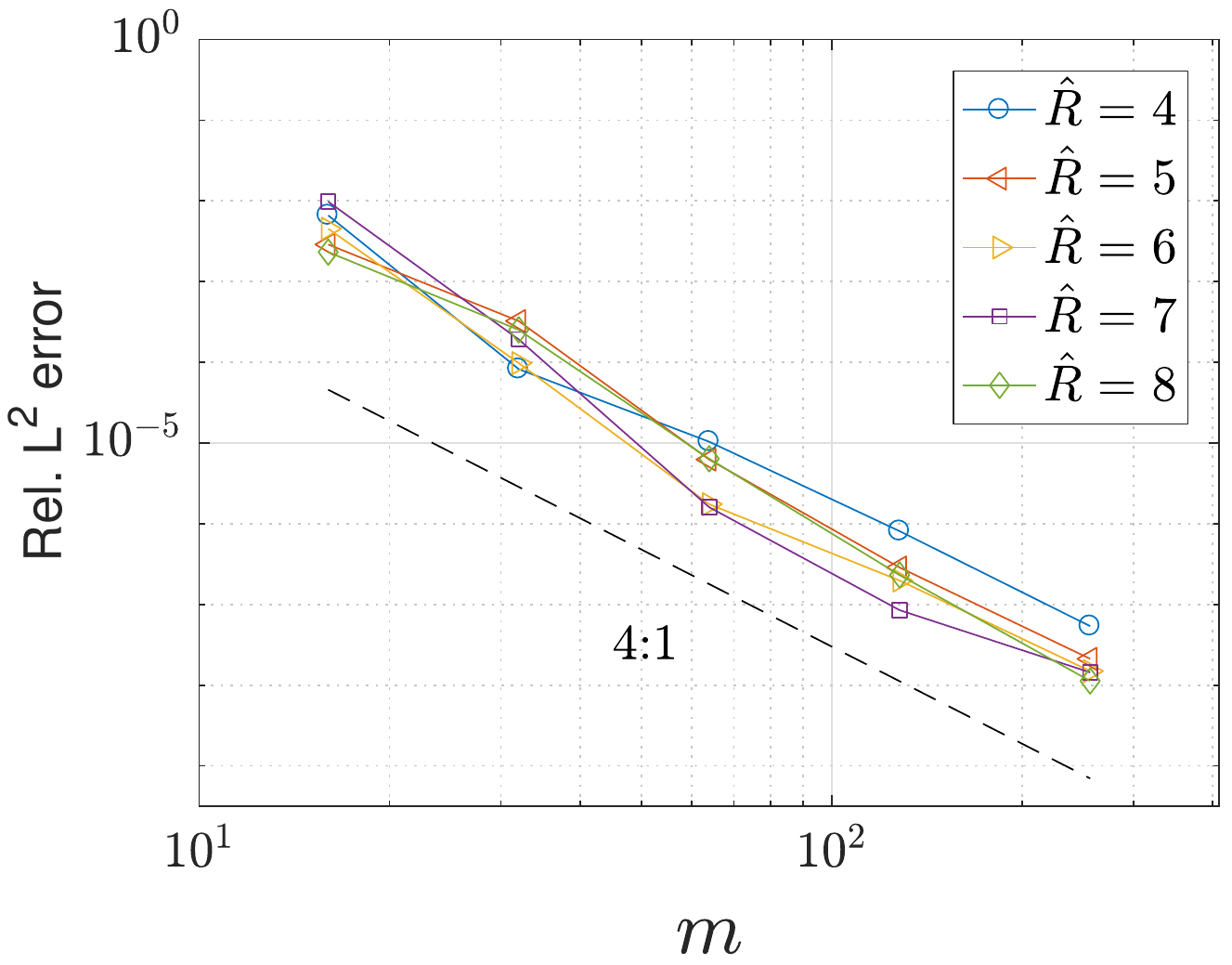}}
 	\subfigure[$n=4,p=6$]
 	{\includegraphics[width=0.32\textwidth,clip,keepaspectratio,angle=0]{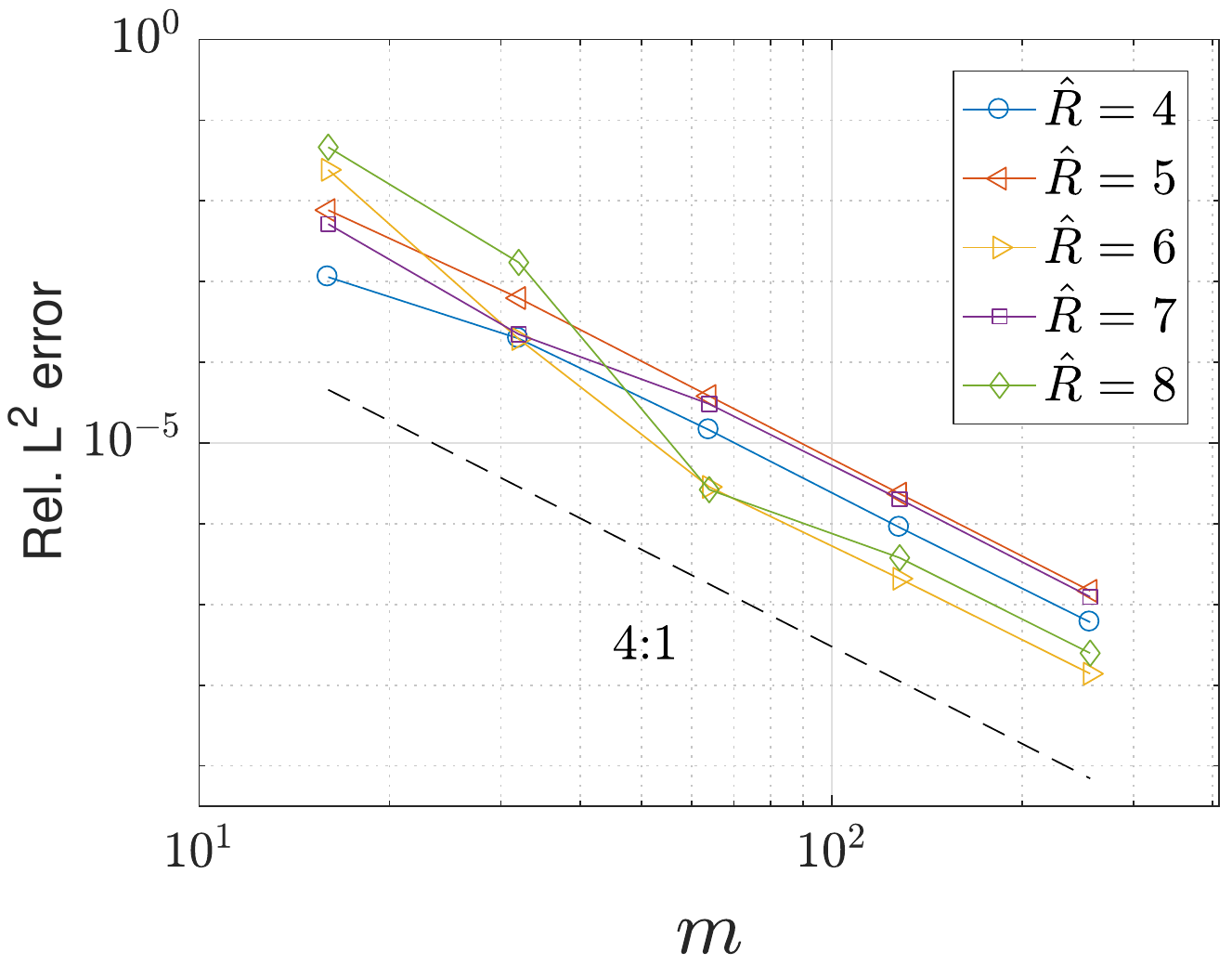}}
 	\subfigure[$n=6,p=2$]
 	{\includegraphics[width=0.32\textwidth,clip,keepaspectratio,angle=0]{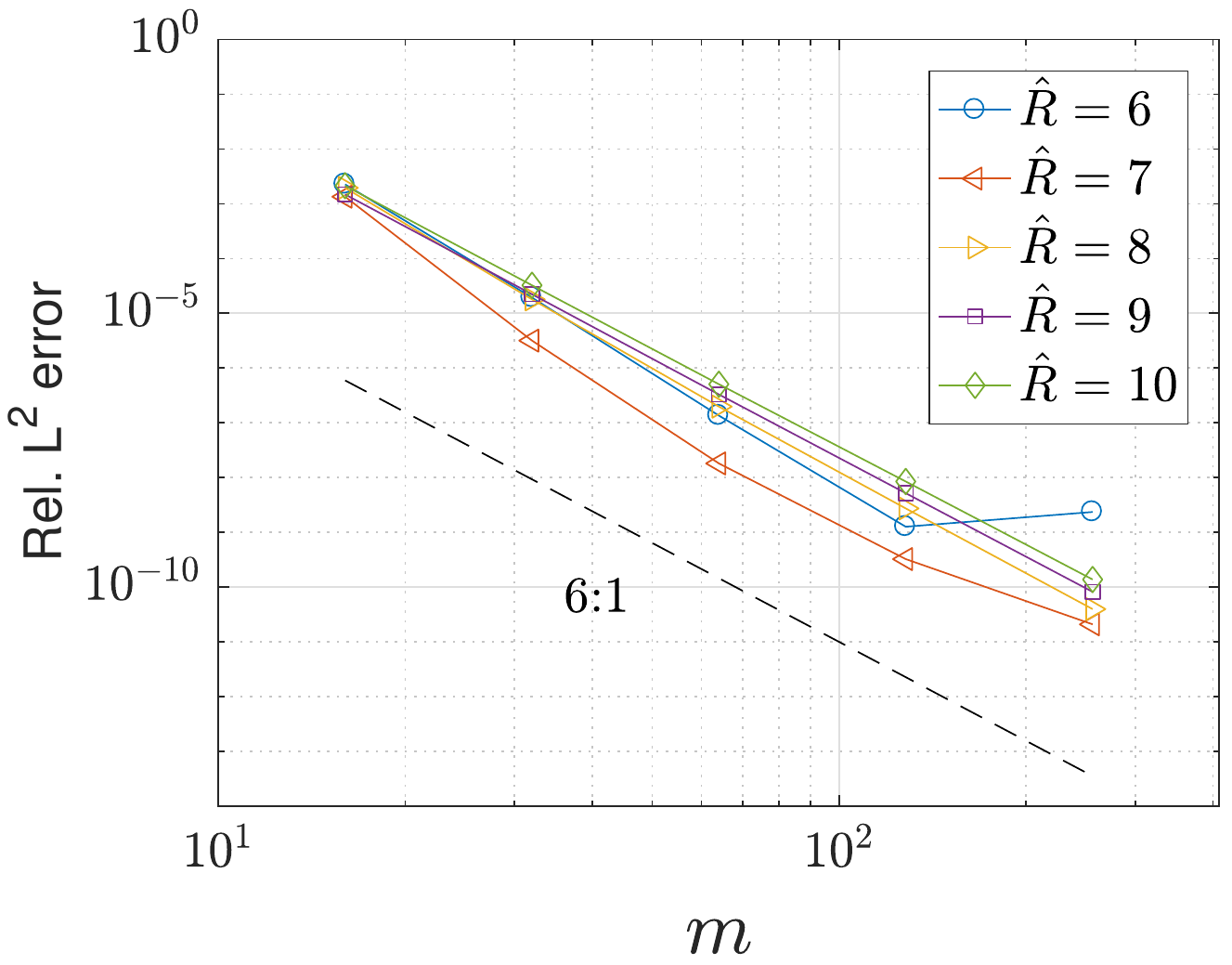}}	
 	\subfigure[$n=6,p=4$]
 	{\includegraphics[width=0.32\textwidth,clip,keepaspectratio,angle=0]{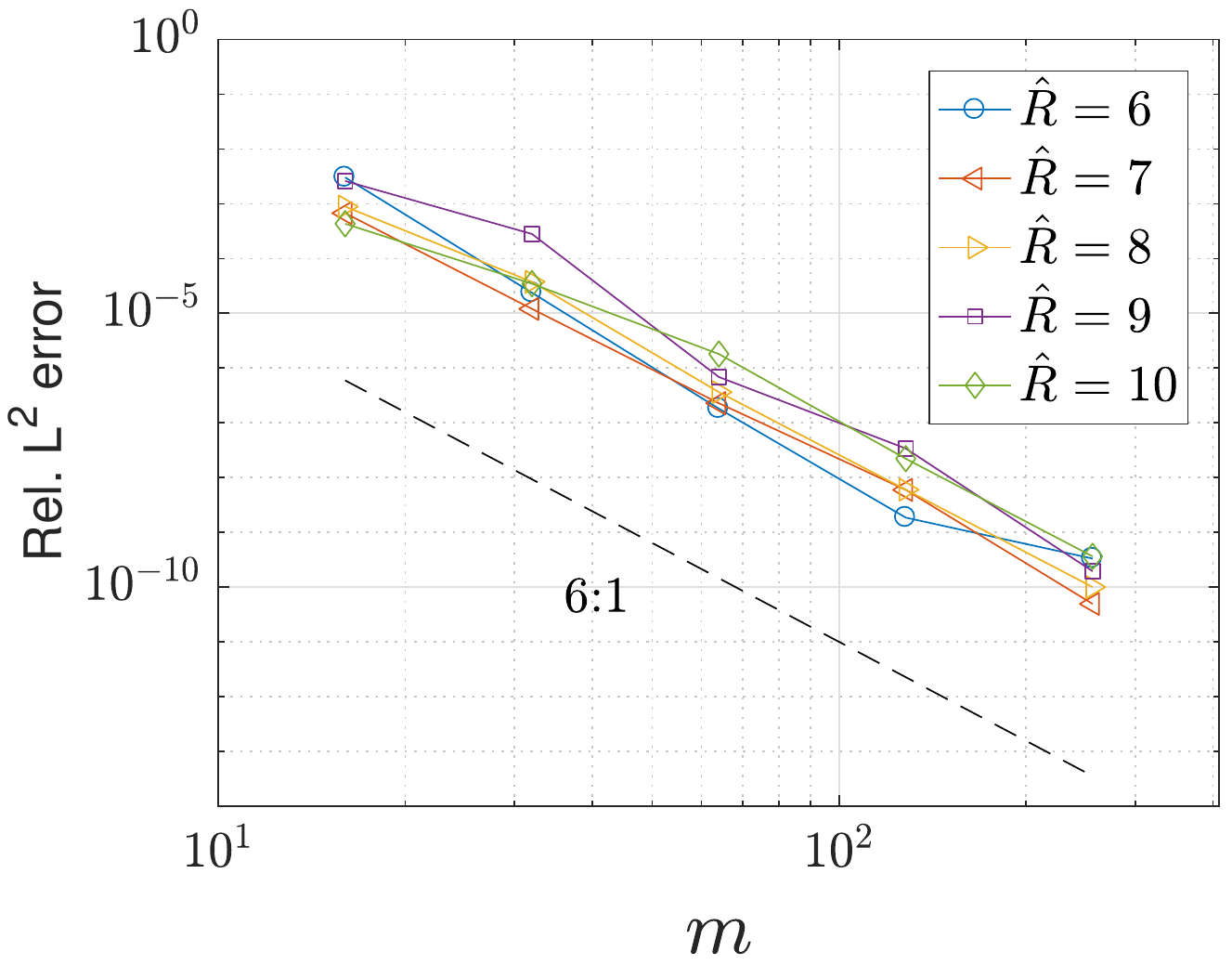}}
 	\subfigure[$n=6,p=6$]
 	{\includegraphics[width=0.32\textwidth,clip,keepaspectratio,angle=0]{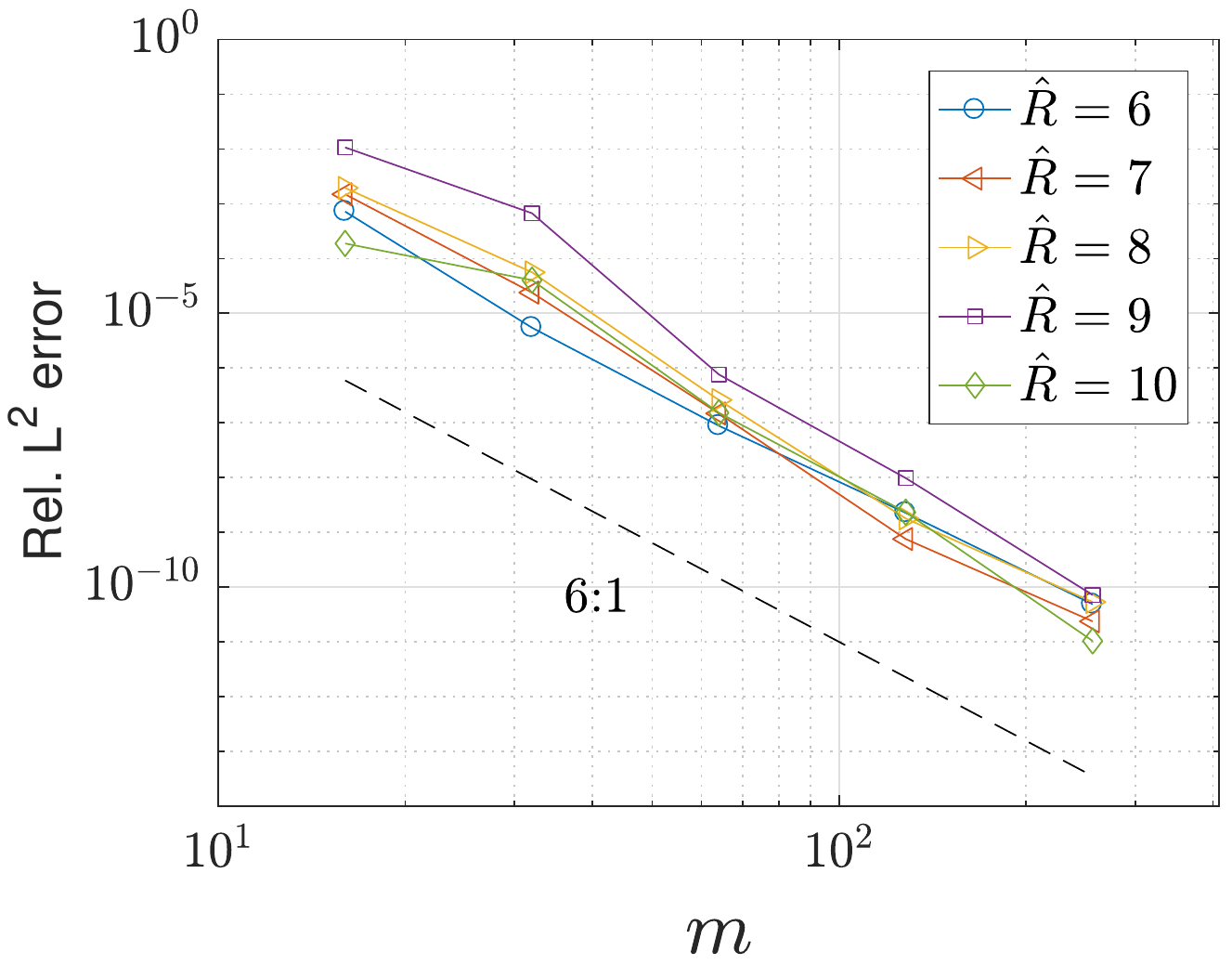}}
 	\caption{Convergence of the relative $L^2$  error   for a 1D Helmholtz problem with essential boundary conditions solved  for different values of $n$ and $p$and a truncation tolerance $\epsilon=10^{-11}$.
 	} 
 	\label{fig:conv_1d_ess}
 \end{figure}

\begin{figure} 
 	\centering
 	\subfigure[$n=2,p=2$]
 	{\includegraphics[width=0.32\textwidth,clip,keepaspectratio,angle=0]{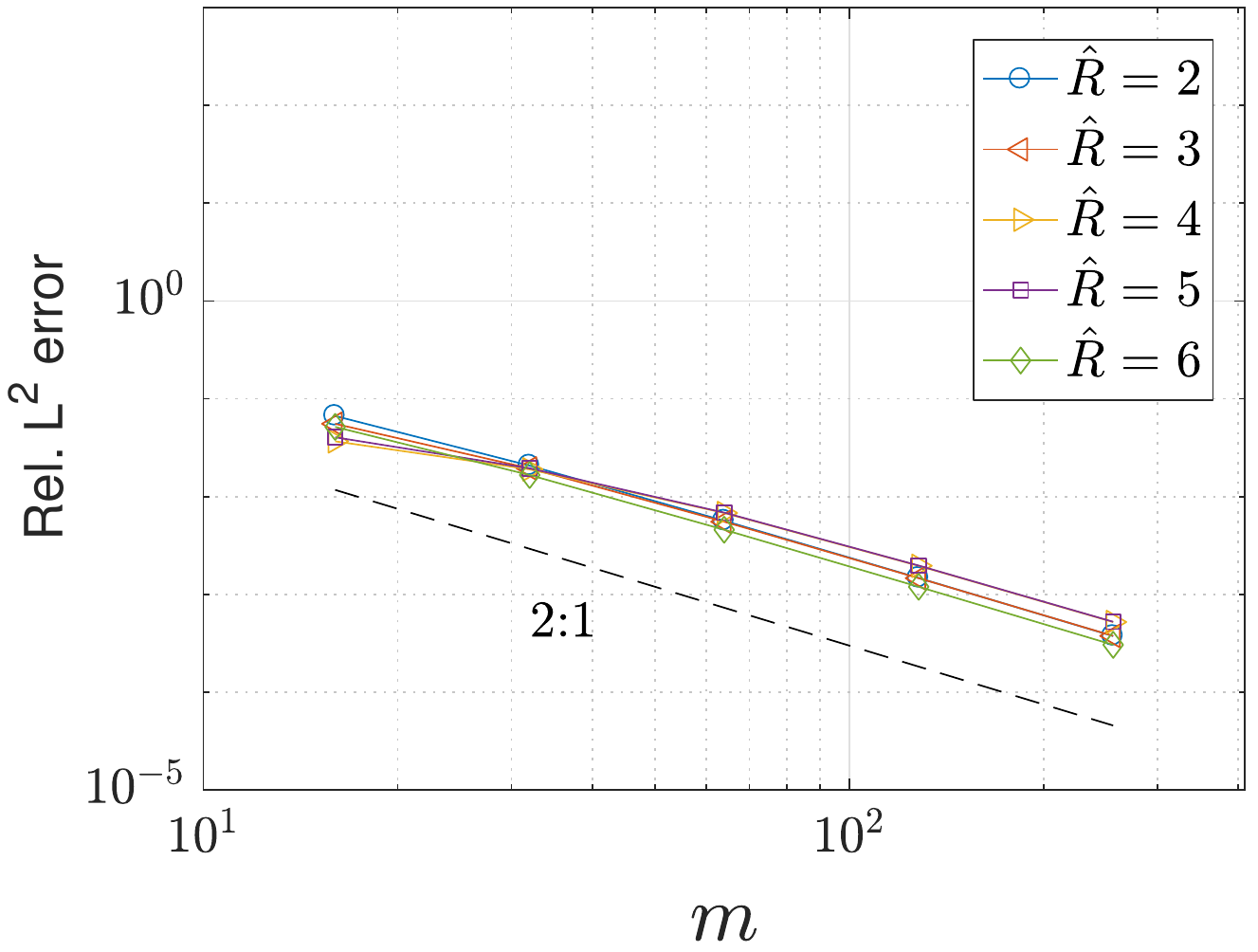}}
 	\subfigure[$n=2,p=4$]
 	{\includegraphics[width=0.32\textwidth,clip,keepaspectratio,angle=0]{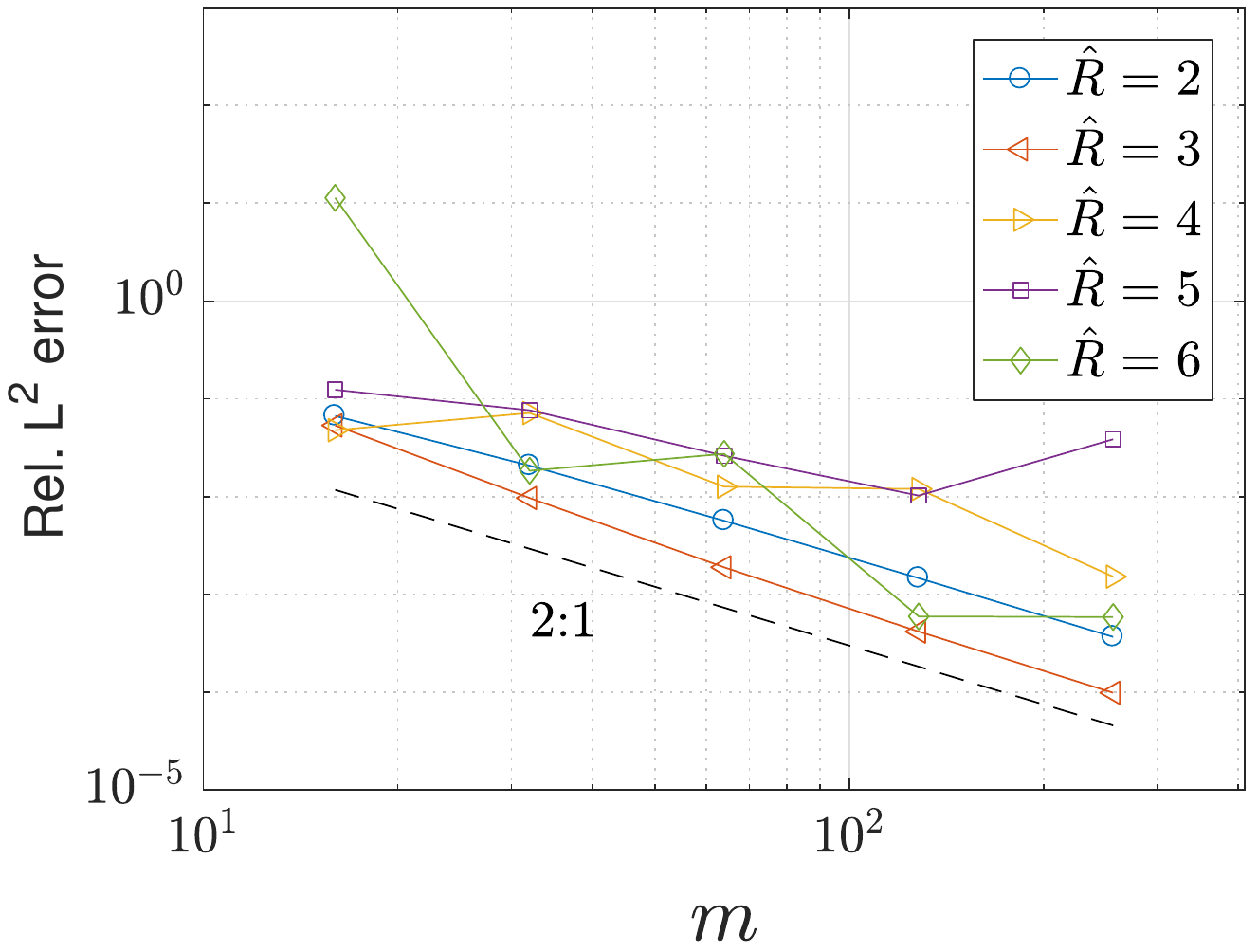}}
 	\subfigure[$n=2,p=6$]
 	{\includegraphics[width=0.32\textwidth,clip,keepaspectratio,angle=0]{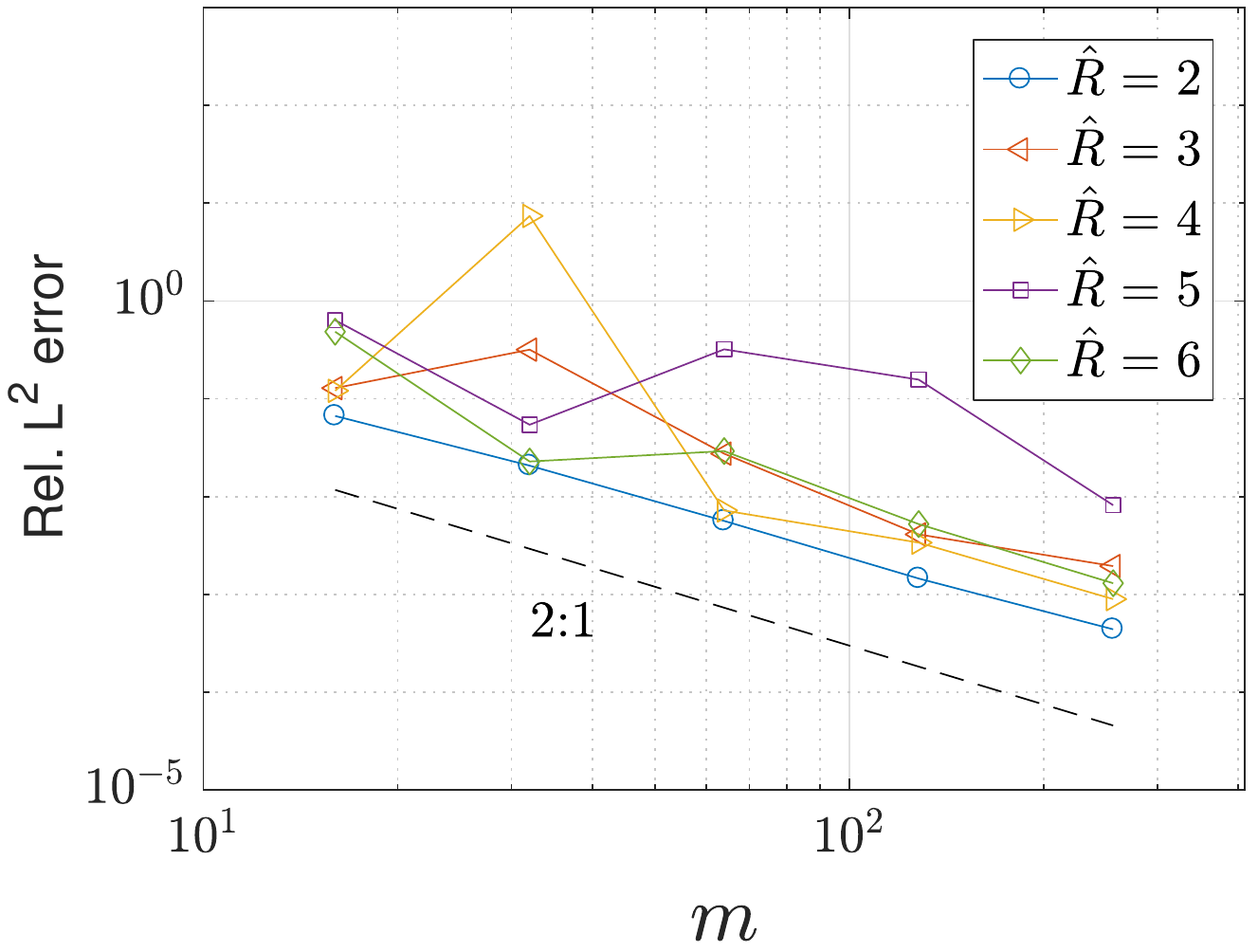}}
 	\subfigure[$n=4,p=2$]
 	{\includegraphics[width=0.32\textwidth,clip,keepaspectratio,angle=0]{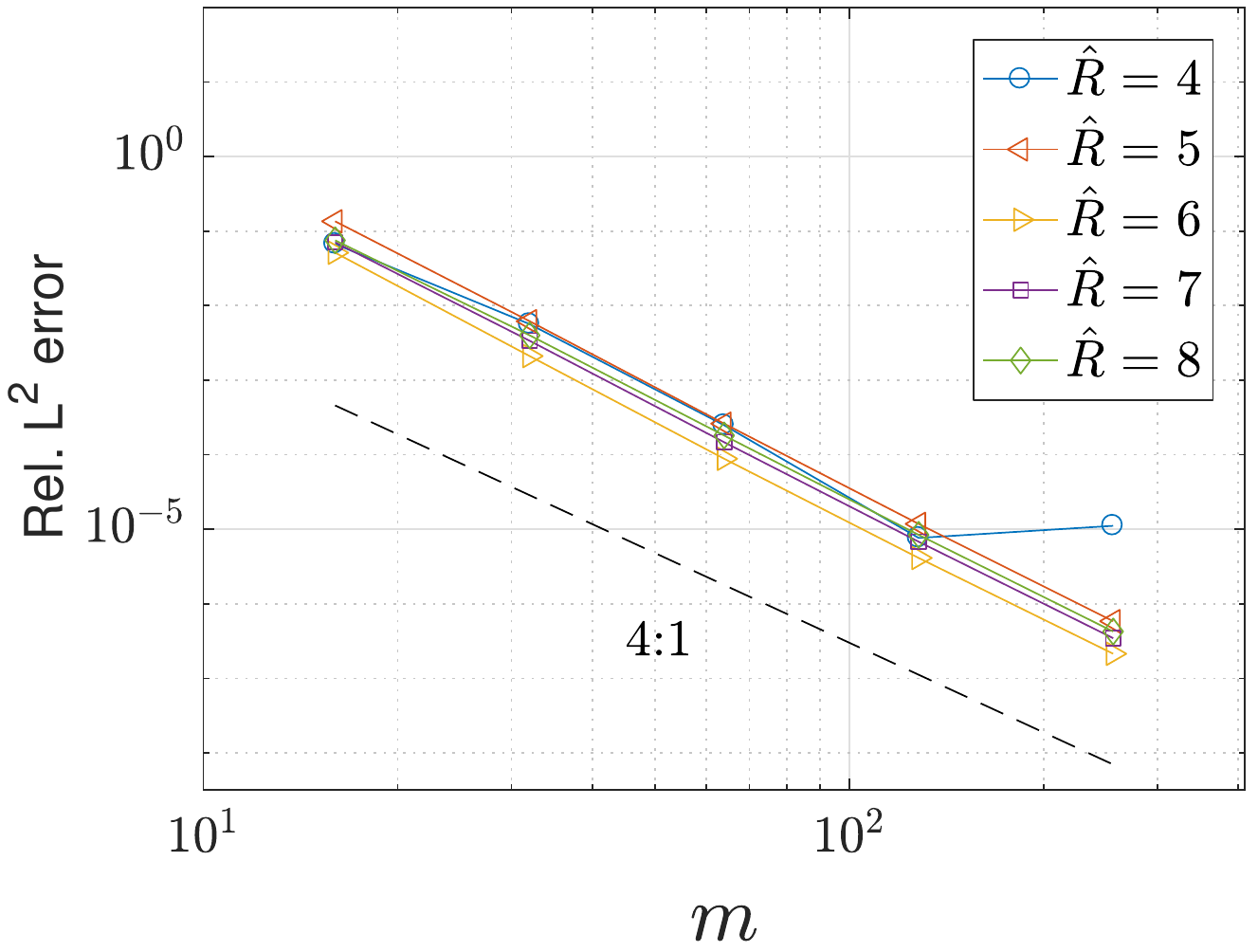}}	
 	\subfigure[$n=4,p=4$]
 	{\includegraphics[width=0.32\textwidth,clip,keepaspectratio,angle=0]{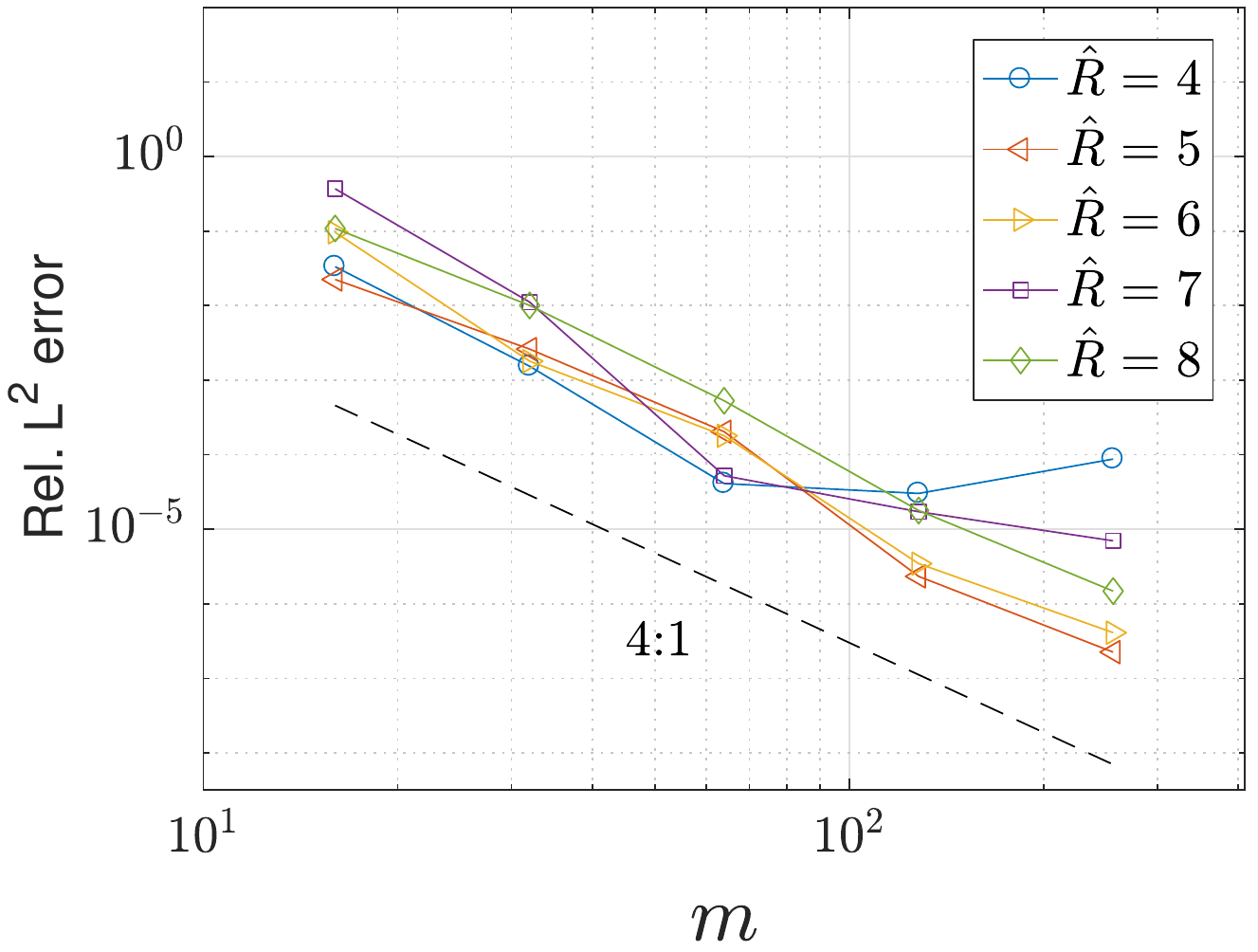}}
 	\subfigure[$n=4,p=6$]
 	{\includegraphics[width=0.32\textwidth,clip,keepaspectratio,angle=0]{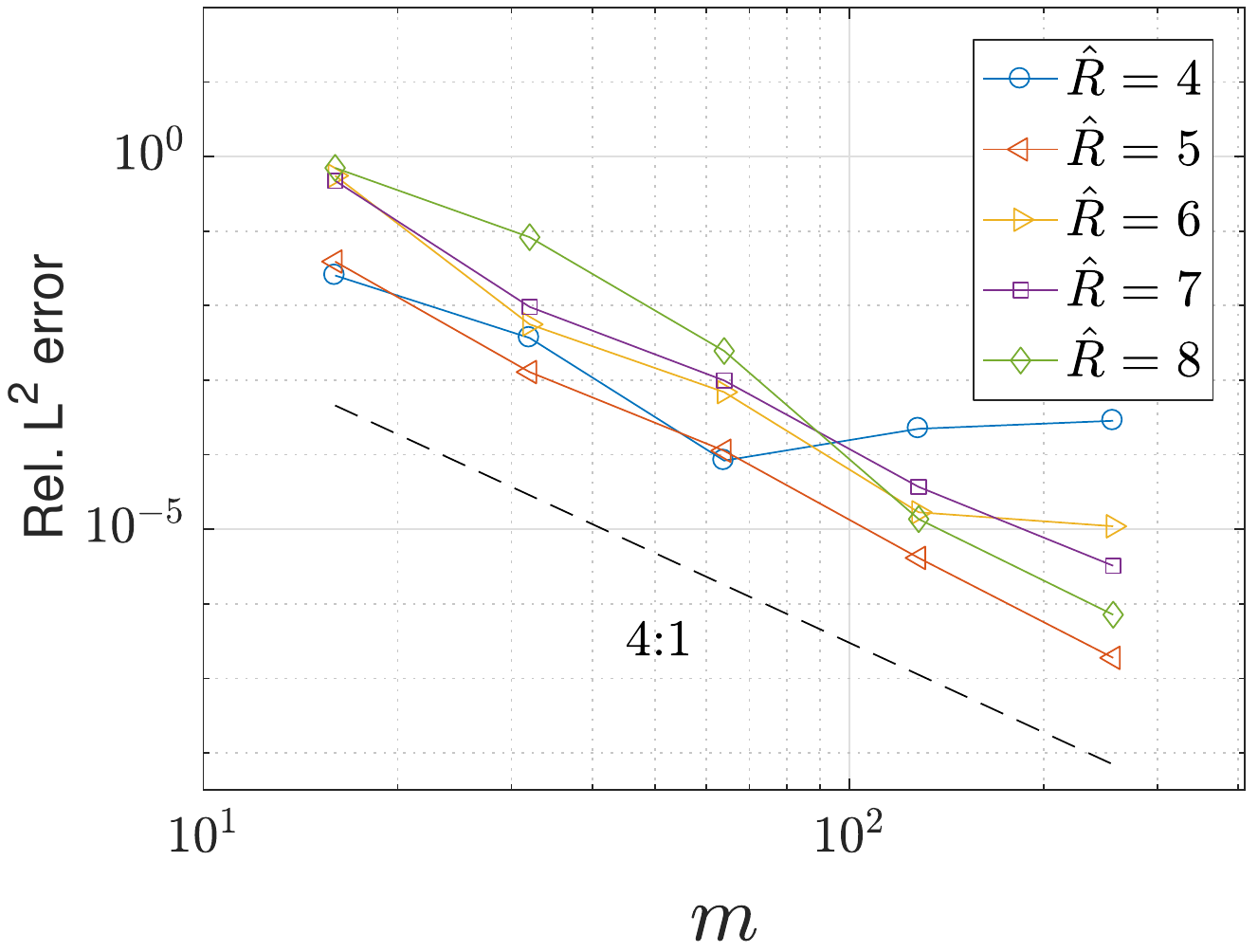}}
 	\subfigure[$n=6,p=2$]
 	{\includegraphics[width=0.32\textwidth,clip,keepaspectratio,angle=0]{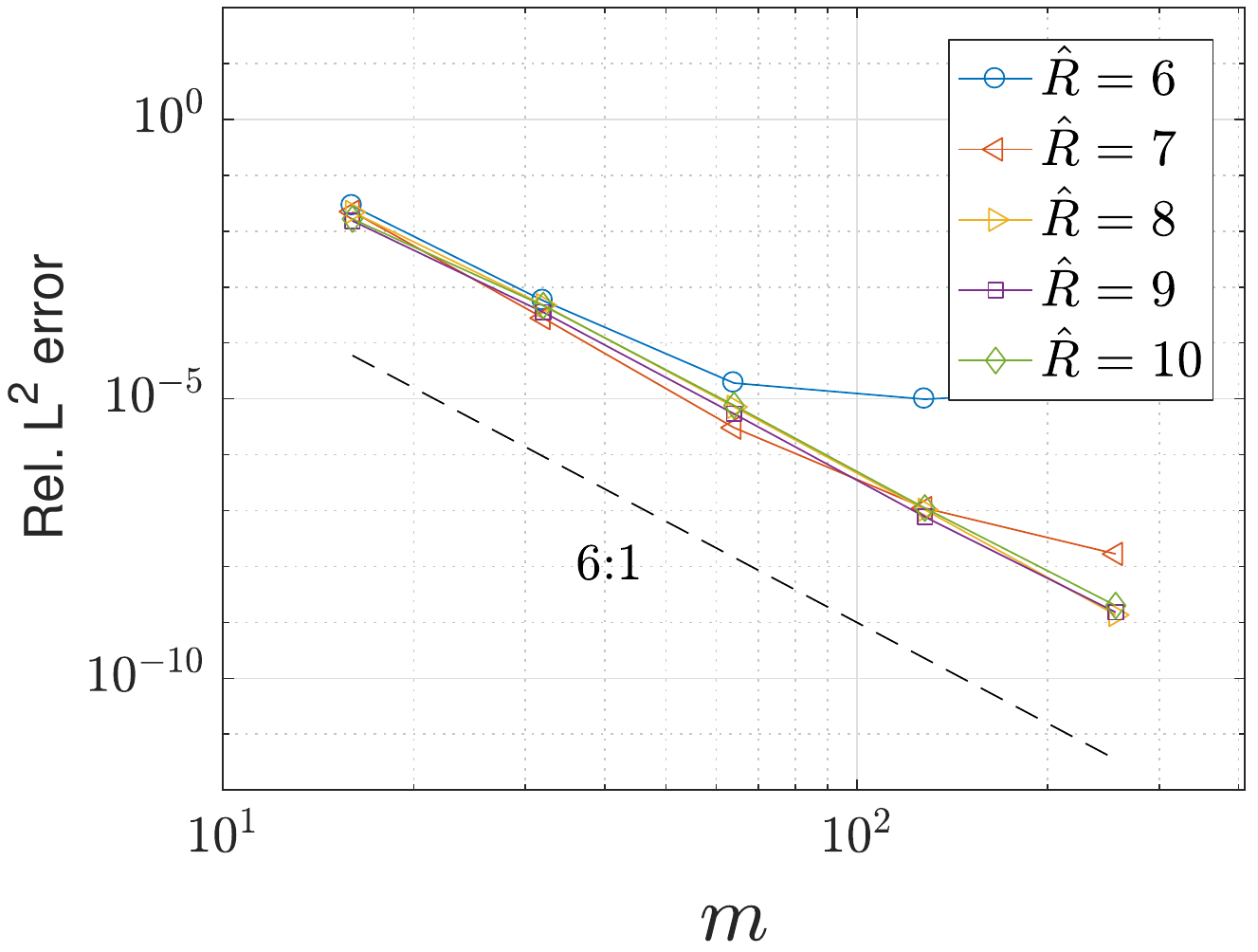}}	
 	\subfigure[$n=6,p=4$]
 	{\includegraphics[width=0.32\textwidth,clip,keepaspectratio,angle=0]{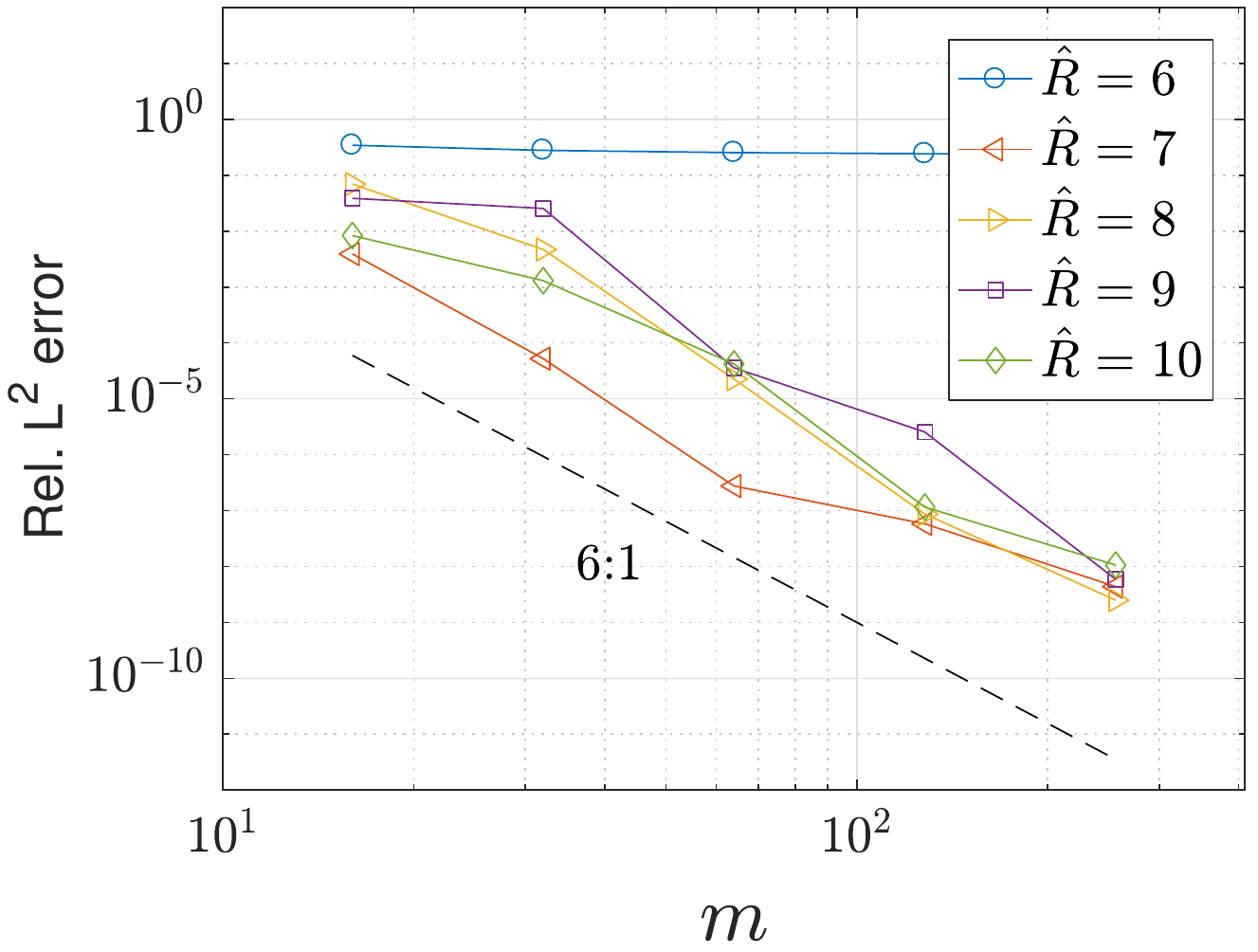}}
 	\subfigure[$n=6,p=6$]
 	{\includegraphics[width=0.32\textwidth,clip,keepaspectratio,angle=0]{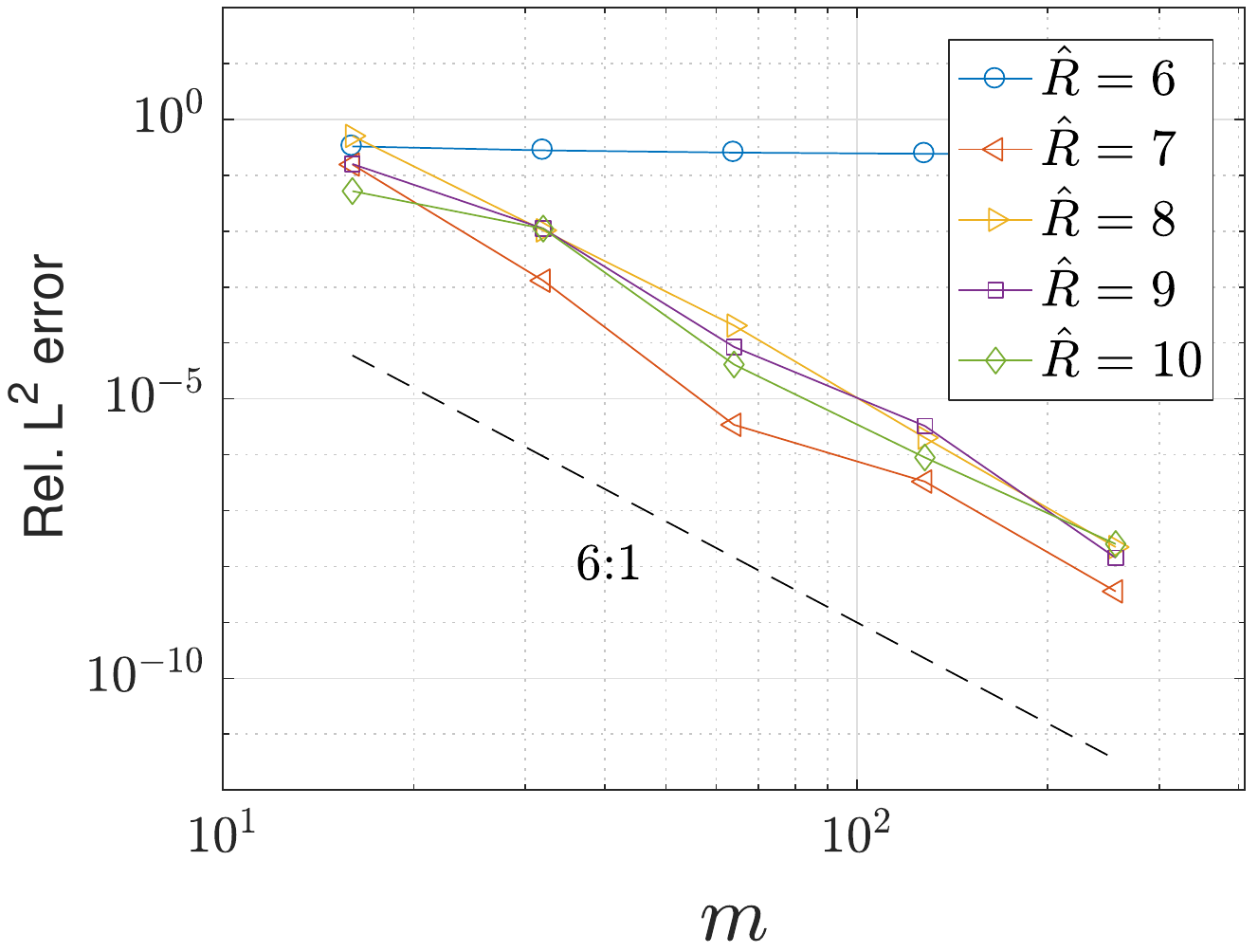}}
 	\caption{Convergence of the relative $L^2$  error       for a 1D Helmholtz problem with natural boundary conditions solved  for different values of $n$ and $p$ and a truncation tolerance $\epsilon=10^{-11}$.
 	} 
 	\label{fig:conv_1d_nat}
 \end{figure}

A one-dimensional problem, stated by the following Helmholtz equation is considered:
\begin{equation}
\Delta u(x) + u(x) = b (x), \;\;\; x \in [-1;1],
\label{eq:Helmholtz1}
\end{equation}
where $b(x)$ is a source term compatible with the following solution 
\begin{equation}
 u(x)=\sin(3x)\exp(x)+\tan^{-1}(x)+\cosh(x).
 \end{equation}
This PDE is solved together with suitable boundary conditions to recover $u(x)$.
Two sets of boundary-value problems are considered, with either  essential or natural boundary conditions, and the corresponding convergence curves of the relative discrete $L^2$ error,   $E_r(u,u^h)$, as a function of the number of nodes $m$ are plotted  in Fig.~\ref{fig:conv_1d_ess} and 
 Fig.~\ref{fig:conv_1d_nat} respectively.
 
 
The convergence study considers HOLMES approximants of different orders ($n=2,4,6$) and different values of the locality parameters $\gamma$ and $p$.
As discussed in Section~\ref{section:holmes}, for each $p$ the values of $\gamma$ are chosen such that values of the normalized support size from $\hat{R}=n$ to $\hat{R}=n+4$  are obtained, for a fixed truncation tolerance of $\epsilon=10^{-11}$.
It has to be remarked that, for HOLMES of a given order $n$,   at least $n+1$ nodes are necessary in the computation of the basis functions for problem \eqref{HOLMES} to be well-posed.
Therefore, the nodes on the boundary need $\hat{R}  \geq n$ to verify this condition.

Analyzing the results relative to essential boundary conditions in Fig.~\ref{fig:conv_1d_ess}, it can be observed how the curves converge in all the cases, but the results become slightly worse for higher values of the locality norm $p$.
This situation becomes more evident for the case of natural boundary conditions in Fig.~\ref{fig:conv_1d_nat}, where for $p=2$ the expected convergence  is obtained, whereas for higher $p$ the curves still converge but with a much more irregular behavior.
The only exception for $\hat{R}  = n$, where a few curves do not converge when $n=4$ and $n=6$.

As mentioned earlier, in the context of Galerkin formulations the condition  $p>n$ is sufficient  for the convergence of the method since it ensures  that the basis functions have a proper exponential decay \cite{bompadre2012convergent}.  Here, converging solutions are obtained for $p \leq n$.
Furthermore, the numerical results presented in this section suggest that HOLMES collocation has a better convergence behavior for $p=2$.
This result can be explained by observing the basis functions derivatives plotted in Fig.~\ref{fig:holmes_fi}, which are more irregular for higher values of $p$ particularly close to the boundary.

\begin{figure} 
	\centering
	\subfigure[$\epsilon=10^{-8}$]
	{\includegraphics[width=0.32\textwidth,clip,keepaspectratio,angle=0]{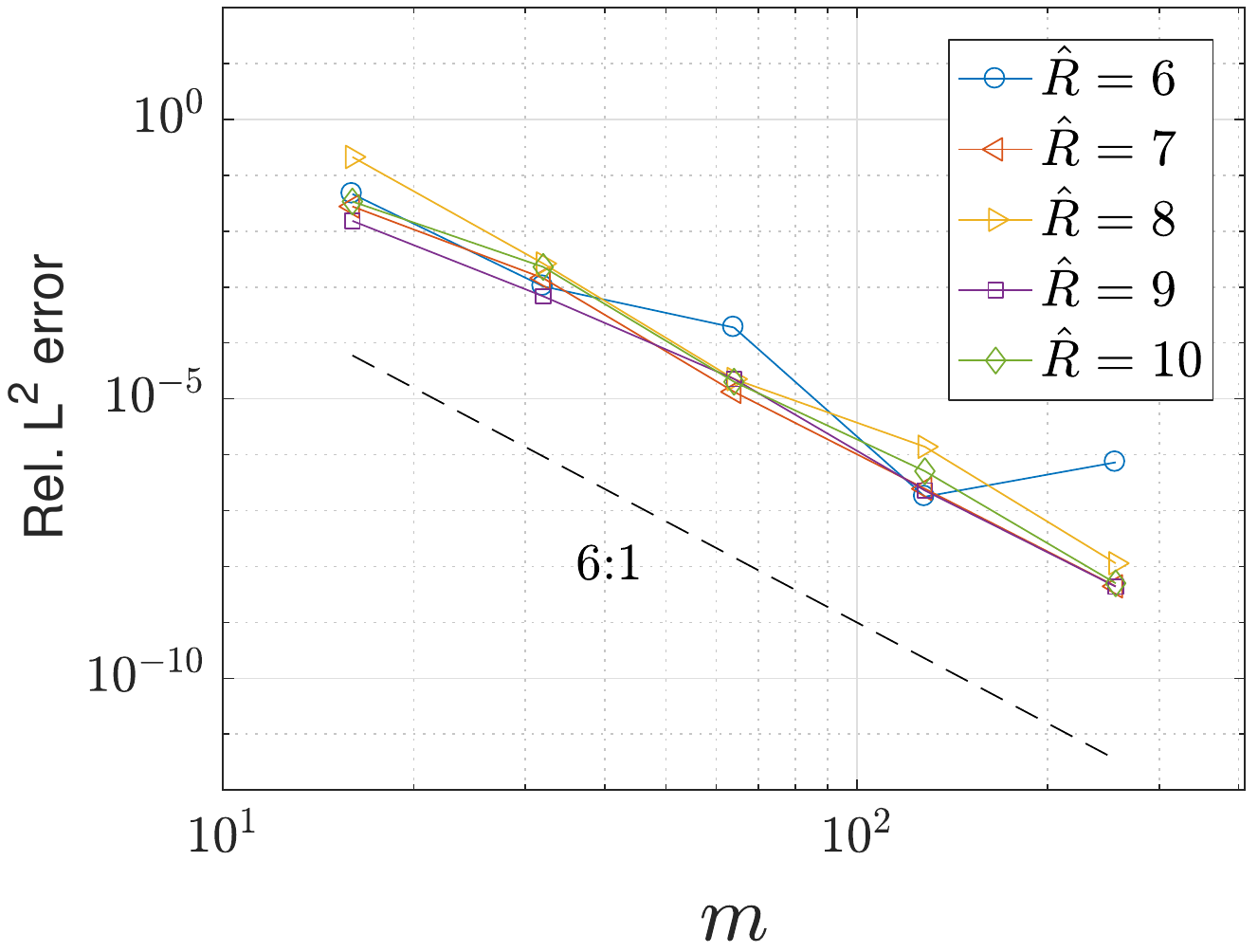}}
	\subfigure[$\epsilon=10^{-11}$]
	{\includegraphics[width=0.32\textwidth,clip,keepaspectratio,angle=0]{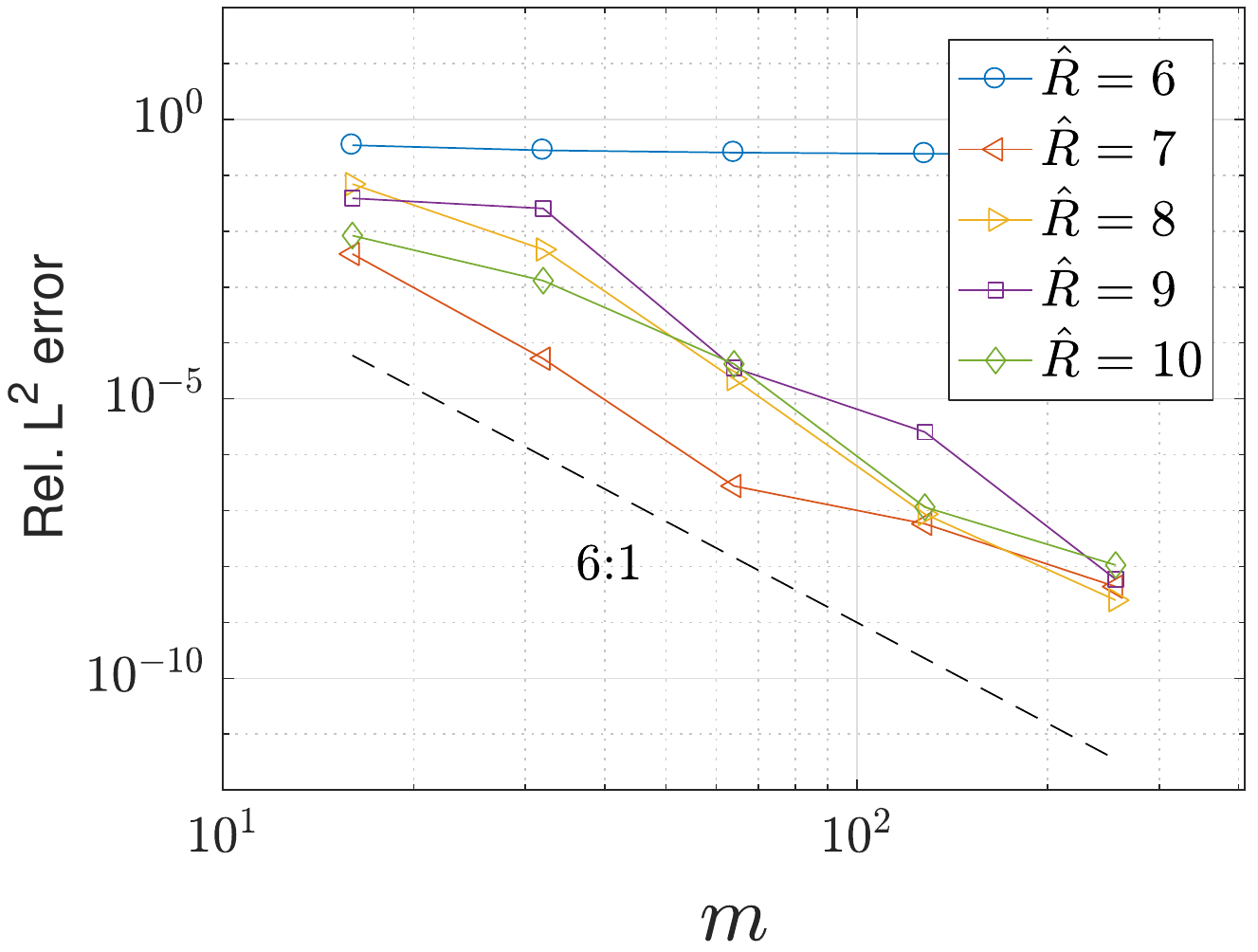}}
	\subfigure[$\epsilon=10^{-14}$]
	{\includegraphics[width=0.32\textwidth,clip,keepaspectratio,angle=0]{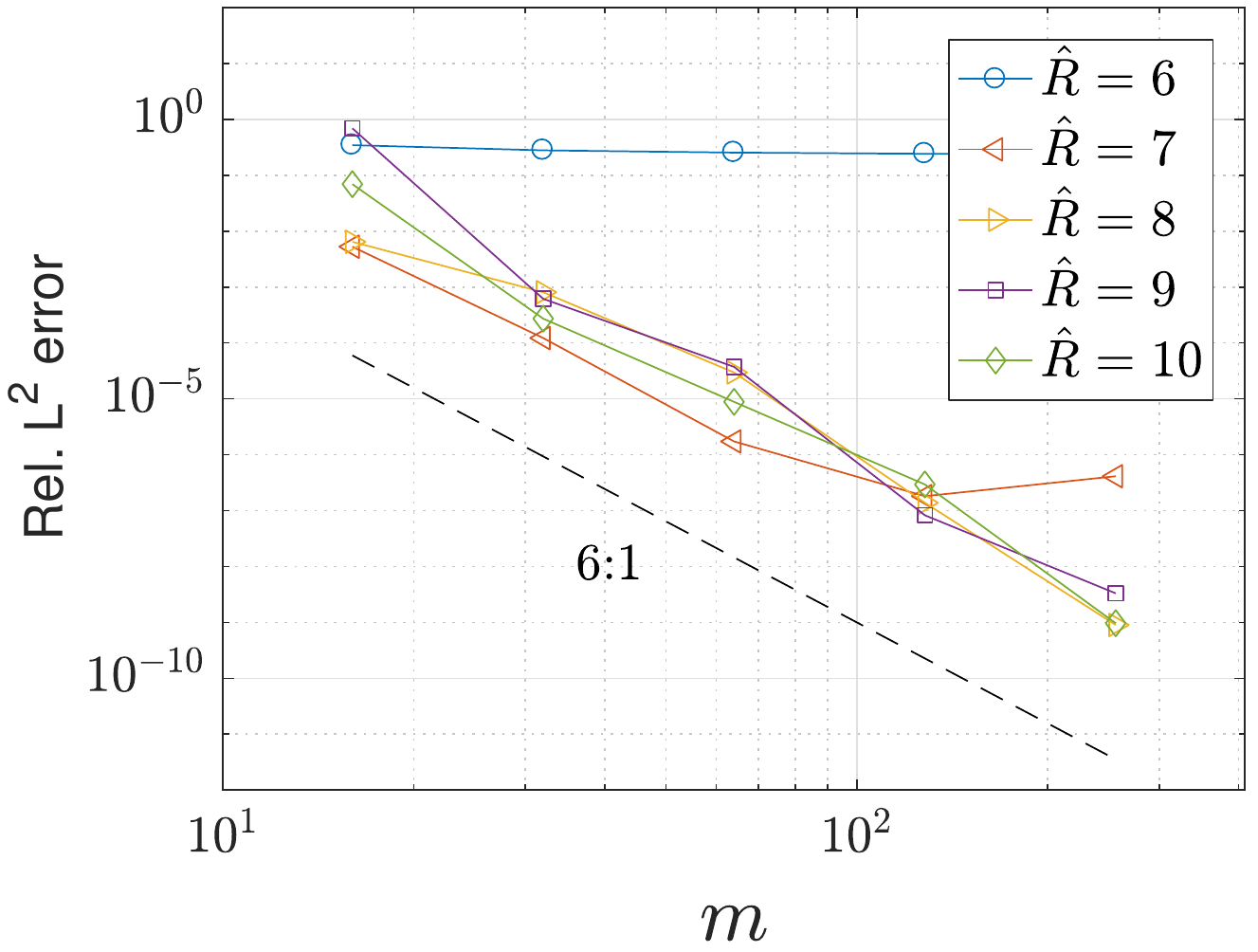}}
	\caption{
		Convergence of the relative $L^2$  error   for a 1D Helmholtz problem with natural boundary conditions, solved for  $n=6$, $p=4$ and different values of the truncation tolerance $\epsilon$.
	} 
	\label{fig:conv_1d_tol}
\end{figure}

A possible explanation for the non-converging curves in the case of $\hat{R}  = n$ can be that the corresponding values of $\gamma$  do not give enough overlap to the basis functions and, therefore, for each collocation node the values of the basis functions of the furthest nodes and their relative second derivatives are too small.
Indeed, it is worth noting that the numerical truncation, defined through \eq{trunc}, is only an approximation of the decay of the basis functions and, in addition, the behavior of the second derivatives is slightly different, particularly for higher values of $p$.
This can also explain why for $n=6$ the convergence curve stagnates around $10^{-5}$ for $p=2$ but does not converge for higher values of $p$. 
This consideration is confirmed by the convergence curves in Fig.~\ref{fig:conv_1d_tol}, where the case of $n=6$ and $p=4$ from Fig.~\ref{fig:conv_1d_nat}h is analyzed for different values of the truncation tolerance $\epsilon$. It can be observed in Fig.~\ref{fig:conv_1d_tol}a how a higher tolerance, which corresponds to lower values of $\gamma$, is indeed changing the behavior of the non-converging curve for $\hat{R}  = 6$.
However, on using higher values of $\epsilon$, the truncation error may affect the bottom part of the convergence curves with an effect which is difficult to estimate. Taking this into account,   the value of $\epsilon=10^{-11}$ is still  used in   the following numerical examples.

\begin{figure} 
	\centering
	\subfigure[Essential BCs]
	{\includegraphics[width=0.49\textwidth,clip,keepaspectratio,angle=0]{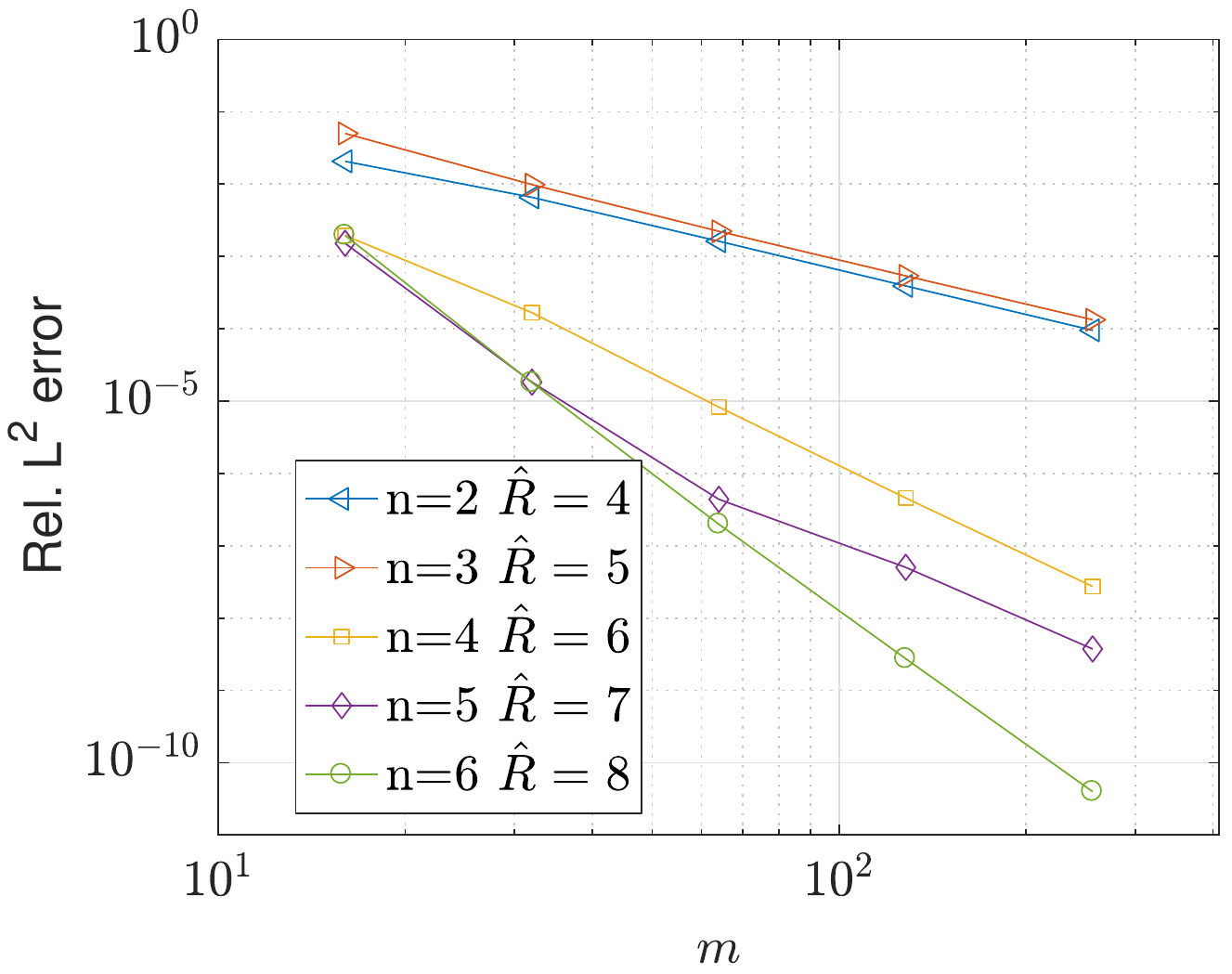}}
	\subfigure[Natural BCs]
	{\includegraphics[width=0.49\textwidth,clip,keepaspectratio,angle=0]{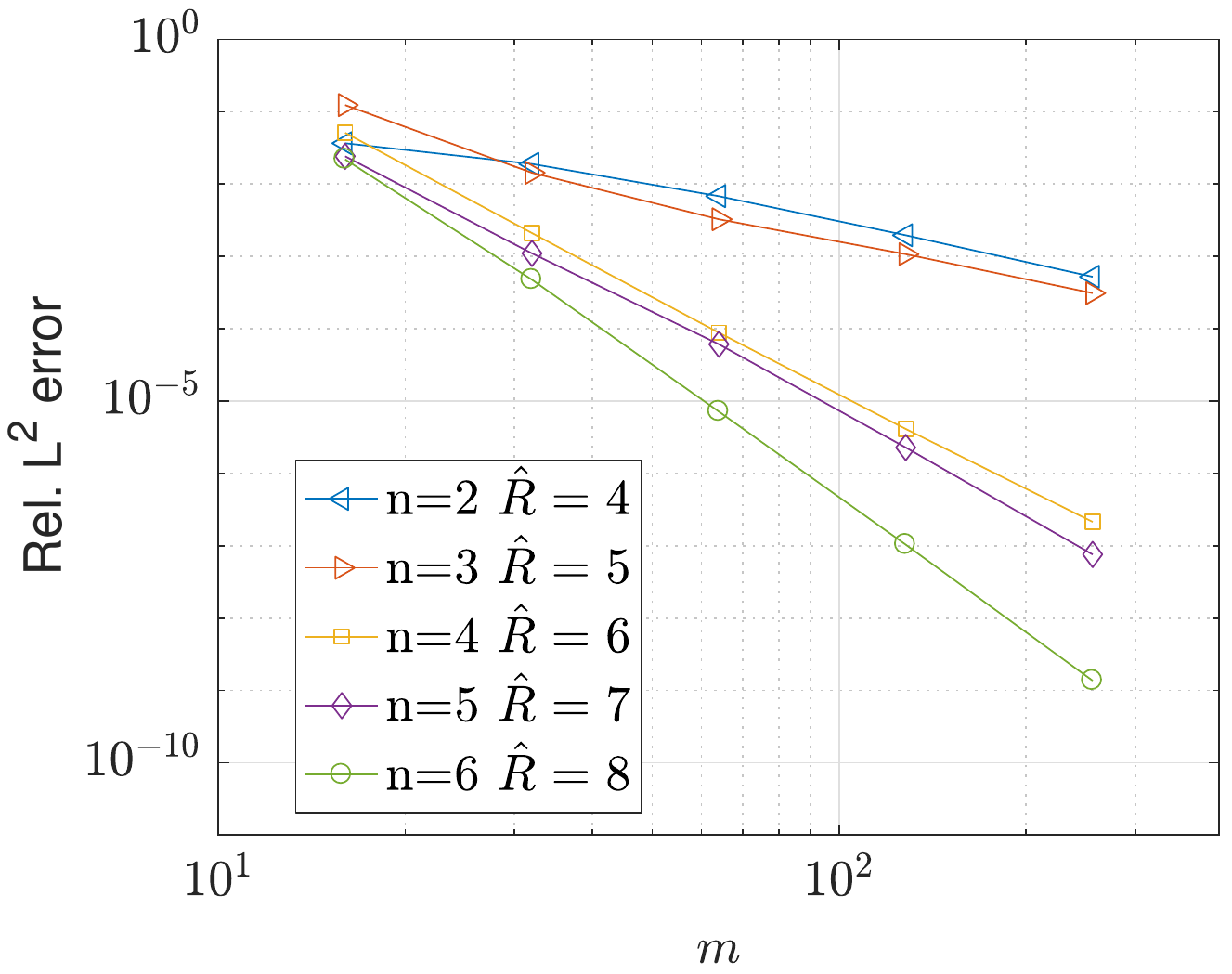}}
	\caption{
		Convergence of  the relative $L^2$  error  for a 1D Helmholtz problem, with different types of boundary conditions, solved   for different values of $n$ and $p=2$.
	} 
	\label{fig:conv_1d_tutt}
\end{figure}

In all the converging curves, the expected convergence rate of $n$ is obtained.
Similar results, in terms of a higher irregularity with higher values of $p$, are  also obtained for odd values ($n=3$ and $n=5$), which also exhibit the expected convergence rates of $n-1$.
A summarizing convergence plot is presented in  Fig.~\ref{fig:conv_1d_tutt} with HOLMES of order $n$ from 2 to 6 and the two types of boundary conditions. 
\red{For this plot, a value of $\hat{R}  = n+2$ is chosen for all the curves. It is worth noting that, according to Figures~\ref{fig:conv_1d_ess} and  ~\ref{fig:conv_1d_nat}, the performance of HOLMES approximants is in general strongly dependent on the locality parameter $\gamma$, with lower values of $\gamma$  corresponding  to a higher support size $\hat{R}$. For each specific case, an optimal value of $\gamma$ exists, below which the accuracy is anyway decreasing. Indeed, based on the curves in the figures, there not seem to be a single optimal value of $\hat{R}$ for all the configurations. Therefore, $\hat{R}  = n+2$ is chosen. 
}

In  Fig.~\ref{fig:conv_1d_tutt} it can be observed how the odd orders have a very similar behavior to the corresponding $n-1$ even orders but, remarkably, the computational times to compute the basis functions  are   higher due to the higher number of constraints to be imposed in problem \eqref{HOLMES}. 

\subsection{Acoustic Helmholtz problems}

\begin{figure} 
	\centering
	\subfigure[ ]
	{\includegraphics[width=0.33\textwidth,clip,keepaspectratio,angle=0]{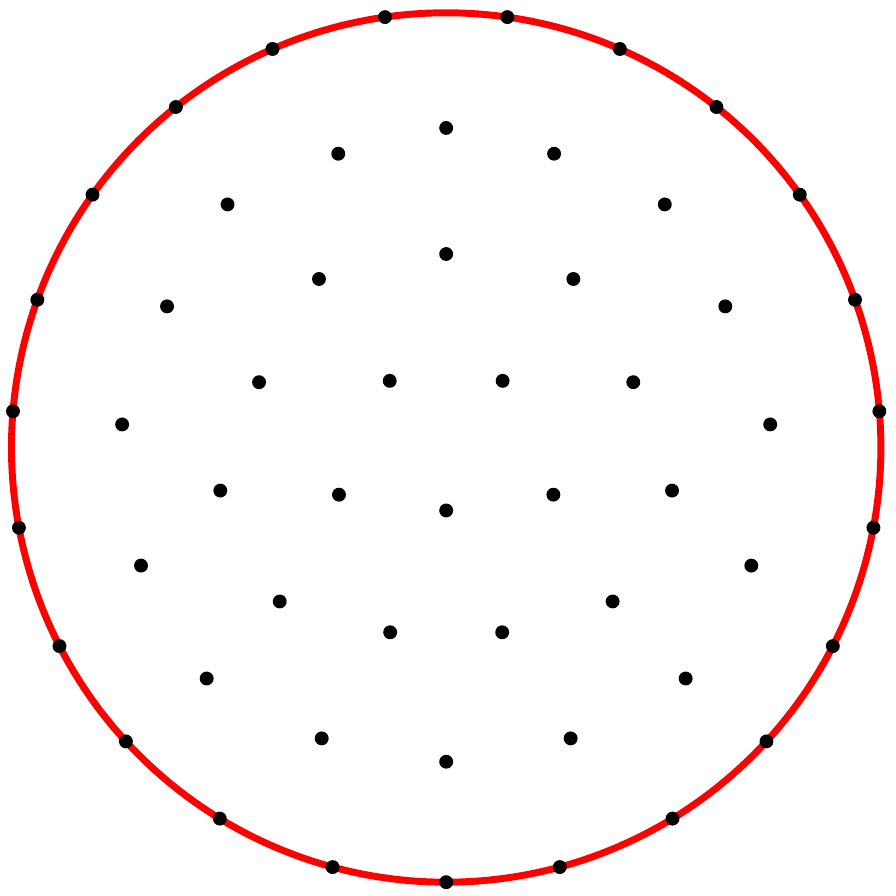}}\;\;\;
	\subfigure[  ]
	{\includegraphics[width=0.33\textwidth,clip,keepaspectratio,angle=0]{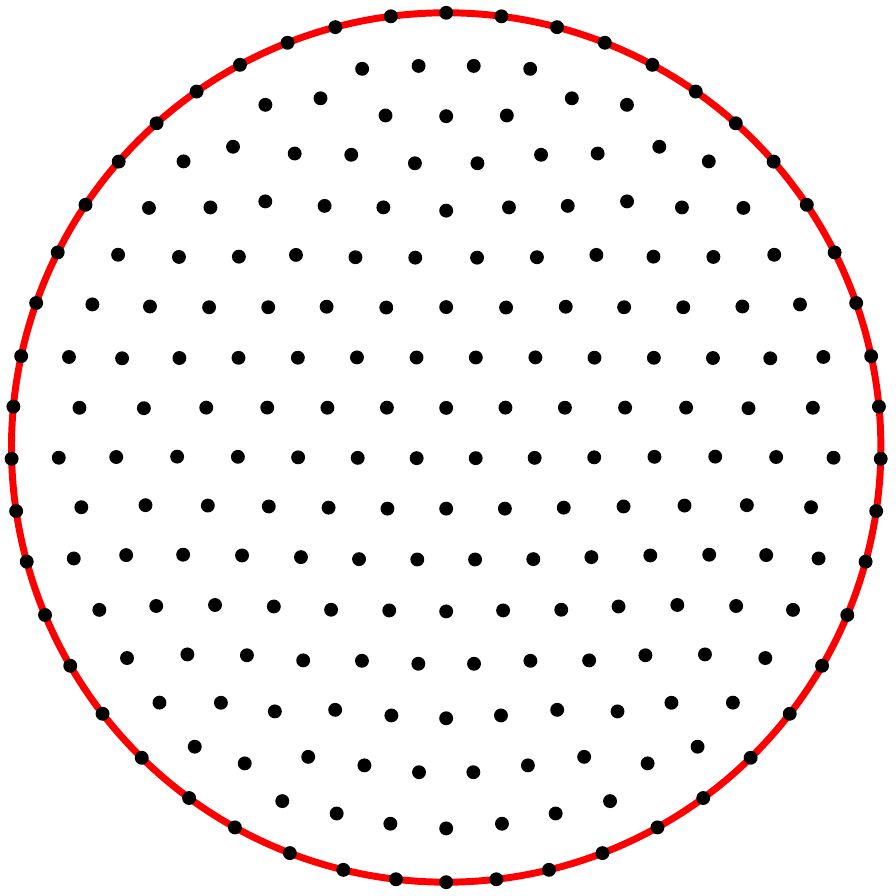}}
	\subfigure[  ]
	{\includegraphics[width=0.33\textwidth,clip,keepaspectratio,angle=0]{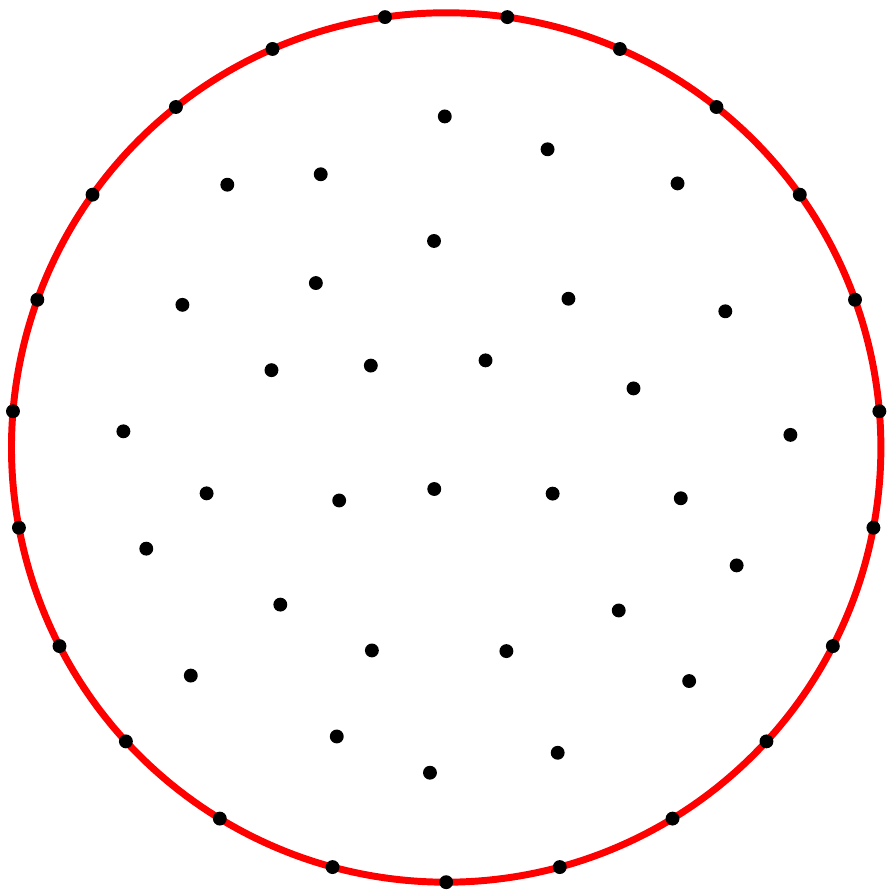}}\;\;\;
	\subfigure[  ]
	{\includegraphics[width=0.33\textwidth,clip,keepaspectratio,angle=0]{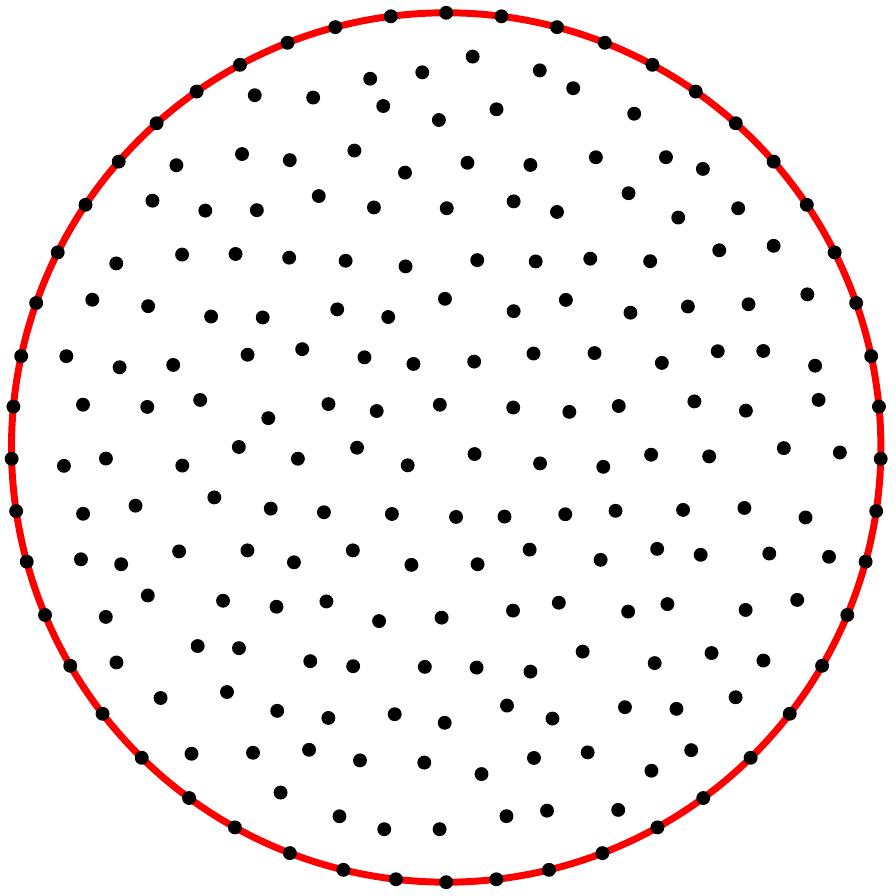}}
	{\includegraphics[width=0.33\textwidth,clip,keepaspectratio,angle=0]{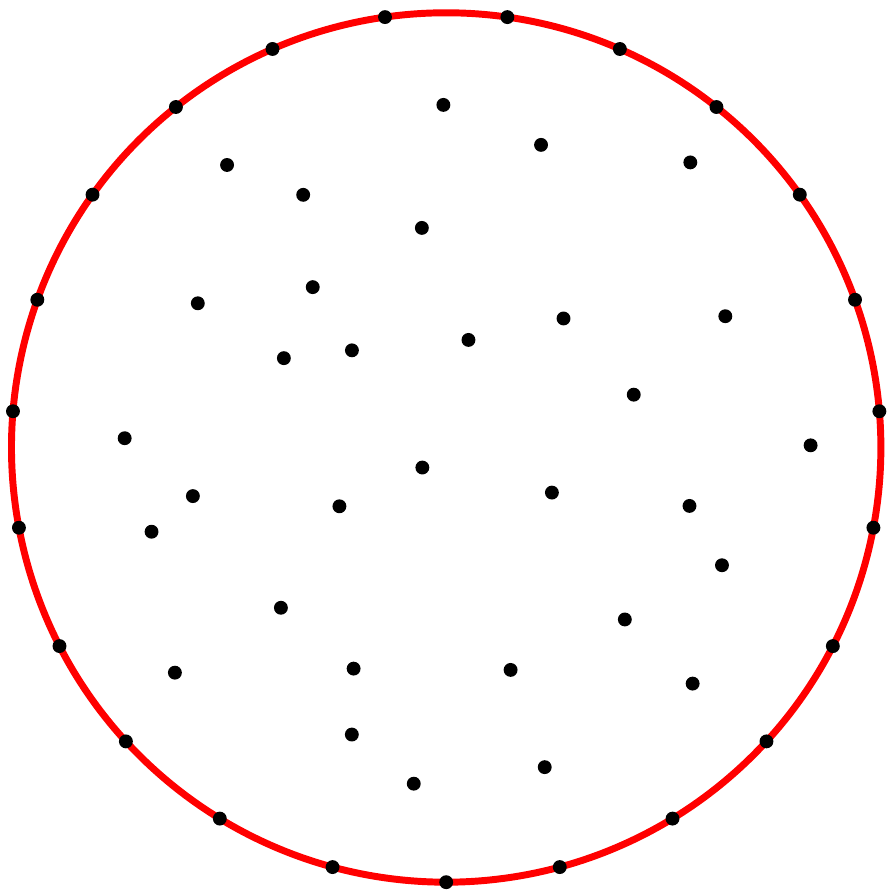}}\;\;\;
	\subfigure[  ]
	{\includegraphics[width=0.33\textwidth,clip,keepaspectratio,angle=0]{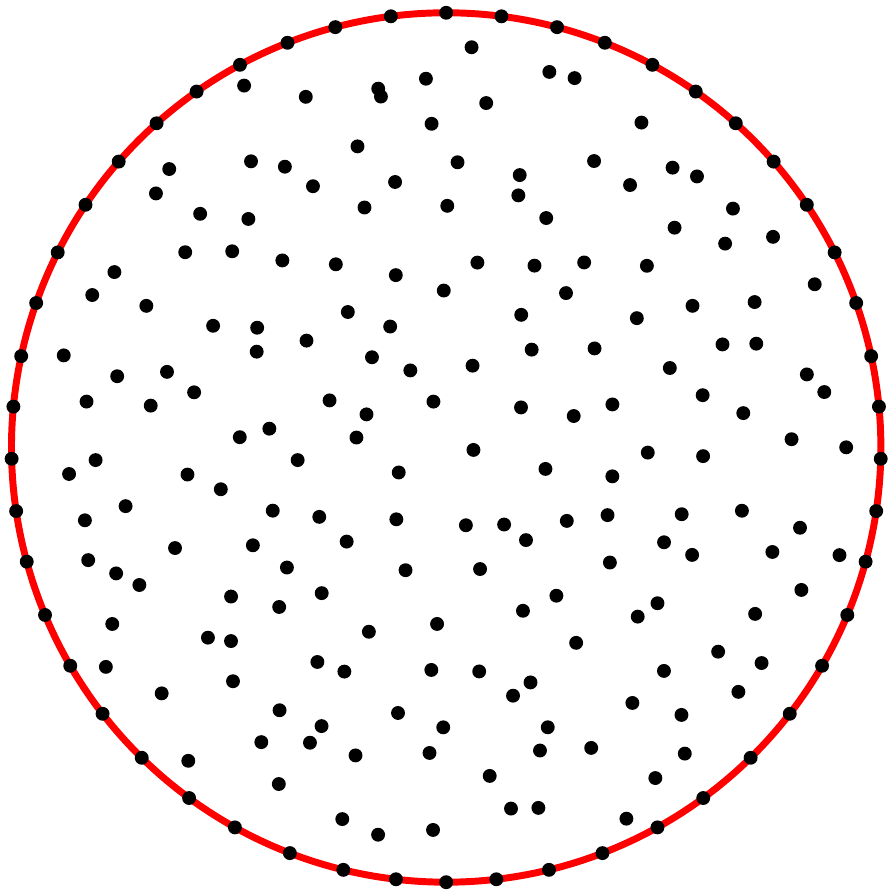}}
	\caption{Nodal discretizations for the Helmholtz problem on a circular domain. The first two grids (a) and (b), with 55 and 229 nodes respectively, are generated with the Matlab mesher \textit{distmesh}. To obtain irregular distributions, the internal nodes are moved by adding a random displacement perturbation  whose components are in the range of $[-0.2h\;0.2h]$ in (c) and (d) and  $[-0.4h\;0.4h]$ in (e) and (f).
	} 
	\label{fig:meshcerchio}
\end{figure}

This section considers the propagation of acoustic waves, described by the following Helmholtz problem:
\begin{equation}
\label{Helmholtz}
\begin{cases}
\Delta p +k^2 p =0  & \text{in } \Omega \\ 
\frac{\jim{j}}{\rho_0 \omega} \nabla p \cdot \bm{n} = \bar{v}_n & \text{on } \partial\Omega,
\end{cases}
\end{equation}
where $p=p(\vx)$ is the acoustic pressure field, $\rho_0$ is the fluid-medium   density, $\jim{j}$ is the imaginary unit and $k$ is the wavenumber, which is related to the angular frequency $\omega$ and the wavelength $\lambda$ by the speed of sound $c$, through the following equations:
\begin{equation}
k=\frac{\omega}{c}=\frac{2 \pi}{\lambda}.
\end{equation} 

The problem is first solved in two dimensions in a circular domain, where it describes the acoustic field generated within a cylinder with infinite vibrating walls. 
The analytical expression of the pressure is   given by 
\begin{equation}
p(r)=-\bar{v}_n \dfrac{J_0(kr)}{\frac{j}{\rho_0 \omega }k J_1(kR)},
\end{equation}
where $J_0$ and $J_1$ are the Bessel functions of the first kind of order $0$ and $1$ respectively \cite{Greco2017d}.
The following parameters are considered: $R=\SI{1}{m}$,
 $\rho_0 =\SI{1.21}{ kg/m^3}$, $c= \SI{343}{ m/s} $, $\bar{v}_n= \SI{1}{m/s} $
 and the problem is solved for a wavelength $\lambda=\SI{2.5}{m}$,  corresponding to $\omega = \SI{862.05}{rad/s} $.
The computational grid on the circular domain is generated using the MATLAB finite element mesh generator \textit{distmesh}, which provides   very regular nodal distributions \cite{distmesh}, as outlined in Fig.~\ref{fig:meshcerchio}~a-b.

\begin{figure} 
	\centering
	\subfigure[$n=4,p=2$]
	{\includegraphics[width=0.32\textwidth,clip,keepaspectratio,angle=0]{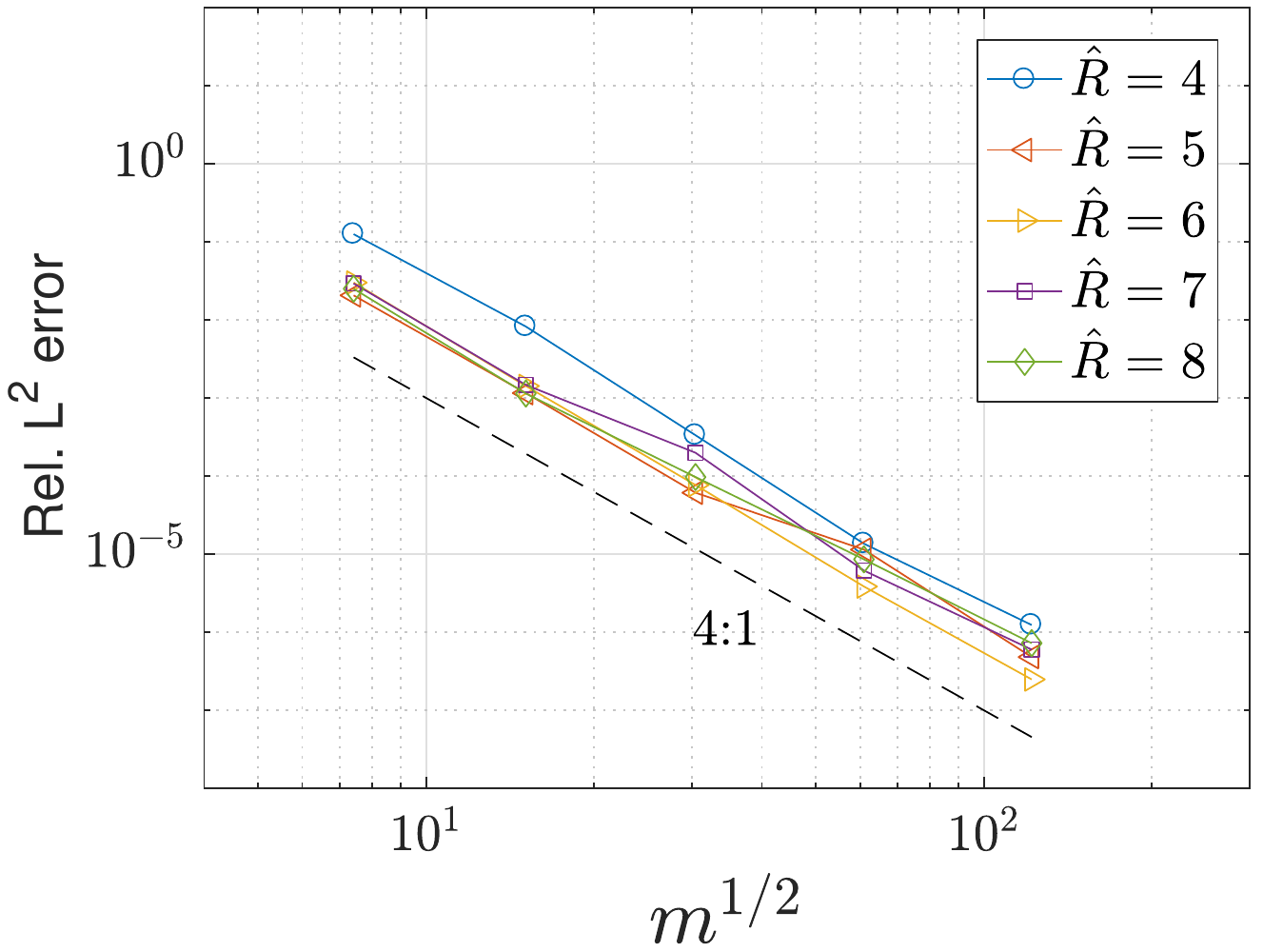}}
	\subfigure[$n=4,p=4$]
	{\includegraphics[width=0.32\textwidth,clip,keepaspectratio,angle=0]{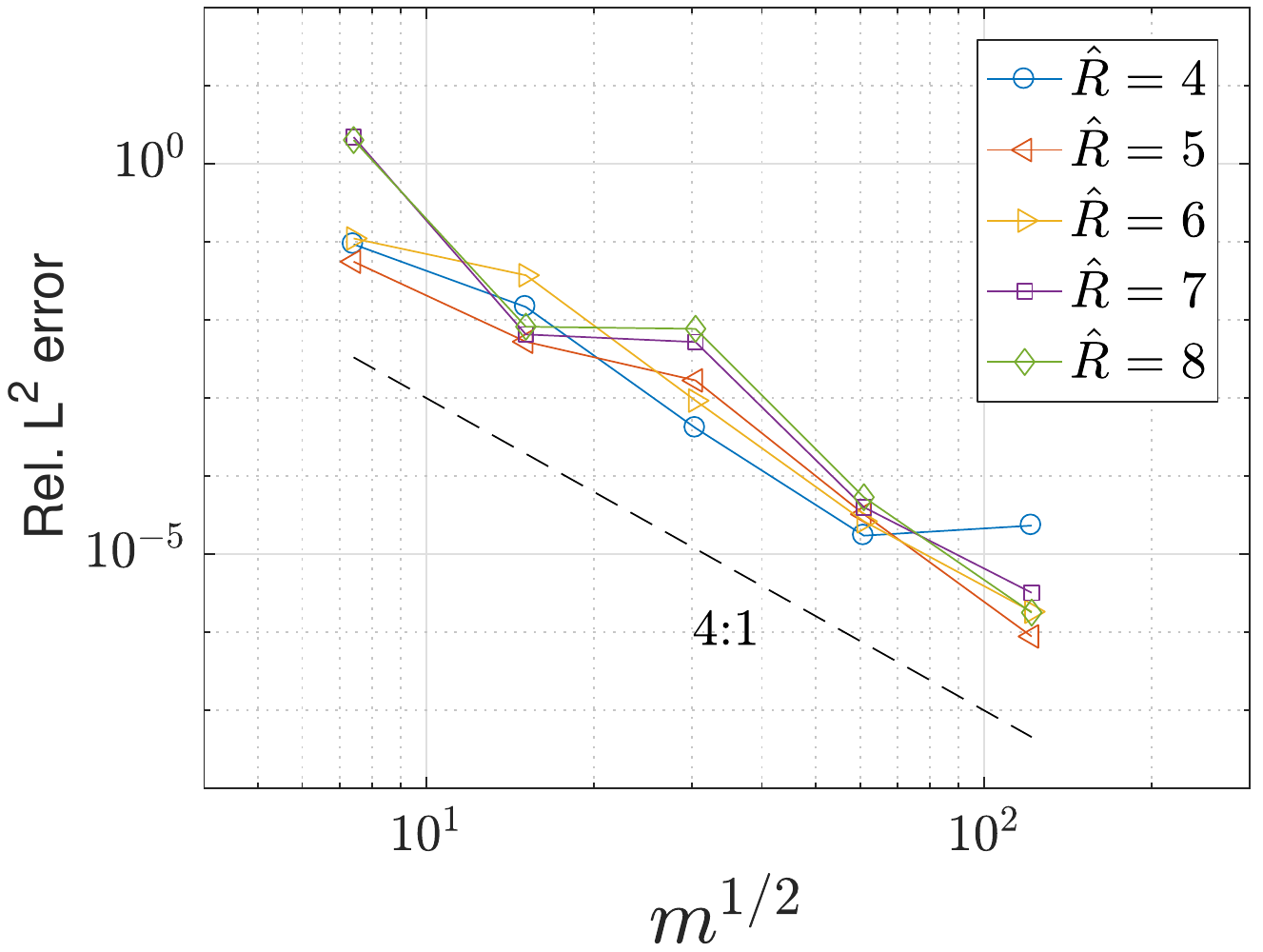}}
	\subfigure[$n=4,p=6$]
	{\includegraphics[width=0.32\textwidth,clip,keepaspectratio,angle=0]{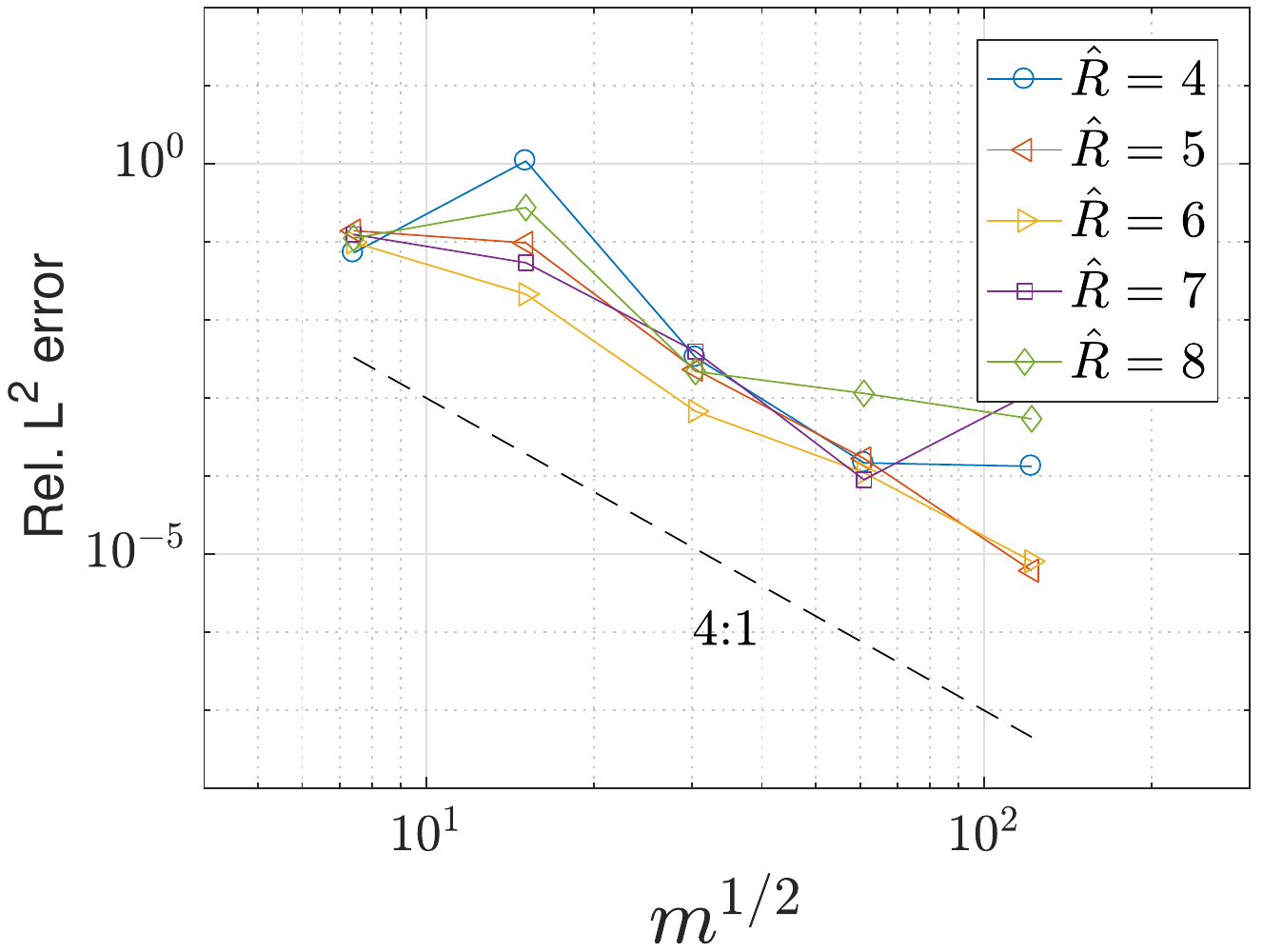}}
	\caption{
		Convergence of  the relative $L^2$  and $H^1$ error   for the 2D acoustic Helmholtz problem  solved  for $n=4$ and different values of and $p$.
	} 
	\label{fig:conv_2d_he}
\end{figure}

\begin{figure} 
	\centering
	\subfigure[ ]
	{\includegraphics[width=0.4\textwidth,clip,keepaspectratio,angle=0]{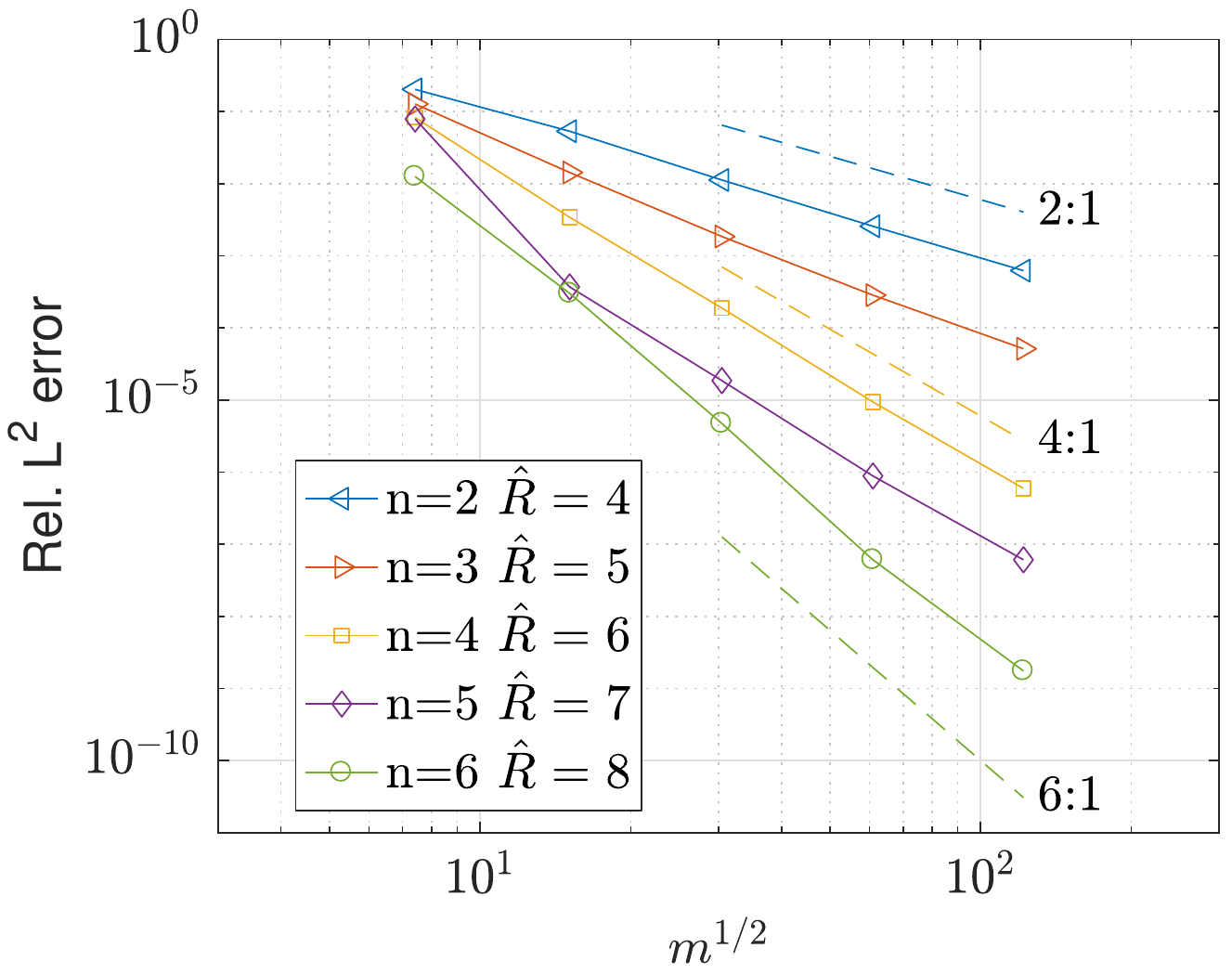}}
			\subfigure[ ]
	{\includegraphics[width=0.4\textwidth,clip,keepaspectratio,angle=0]{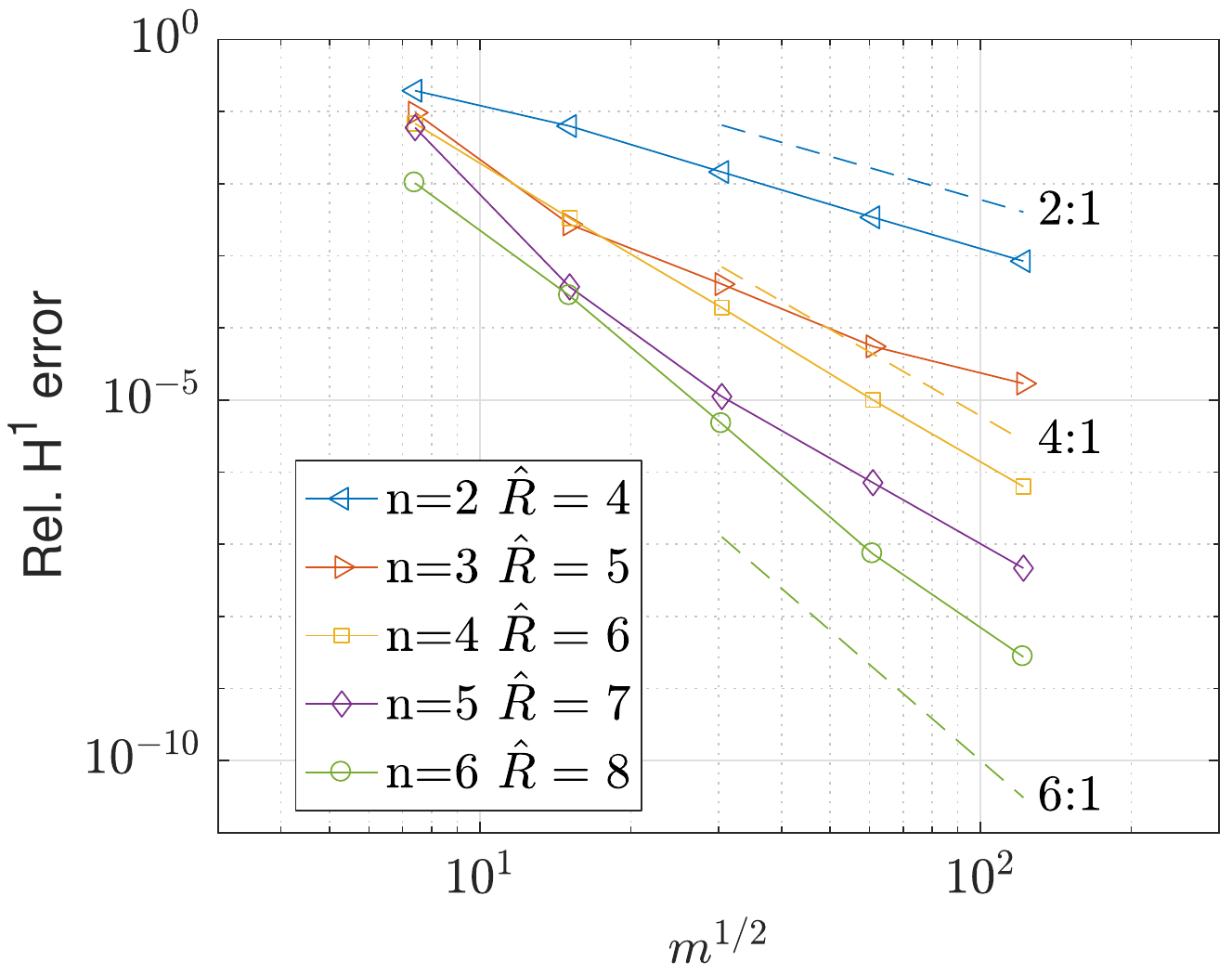}}
	\subfigure[ ]
	{\includegraphics[width=0.4\textwidth,clip,keepaspectratio,angle=0]{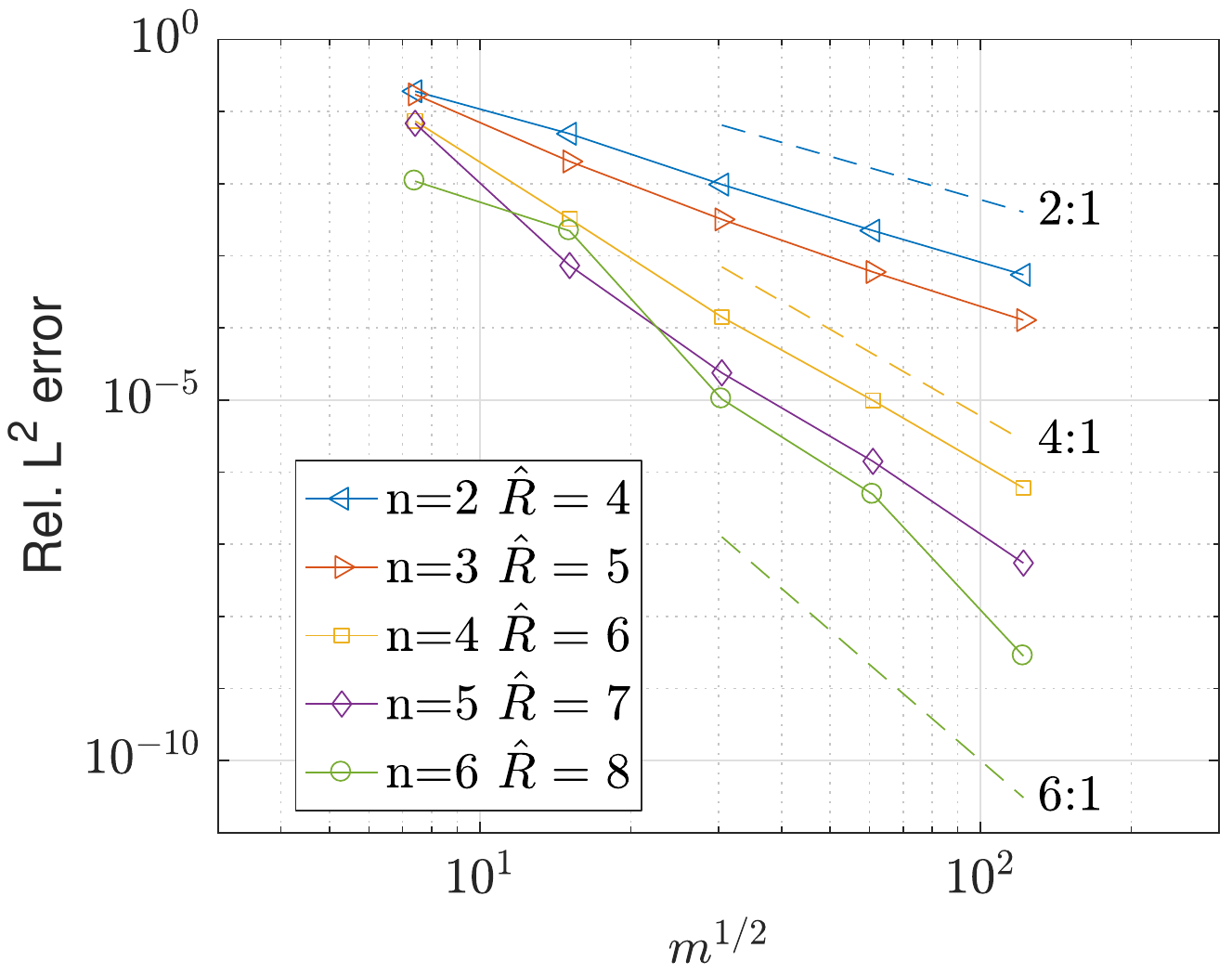}}
			\subfigure[ ]
	{\includegraphics[width=0.4\textwidth,clip,keepaspectratio,angle=0]{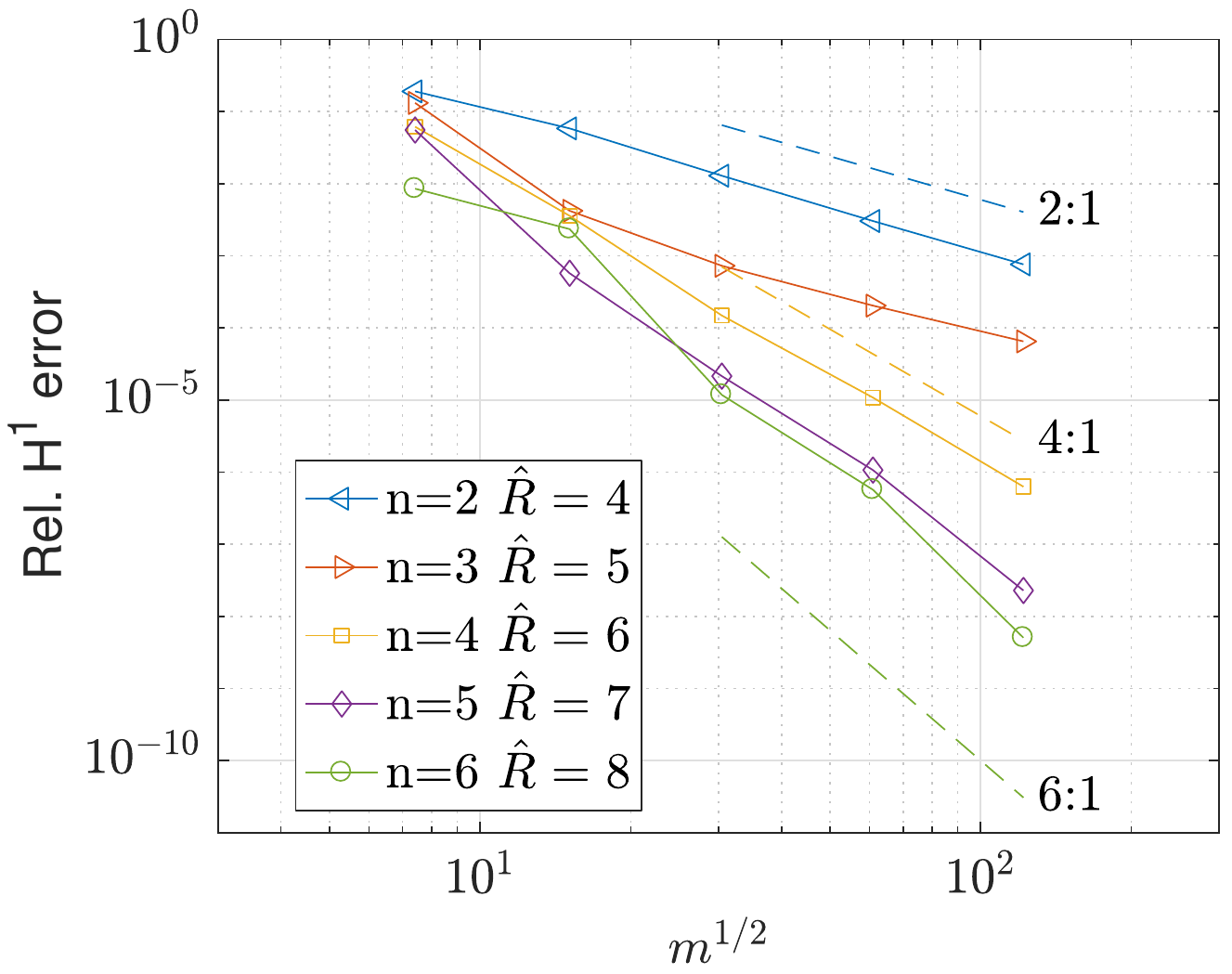}}
	\subfigure[ ]
	{\includegraphics[width=0.4\textwidth,clip,keepaspectratio,angle=0]{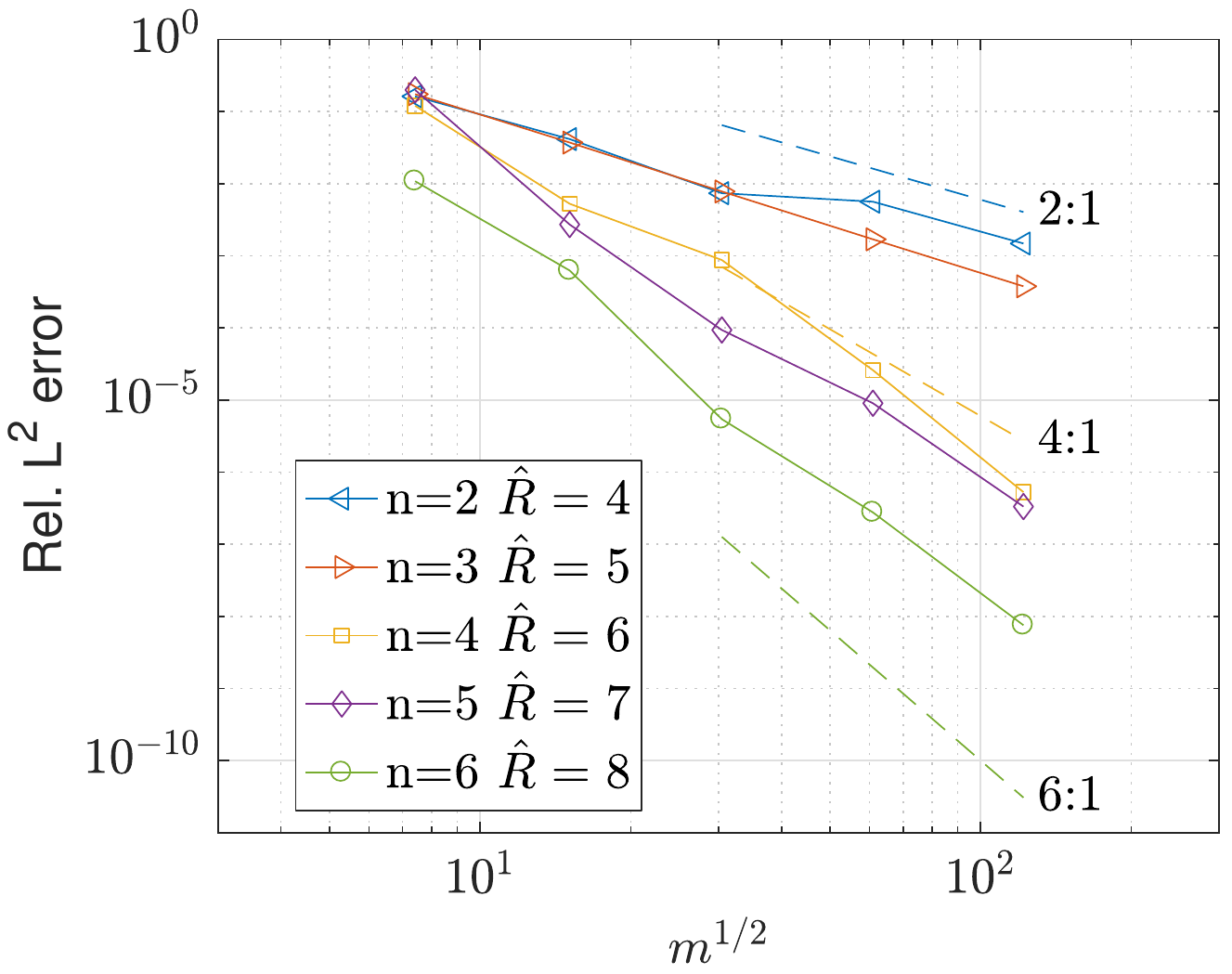}}
			\subfigure[ ]
	{\includegraphics[width=0.4\textwidth,clip,keepaspectratio,angle=0]{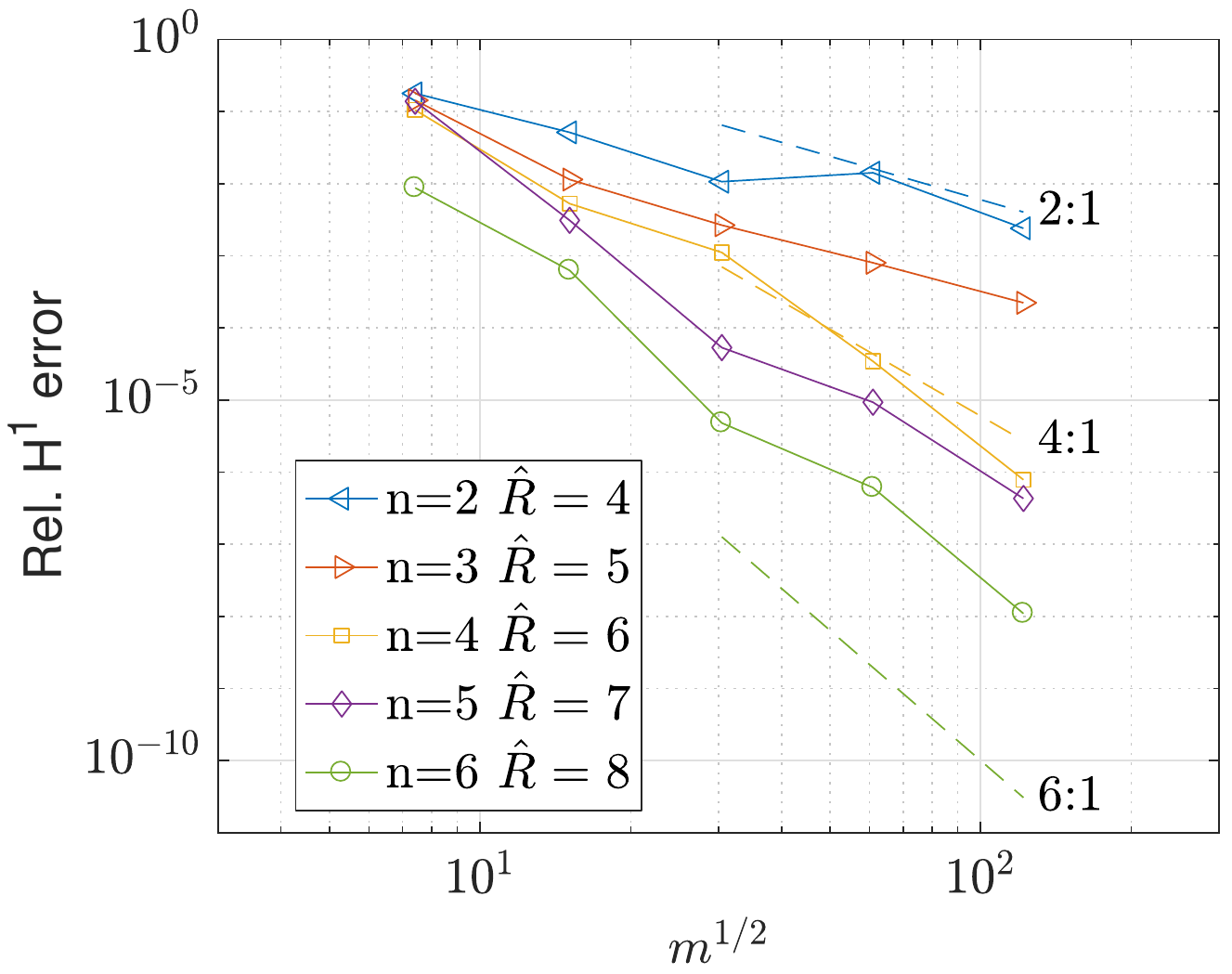}}
		\caption{
		Convergence of the relative $L^2$  and $H^1$ error  for the 2D acoustic Helmholtz problem  solved  for different values of $n$ and   $p=2$. Regular nodal distributions are considered in (a) and (b), while a perturbation of $0.2h$ is considered in (c) and (d) and a perturbation of $0.4h$ is considered in (e) and (f).  }
	\label{fig:conv_2d_he_tu}
\end{figure}

Convergence curves of the relative discrete $L^2$ error, $E_r(u,u^h)$, in function of the square root of the number of nodes $m^{1/2}$ are plotted in Fig.~\ref{fig:conv_2d_he}, for HOLMES approximants of order $n=4$ and different values of $p$ and $\hat{R}$.
This two-dimensional application confirms that a more regular convergence behavior is obtained for  $p=2$ and similar results, not included for the sake of brevity, are found for HOLMES of different orders.
In Fig.~\ref{fig:conv_2d_he_tu}a, orders  from 2 to 6 with $p=2$ are considered and are shown to test the expected convergence rates for the $L^2$ error. 
Unlike the one-dimensional case, HOLMES of odd order show better accuracy than the corresponding $n-1$ even orders, but the asymptotic rates still verify the theoretical prediction of a slope given by $n-1$.
\begin{MyColorPar}{red}
Fig.~\ref{fig:conv_2d_he_tu}b considers also the relative error in the discrete $H^1$ semi-norm, which is defined according to \eq{eq:er} as $E_r(\nabla u,\nabla u^h)$.
 The expected convergence rates, equal to those of  the $L^2$ error, are observed.  

To study the performance of HOLMES collocation on irregular nodal distributions, the grids of the convergence study are perturbed by adding a random displacement to the internal nodes, as shown in Fig.~\ref{fig:meshcerchio}~c-f. 
Two cases are considered with an added displacement whose components are in the range of $[-0.2h\;0.2h]$ and $[-0.4h\;0.4h]$ respectively.
The corresponding convergence curves for the relative discrete $L^2$ and $H^1$ errors are reported in Fig.~\ref{fig:conv_2d_he_tu}~c-f.
Although a slight decrease of accuracy can be observed, the curves  converge with the expected rates   in both cases, which confirms the validity of the collocation approach when using irregular nodal distributions. Such configurations can arise for instance in dynamical large deformation problems.
\end{MyColorPar}

The Helmholtz problem stated by \eq{Helmholtz} is studied also in three dimensions to compute the acoustic field inside a sphere with vibrating walls of radius $R=\SI{1}{m}$. The same acoustic parameters and wavelength of the 2D case are considered.
In three dimensions, the analytical solution of the problem is given
by
\begin{equation}
p(r)= -\jim{j} \bar{v}_n \rho_0 \omega \frac{e^{-\jim{j}k (r-1)} \left(e^{2 \jim{j}k r}-1\right)}{\left[e^{2
		\jim{j}k} \left(\jim{j}k-1\right)+\jim{j}k+1\right] r}.
\end{equation}

%

\begin{figure}  
	\centering
	\subfigure[ ]
{\includegraphics[width=0.4\textwidth,clip,keepaspectratio,angle=0]{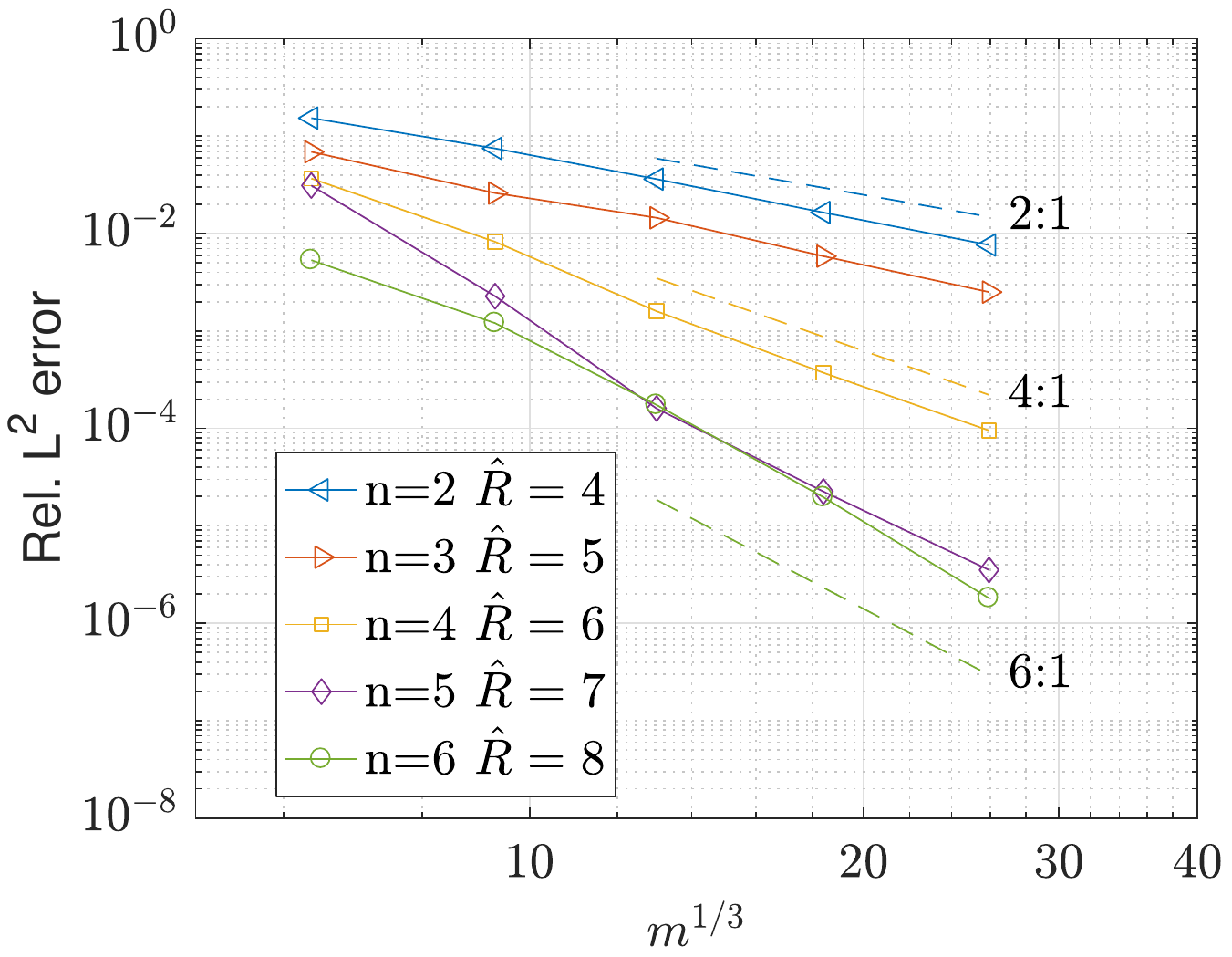}}
\subfigure[ ]
{\includegraphics[width=0.4\textwidth,clip,keepaspectratio,angle=0]{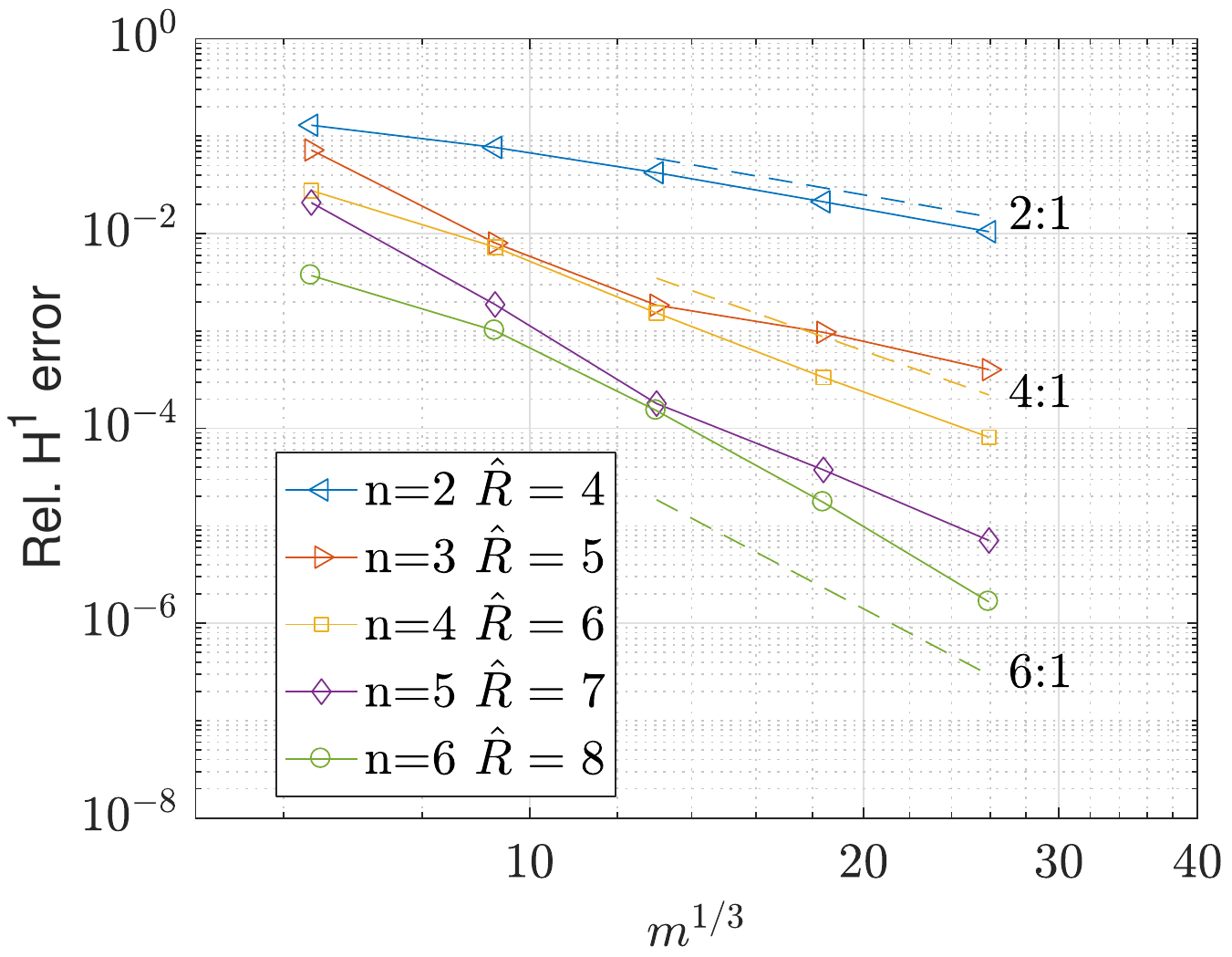}}
	\caption{
		Convergence of the relative $L^2$  and $H^1$ error  for the 3D acoustic Helmholtz problem solved  for different values of $n$ and $p=2$.} 
	\label{fig:conv_3d_he_tu}
\end{figure}

The convergence results are reported in Fig.~\ref{fig:conv_3d_he_tu} for HOLMES of  orders $n$ from 2 to 6 and a locality norm $p=2$.
A similar convergence behavior to  the 2D case is  obtained  and the expected convergence rates are verified for both the $L^2$ and the $H^1$ error.

\subsection{Linear elasticity}

\subsubsection{Infinite elastic plate with hole}

\begin{figure}  [t]
	\centering{
		
	\immediate\write18{svgmodified plate 1 drawing}
	\def\svgscale{.4}
\begingroup%
  \makeatletter%
  \providecommand\color[2][]{%
    \errmessage{(Inkscape) Color is used for the text in Inkscape, but the package 'color.sty' is not loaded}%
    \renewcommand\color[2][]{}%
  }%
  \providecommand\transparent[1]{%
    \errmessage{(Inkscape) Transparency is used (non-zero) for the text in Inkscape, but the package 'transparent.sty' is not loaded}%
    \renewcommand\transparent[1]{}%
  }%
  \providecommand\rotatebox[2]{#2}%
  \newcommand*\fsize{\dimexpr\f@size pt\relax}%
  \newcommand*\lineheight[1]{\fontsize{\fsize}{#1\fsize}\selectfont}%
  \ifx\svgwidth\undefined%
    \setlength{\unitlength}{548.27440103bp}%
    \ifx\svgscale\undefined%
      \relax%
    \else%
      \setlength{\unitlength}{\unitlength * \real{\svgscale}}%
    \fi%
  \else%
    \setlength{\unitlength}{\svgwidth}%
  \fi%
  \global\let\svgwidth\undefined%
  \global\let\svgscale\undefined%
  \makeatother%
  \begin{picture}(1,0.44065526)%
    \lineheight{1}%
    \setlength\tabcolsep{0pt}%
    \put(0,0){\includegraphics[width=\unitlength,page=1]{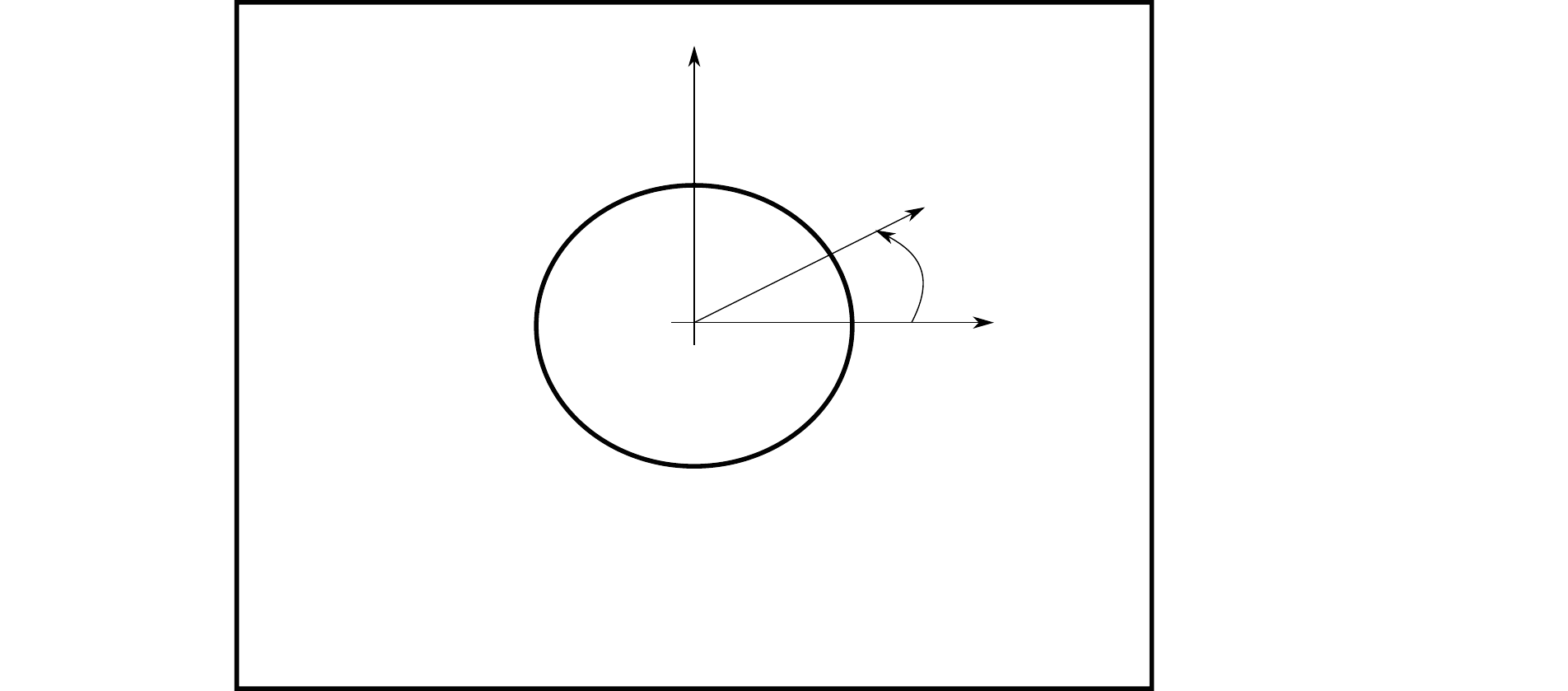}}%
    \put(0.64127224,0.22032761){\color[rgb]{0,0,0}\makebox(0,0)[lt]{\lineheight{0}\smash{\begin{tabular}[t]{l}$x$\end{tabular}}}}%
    \put(0.46034094,0.40417718){\color[rgb]{0,0,0}\makebox(0,0)[lt]{\lineheight{0}\smash{\begin{tabular}[t]{l}$y$\end{tabular}}}}%
    \put(0.59603941,0.31517063){\color[rgb]{0,0,0}\makebox(0,0)[lt]{\lineheight{0}\smash{\begin{tabular}[t]{l}$r$\end{tabular}}}}%
    \put(0.60333503,0.27139693){\color[rgb]{0,0,0}\makebox(0,0)[lt]{\lineheight{0}\smash{\begin{tabular}[t]{l}$\theta$\end{tabular}}}}%
    \put(0,0){\includegraphics[width=\unitlength,page=2]{plate.pdf}}%
    \put(0.90975094,0.20281813){\color[rgb]{0,0,0}\makebox(0,0)[lt]{\lineheight{0}\smash{\begin{tabular}[t]{l}$T_\infty$\end{tabular}}}}%
  \end{picture}%
\endgroup%

		\caption{Elastic plate with a circular hole under tension.}
		\label{fig:plate}}
\end{figure}

This section considers the well known benchmark problem  of an infinite plate with a circular hole, subject to an uni-axial traction $T_\infty$ on the remote boundary, as outlined in Fig.~\ref{fig:plate}.
The analytical solution for the displacement field is given in polar coordinates as \cite{Timoshenko1934} 
\begin{equation}
\begin{aligned}
u(r,\theta)&=T_\infty\dfrac{R}{8G}     \bigg\{   \dfrac  {r}{R} (\kappa+1)   \cos{\theta}   + \dfrac  {2r}{R}   \big[   (1+\kappa)   \cos{\theta} +\cos{3\theta}       \big]    - \dfrac  {2r^3}{R^3}  \cos{3\theta}   \bigg\}         ,\\
v(r,\theta)&=T_\infty\dfrac{R}{8G}     \bigg\{   \dfrac  {r}{R} (\kappa-3)   \sin{\theta}   + \dfrac  {2r}{R}   \big[   (1-\kappa)   \sin{\theta} +\sin{3\theta}       \big]    - \dfrac  {2r^3}{R^3}  \sin{3\theta}   \bigg\}      ,\\
\end{aligned}
\end{equation}
being $G$ the shear modulus and $\kappa$  the Kolosov constant. In particular, a state of plane stress is assumed, such that it takes the value
\begin{equation}
\kappa=\dfrac  {3- \nu}{1+\nu}.
\end{equation}
The corresponding stress distribution is given by

\begin{equation}
\begin{aligned}
\sigma_x(r,\theta)&=T_\infty\left[ 1 -\dfrac{R^2}{r^2}\left( \dfrac 32 \cos{2\theta}+\cos{4\theta}\right) +\dfrac{3R^4}{2r^4}\cos{4\theta}\right],\\
\sigma_y(r,\theta)&=T_\infty\left[-\dfrac{R^2}{r^2}\left( \dfrac 12 \cos{2\theta}-\cos{4\theta}\right) - \dfrac{3R^4}{2r^4}\cos{4\theta}\right],\\
\tau_{xy}(r,\theta)&=T_\infty\left[-\dfrac{R^2}{r^2}\left( \dfrac 12 \sin{2\theta}+\sin{4\theta}\right) +\dfrac{3R^4}{2r^4}\sin{4\theta}\right].
\end{aligned}
\end{equation}

\begin{MyColorPar}{red}
The problem is solved imposing essential boundary conditions, for a value of  $T_\infty=1$ and taking as material parameters $E=1$ and $\nu=0.3$. A computational  domain $[-4,4] \times [-4,4]$, with an interior circle of radius $R=1$ is considered and, because of symmetry, only a quarter of the plate is studied.    
Two types of discretizations are used: regular computational grids created with the \textit{distmesh} generator and irregular ones obtained by adding a nodal perturbation whose components are in the range of $[-0.2h\;0.2h]$, as outlined in Fig.~\ref{fig:nod_elas}. 

The convergence of the relative $L^2$ error of the displacement and   stress fields in function of the square root of 
the number of nodes $m^{1/2}$ 
are reported in Fig.~\ref{fig:conv_elas}.
It can be observed how the expected convergence rates are obtained for both the displacement and the stress fields.
For the latter, we note that the stress error in the $L^2$ norm is expected to converge as a displacement error in the $H^1$ semi-norm. 
Remarkably, also for this problem, the irregular nodal distribution provides only a small accuracy shift of some curves, without altering the correct convergence rates.

\begin{figure} 
	\centering 
	\subfigure[  ]
	{\includegraphics[width=0.40\textwidth,clip,keepaspectratio,angle=0]{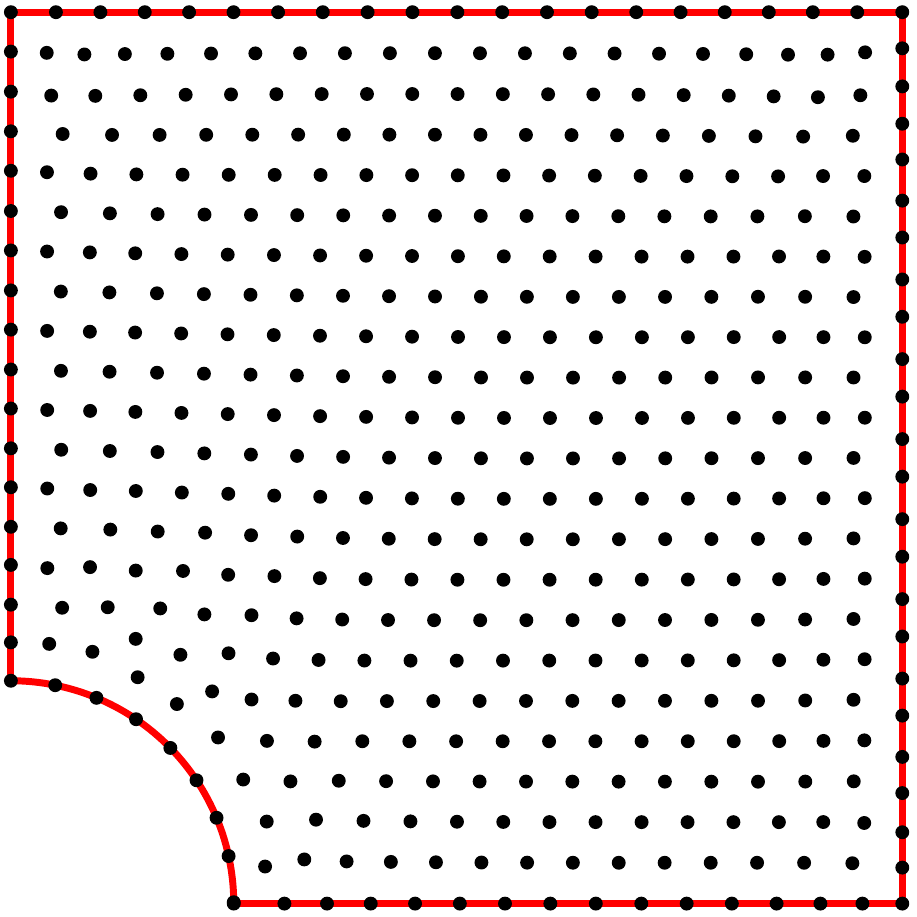}}\;\;\;
	\subfigure[ ]
	{\includegraphics[width=0.40\textwidth,clip,keepaspectratio,angle=0]{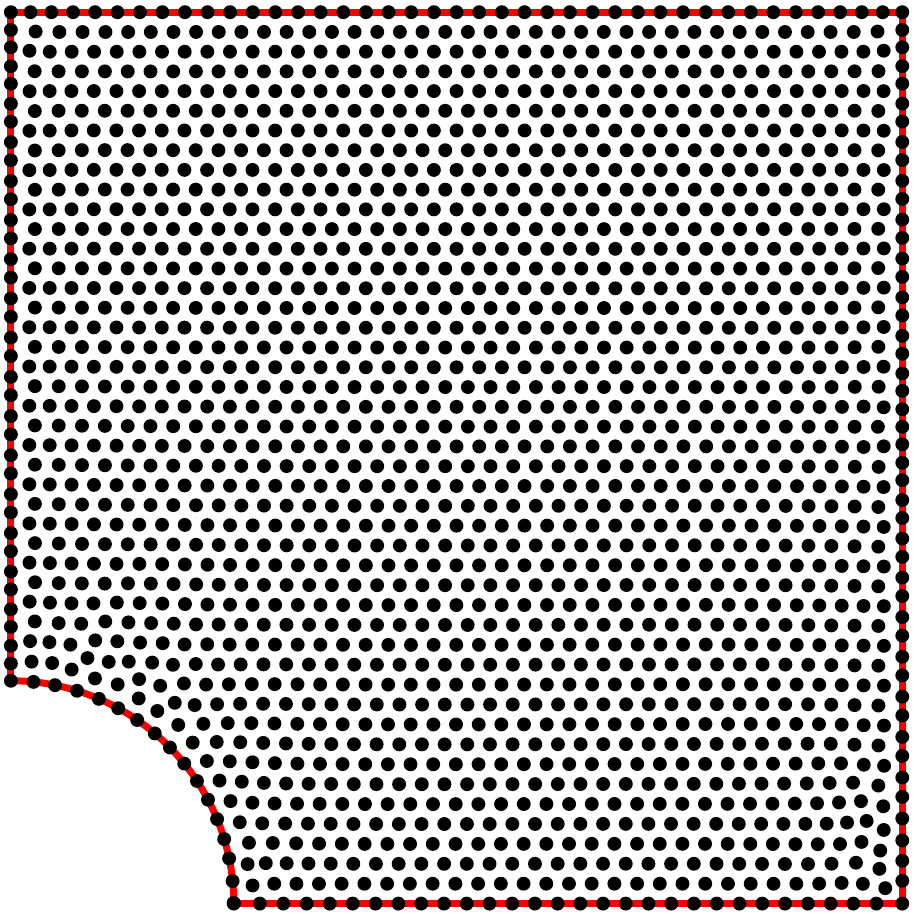}}
	\subfigure[  ]
	{\includegraphics[width=0.40\textwidth,clip,keepaspectratio,angle=0]{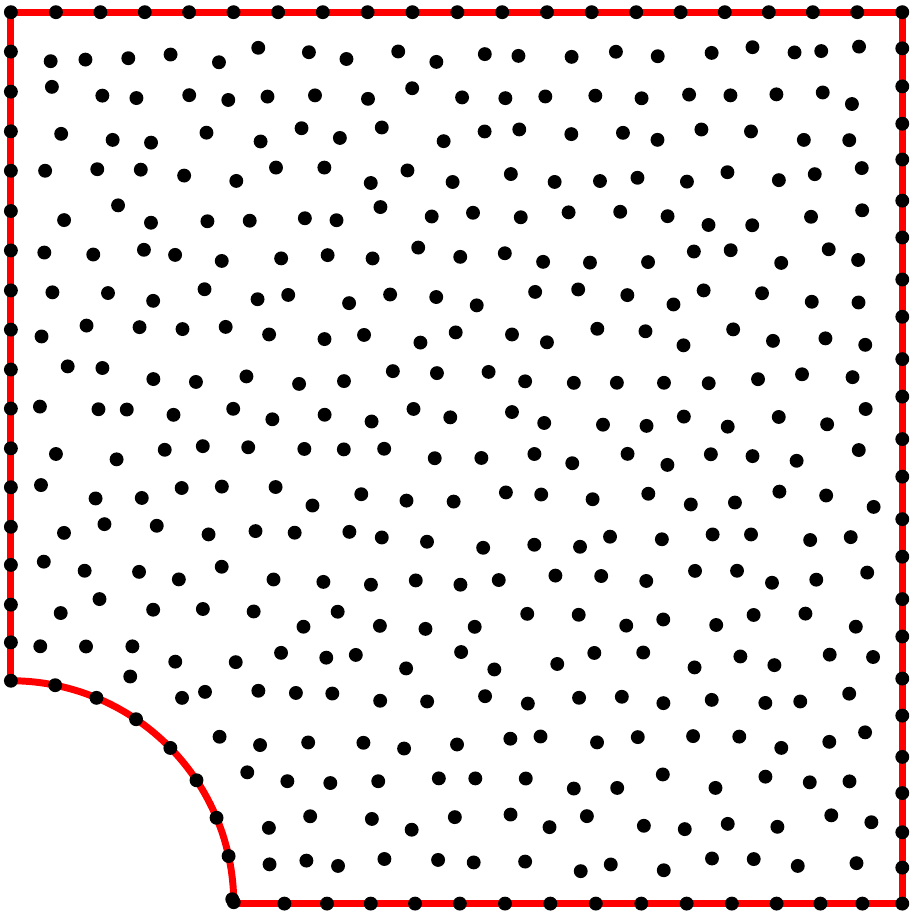}}\;\;\;
	\subfigure[ ]
	{\includegraphics[width=0.40\textwidth,clip,keepaspectratio,angle=0]{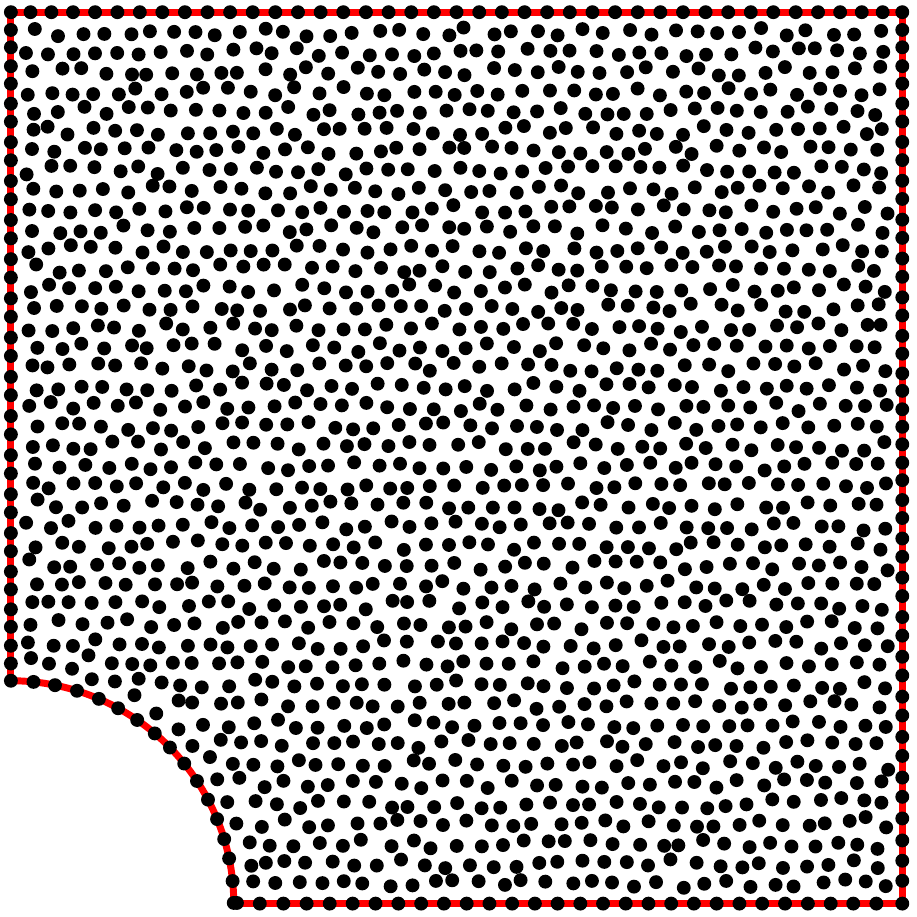}}
	\caption{Nodal discretizations for the elastic plate with hole problem. The first two discretizations (a) and (b), with 456 and 1779 nodes respectively, are generated with the Matlab mesher \textit{distmesh}. To obtain irregular distributions, the internal nodes are moved by adding a random displacement perturbation  whose components are in the range of $[-0.2h\;0.2h]$ in (c) and (d).
	} 
	\label{fig:nod_elas}
\end{figure}

\begin{figure}  
	\centering
	\subfigure[Displacements ]
	{\includegraphics[width=0.49\textwidth,clip,keepaspectratio,angle=0]{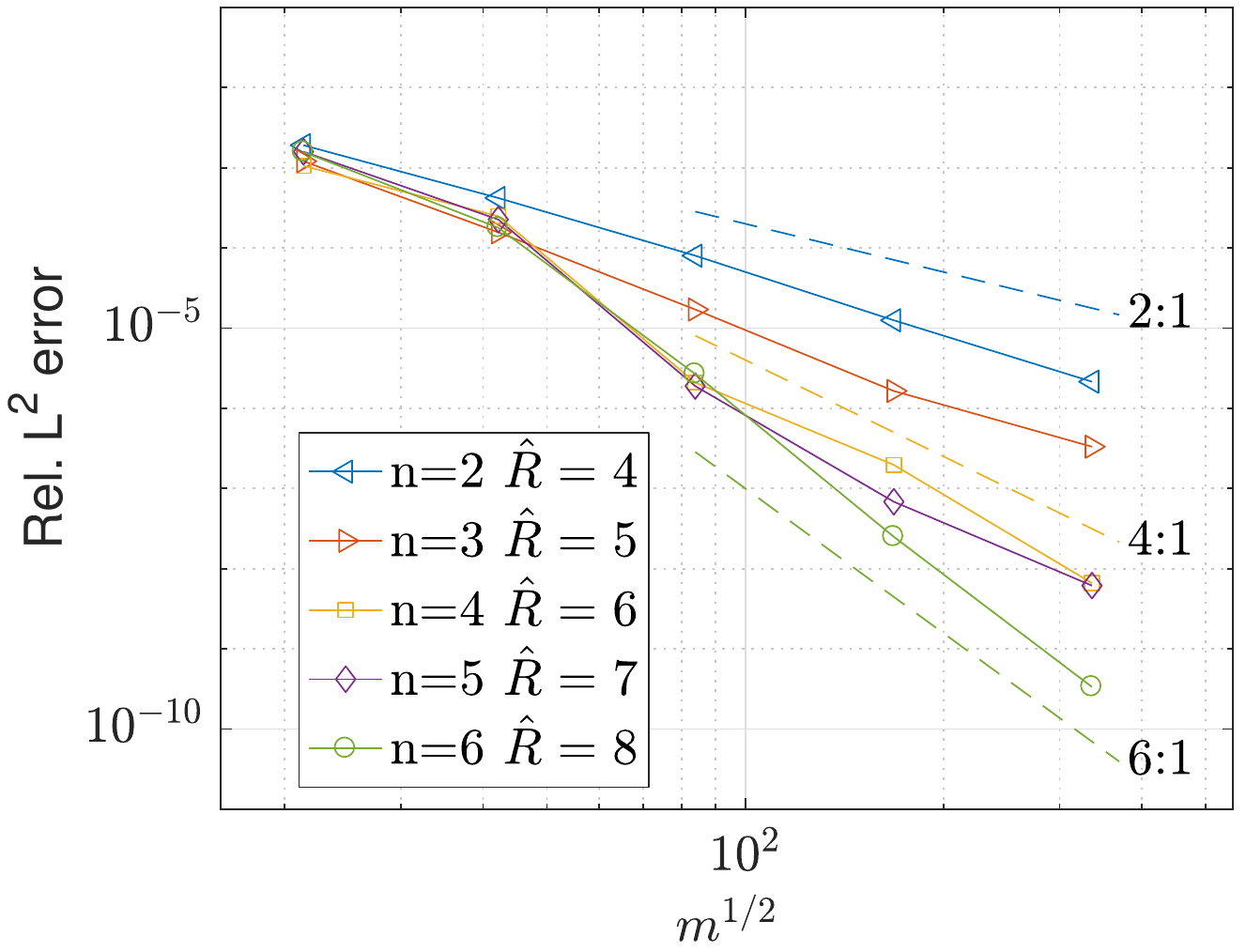}}
		\subfigure[Stress ]
		{\includegraphics[width=0.49\textwidth,clip,keepaspectratio,angle=0]{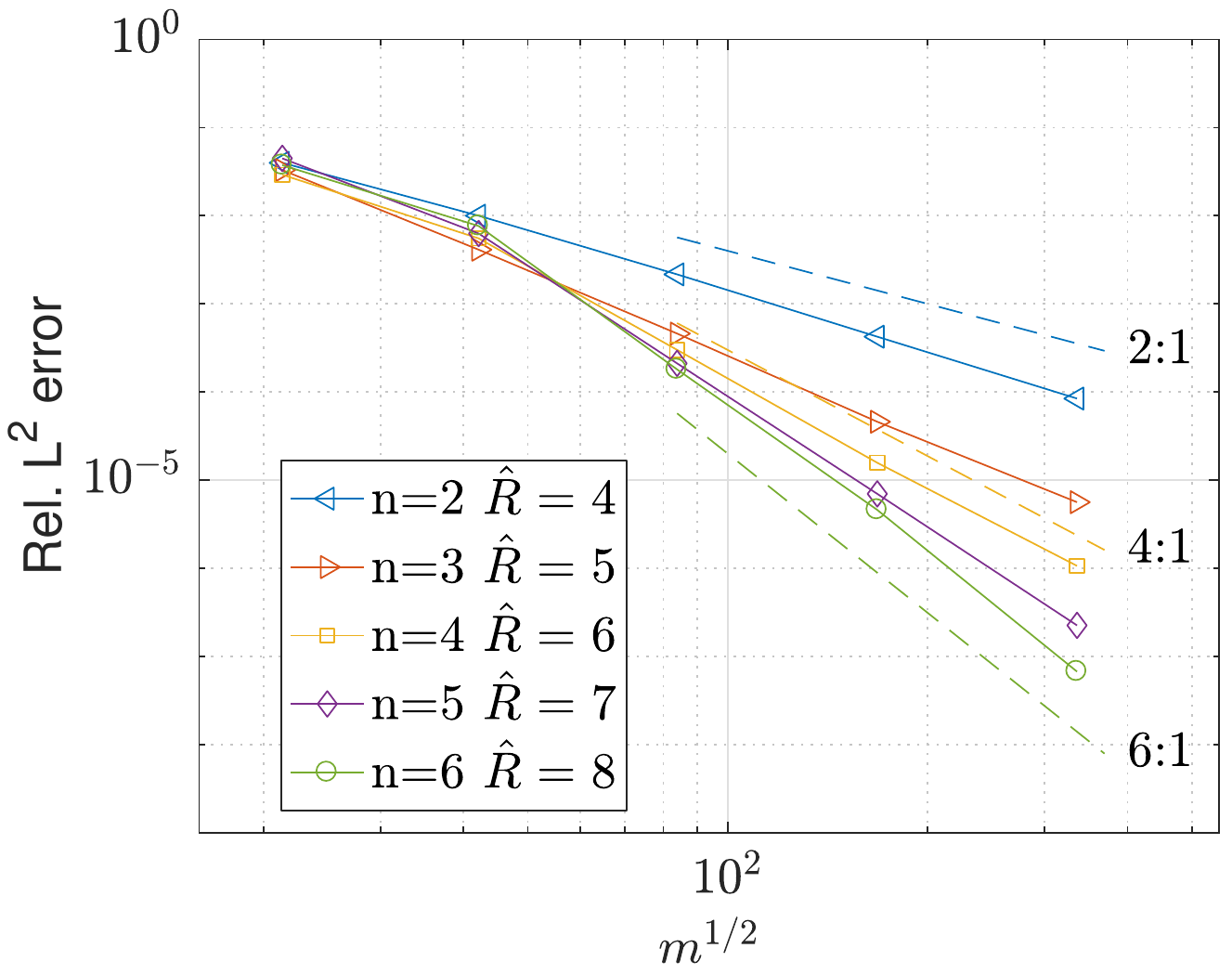}}
			\subfigure[Displacements ]
		{\includegraphics[width=0.49\textwidth,clip,keepaspectratio,angle=0]{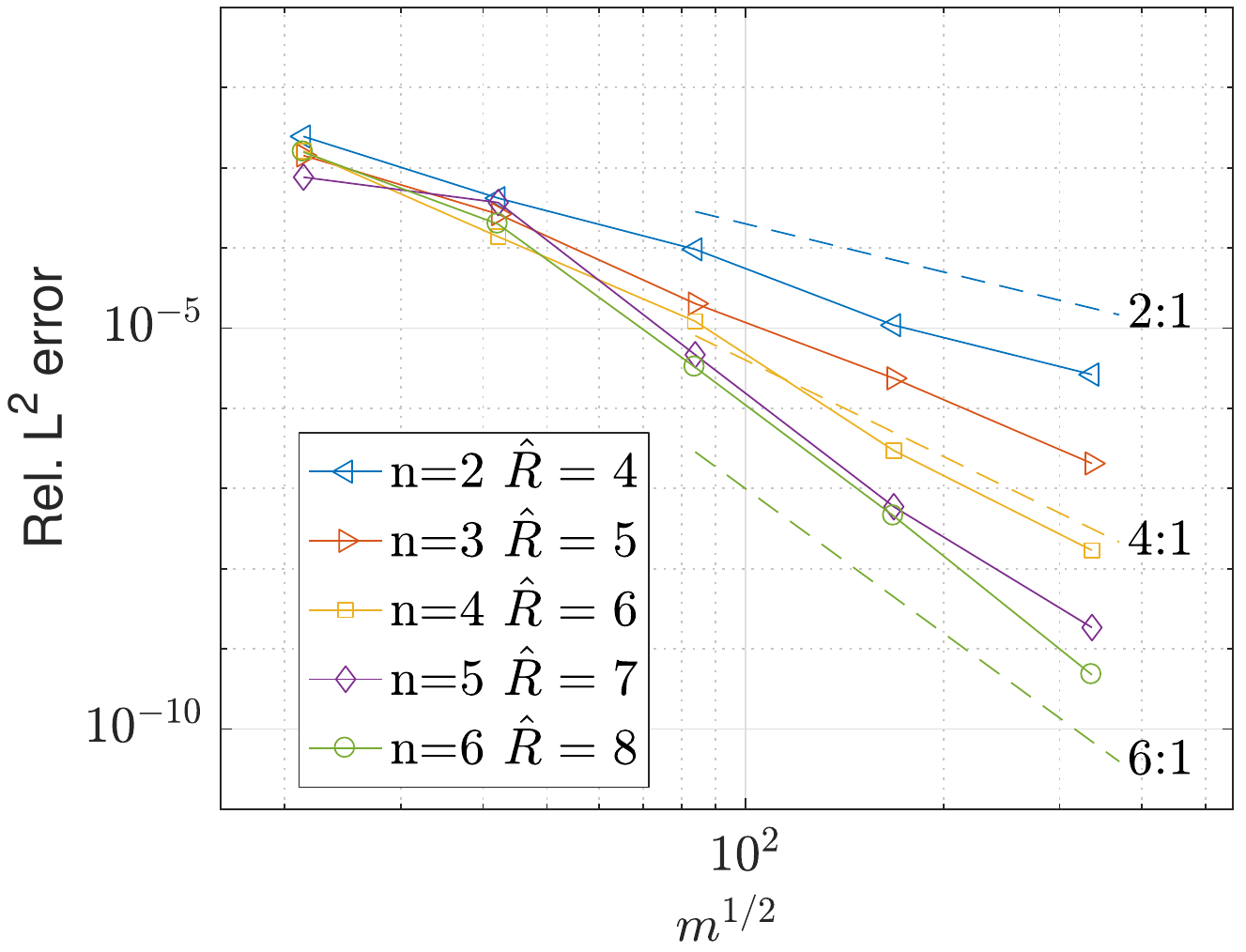}}
		\subfigure[Stress ]
		{\includegraphics[width=0.49\textwidth,clip,keepaspectratio,angle=0]{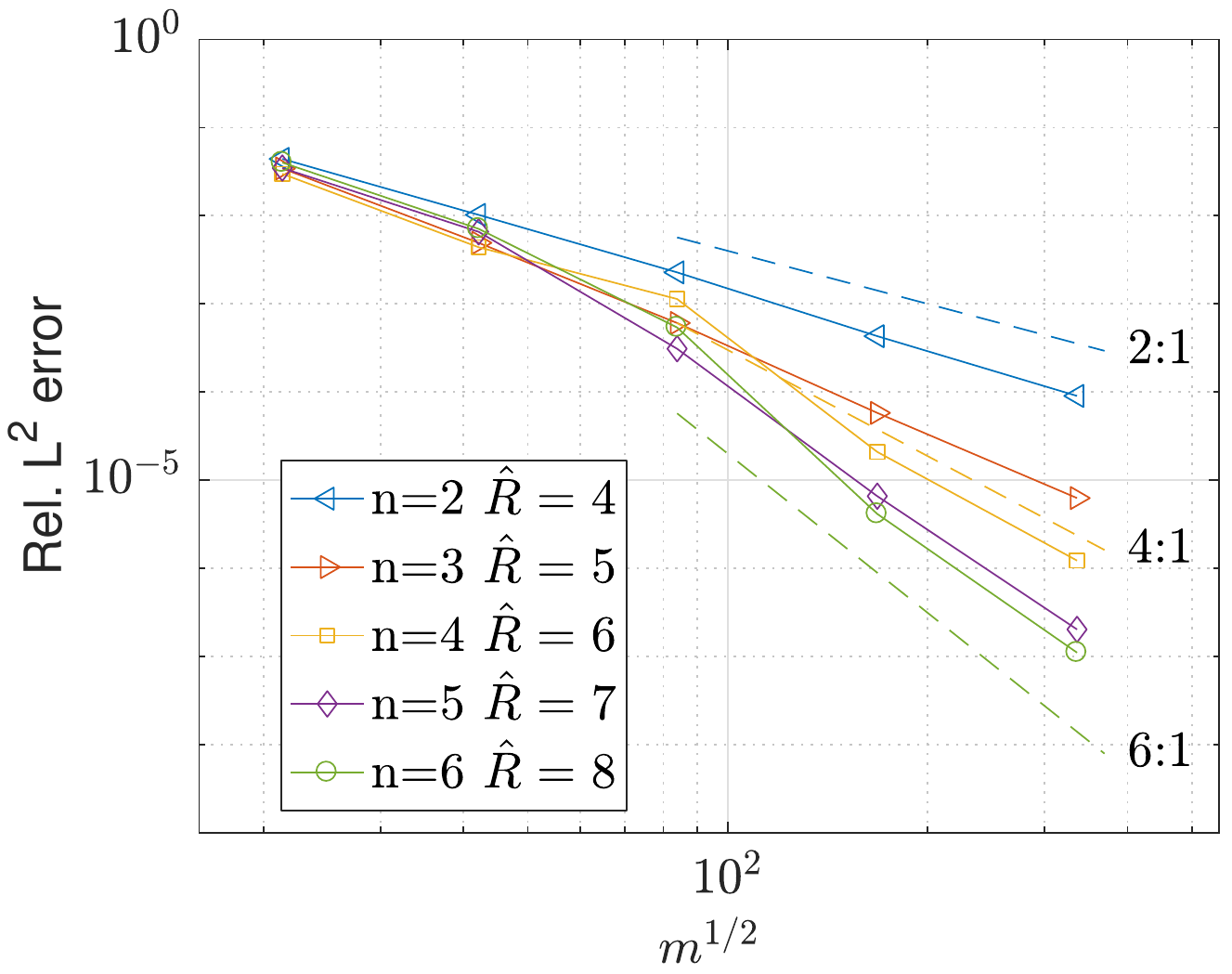}}
	\caption{
		Convergence of the relative $L^2$   error   of the displacements and stress fields for the infinite plate with hole problem, solved   for different values of $n$ and $p=2$.
		Regular nodal distributions are considered in (a) and (b), while a perturbation of $0.2h$ is considered in (c) and (d).
	} 
	\label{fig:conv_elas}
\end{figure}

\subsubsection{   L-shaped plate with stress singularity}

\begin{figure} 
	\centering{
		
	\immediate\write18{svgmodified platelntr 1 drawing}
	\def\svgscale{.4}
\begingroup%
  \makeatletter%
  \providecommand\color[2][]{%
    \errmessage{(Inkscape) Color is used for the text in Inkscape, but the package 'color.sty' is not loaded}%
    \renewcommand\color[2][]{}%
  }%
  \providecommand\transparent[1]{%
    \errmessage{(Inkscape) Transparency is used (non-zero) for the text in Inkscape, but the package 'transparent.sty' is not loaded}%
    \renewcommand\transparent[1]{}%
  }%
  \providecommand\rotatebox[2]{#2}%
  \newcommand*\fsize{\dimexpr\f@size pt\relax}%
  \newcommand*\lineheight[1]{\fontsize{\fsize}{#1\fsize}\selectfont}%
  \ifx\svgwidth\undefined%
    \setlength{\unitlength}{251.92200568bp}%
    \ifx\svgscale\undefined%
      \relax%
    \else%
      \setlength{\unitlength}{\unitlength * \real{\svgscale}}%
    \fi%
  \else%
    \setlength{\unitlength}{\svgwidth}%
  \fi%
  \global\let\svgwidth\undefined%
  \global\let\svgscale\undefined%
  \makeatother%
  \begin{picture}(1,1.78924428)%
    \lineheight{1}%
    \setlength\tabcolsep{0pt}%
    \put(0,0){\includegraphics[width=\unitlength,page=1]{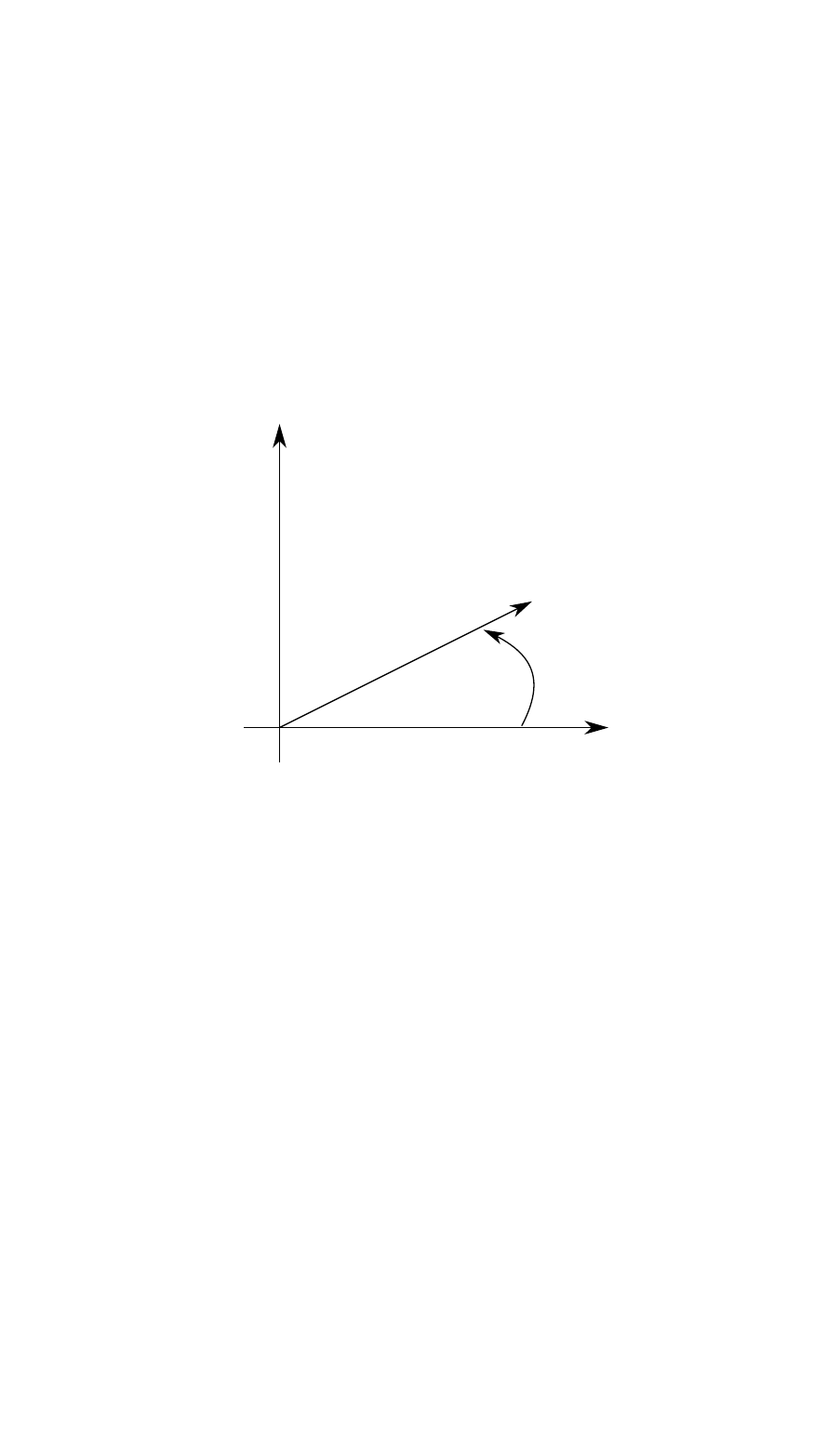}}%
    \put(0.66069898,0.83298591){\color[rgb]{0,0,0}\makebox(0,0)[lt]{\lineheight{0}\smash{\begin{tabular}[t]{l}$x$\end{tabular}}}}%
    \put(0.36384662,1.26629718){\color[rgb]{0,0,0}\makebox(0,0)[lt]{\lineheight{0}\smash{\begin{tabular}[t]{l}$y$\end{tabular}}}}%
    \put(0.6486504,1.07409011){\color[rgb]{0,0,0}\makebox(0,0)[lt]{\lineheight{0}\smash{\begin{tabular}[t]{l}$r$\end{tabular}}}}%
    \put(0.67383966,0.95582706){\color[rgb]{0,0,0}\makebox(0,0)[lt]{\lineheight{0}\smash{\begin{tabular}[t]{l}$\theta$\end{tabular}}}}%
    \put(0,0){\includegraphics[width=\unitlength,page=2]{platelntr.pdf}}%
    \put(0.02295268,0.87803857){\color[rgb]{0,0,0}\makebox(0,0)[lt]{\lineheight{0}\smash{\begin{tabular}[t]{l}$\alpha$\end{tabular}}}}%
  \end{picture}%
\endgroup%

		\caption{Elastic plate with a re-entrant corner.}
		\label{fig:platel}}
\end{figure}

 In order to better investigate the performance of HOLMES collocation   also for   problems with  singularities, this section considers  an elastic body with a re-entrant corner subjected to tractions on remote boundaries, as shown in Fig.~\ref{fig:platel}.
 The analytical solution  in the vicinity of the singular point was  given in \cite{Williams1952} and depends on the particular traction configuration. 
 In this example, we consider a corner angle $\alpha=\pi/2$ and reproduce the Mode I term of the asymptotic expression of the displacement field, which is given by
\begin{equation}
\begin{aligned}
u(r,\theta)&=r^{\lambda_1} \dfrac{1}{2G}     \Big\{  \big[    \kappa    -Q(\lambda_1 +1) \big] \cos{(\lambda_1 \theta)}    -\lambda_1 \cos[(\lambda_1 -2)\theta] \Big\}         ,\\
v(r,\theta)&=r^{\lambda_1} \dfrac{1}{2G}     \Big\{  \big[    \kappa    +Q(\lambda_1 +1)  \big]  \sin{(\lambda_1 \theta)}  +\lambda_1 \sin[(\lambda_1 -2)\theta] \Big\},\\
\end{aligned}
\end{equation}
where $\lambda_1$ is an eigenvalue that determines the order of the singularity and $Q$ is a constant, both depending in general on the corner angle $\alpha$. In particular, for  $\alpha=\pi/2$, their value is given by $\lambda_1=0.544483737$ and $Q=0.543075579$.
The interested reader is referred to \cite{Szabo2011} for a more detailed discussion of the general solution.
The above displacement field is imposed as an essential   boundary condition and the problem is solved assuming a state of plane stress and the material parameters $E=1$ and $\nu=0.3$. The corresponding stress field is given by
\begin{equation}
\begin{aligned}
\sigma_x(r,\theta)&=\lambda_1r^{\lambda_1-1}  \Big\{  \big[   2    -Q(\lambda_1 +1) \big]   \cos{[(\lambda_1 -1)\theta]}     -(\lambda_1-1) \cos[(\lambda_1 -3)\theta] \Big\}  ,\\
\sigma_y(r,\theta)&=\lambda_1r^{\lambda_1-1}    \Big\{  \big[   2    +Q(\lambda_1 +1) \big]   \cos{[(\lambda_1 -1)\theta]}     +(\lambda_1-1) \cos[(\lambda_1 -3)\theta] \Big\} ,\\
\tau_{xy}(r,\theta)&=\lambda_1r^{\lambda_1-1}     \Big\{   Q(\lambda_1 +1)    \sin{[(\lambda_1 -1)\theta]}    +(\lambda_1-1) \sin[(\lambda_1 -3)\theta] \Big\},
\end{aligned}
\end{equation}
and presents a singularity at the corner, which limits the convergence of the numerical solution.
In particular, for   finite element  Galerkin formulations,  it is well known that the energy error converges with a rate equal to $\lambda_1 $,  independently of the approximation polynomial order $n$ \cite{Szabo2011}.
As discussed earlier, a theoretical framework for the numerical analysis of collocation methods has not been completely developed yet.
The presence of singularities further complicates the problem and convergence rates are estimated in the literature  based on numerical studies only \cite{Schillinger2013b,Maurin2018f}.

The plate domain is discretized using   regular and irregular nodal distributions, as  outlined in Fig.~\ref{fig:node_L}.
In this case, the regular distributions are obtained by merging together three tensor product grids, generated in the unit square, and the irregular ones as for the previous elasticity example.
 
The relative convergence studies are reported in  Fig.~\ref{fig:conv_L}, where it can be observed how the stress error converges with a rate of   $\lambda_1 $ for both the regular and irregular grid cases, which is in agreement with the theoretical prediction for Galerkin methods.
Interestingly, the displacement error converges with a faster rate of $\sim 1.3$.
 
\begin{figure} 
	\centering 
	\subfigure[  ]
	{\includegraphics[width=0.40\textwidth,clip,keepaspectratio,angle=0]{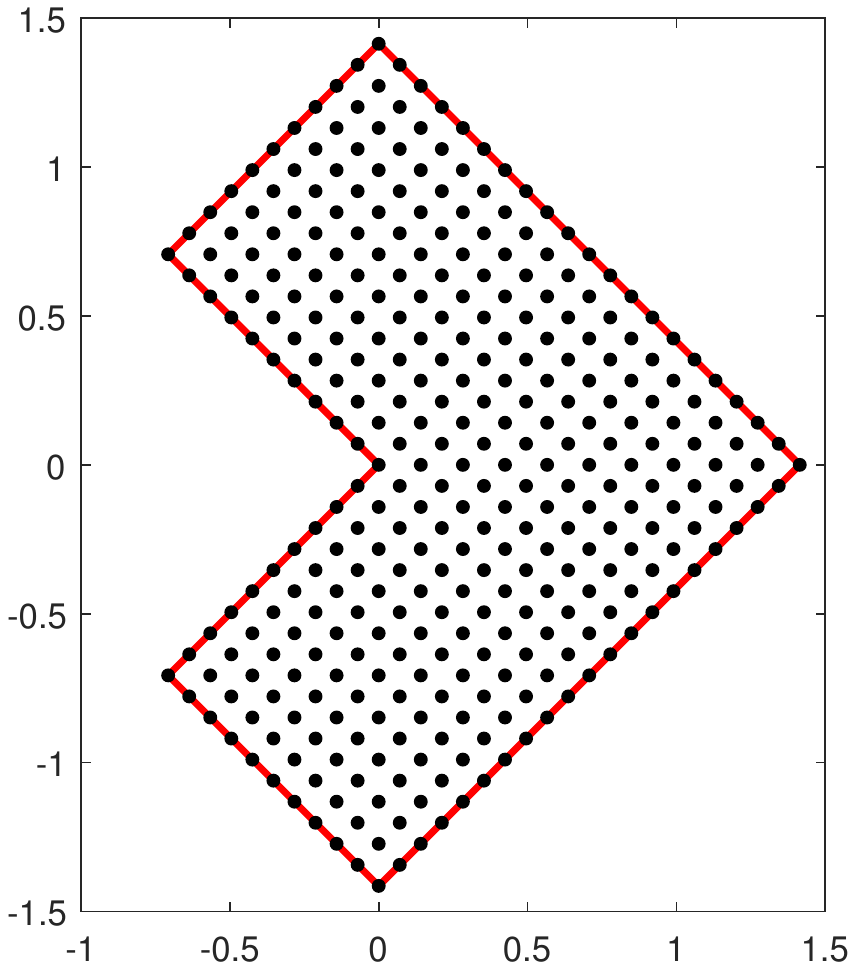}}\;\;\;
	\subfigure[  ]
	{\includegraphics[width=0.40\textwidth,clip,keepaspectratio,angle=0]{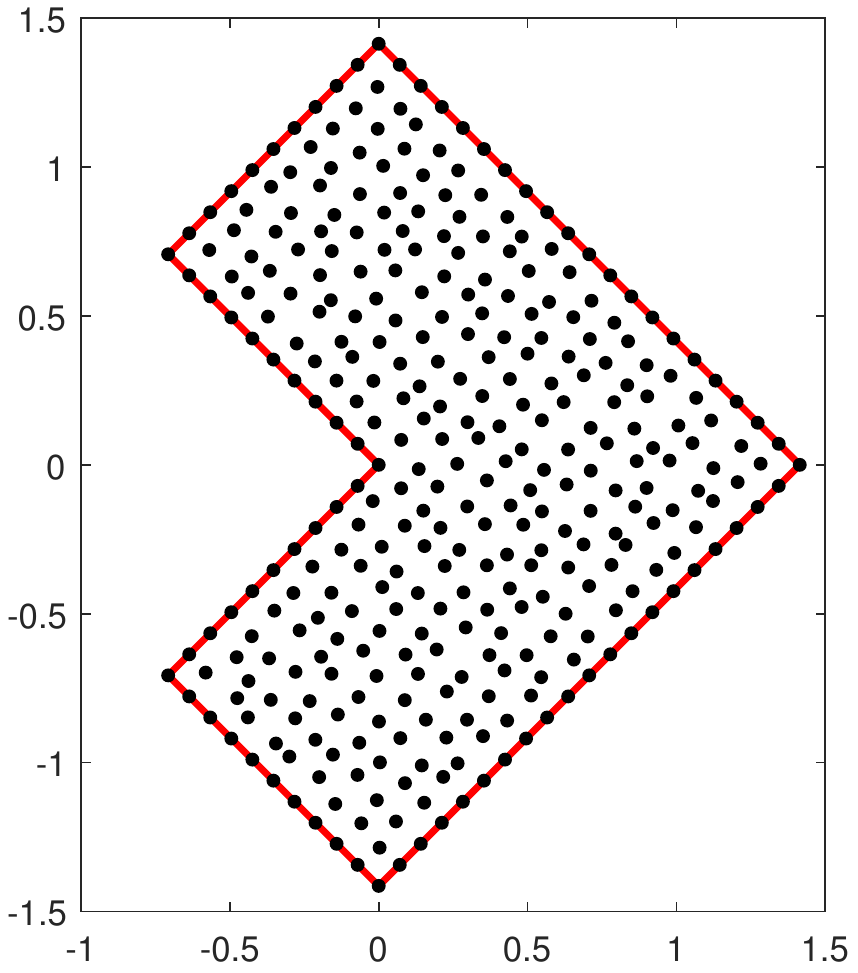}}
	\caption{Nodal discretizations for the L-shaped elastic plate with singularity problem. The discretization in (a) is generated by merging three square tensor product grids. To obtain an irregular distribution, the internal nodes are moved by adding a random   perturbation  whose components are in the range of $[-0.2h\;0.2h]$ in   (b).
	} 
	\label{fig:node_L}
\end{figure}

\begin{figure}  
	\centering
	\subfigure[Displacements ]
	{\includegraphics[width=0.49\textwidth,clip,keepaspectratio,angle=0]{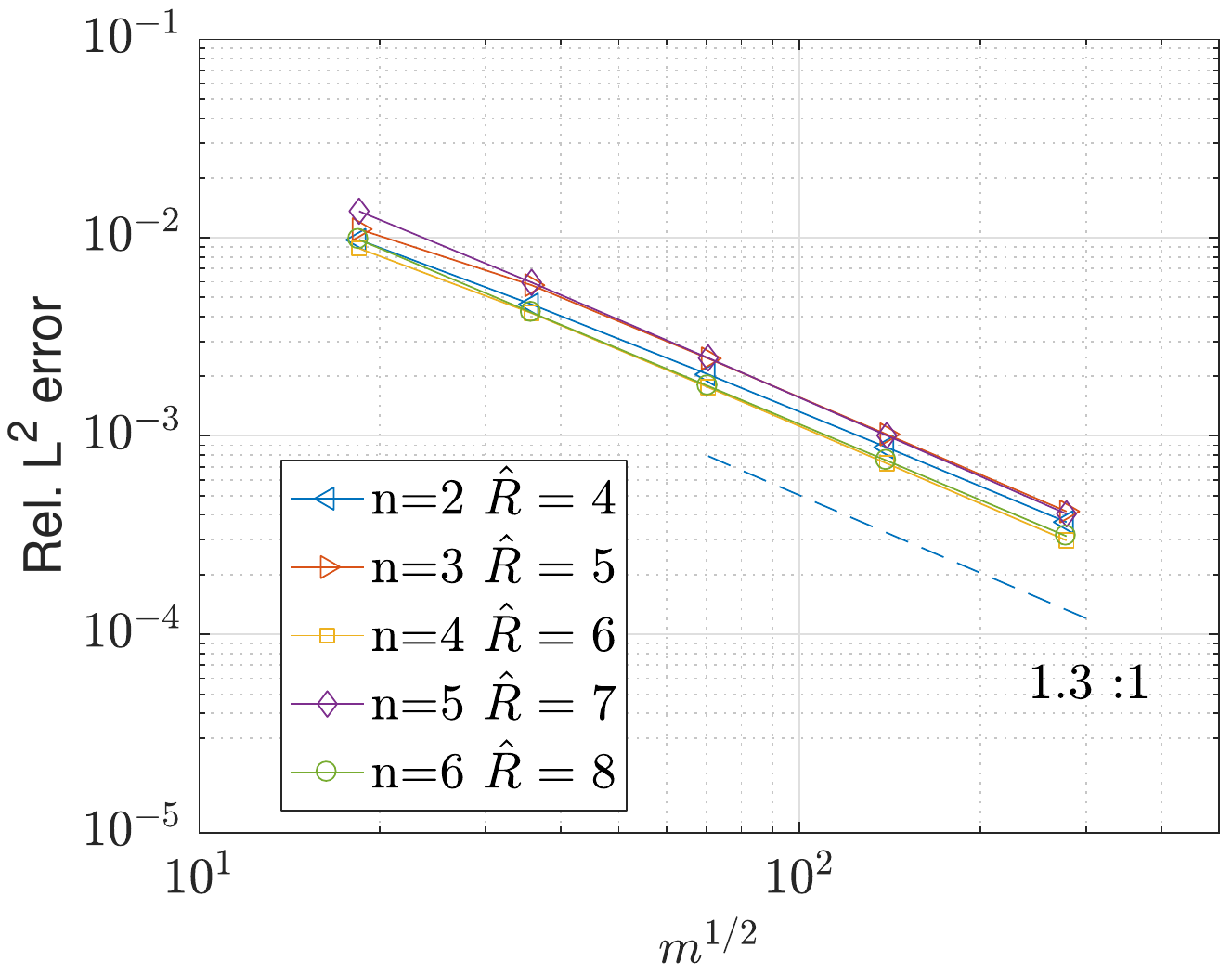}}
	\subfigure[Stress ]
	{\includegraphics[width=0.49\textwidth,clip,keepaspectratio,angle=0]{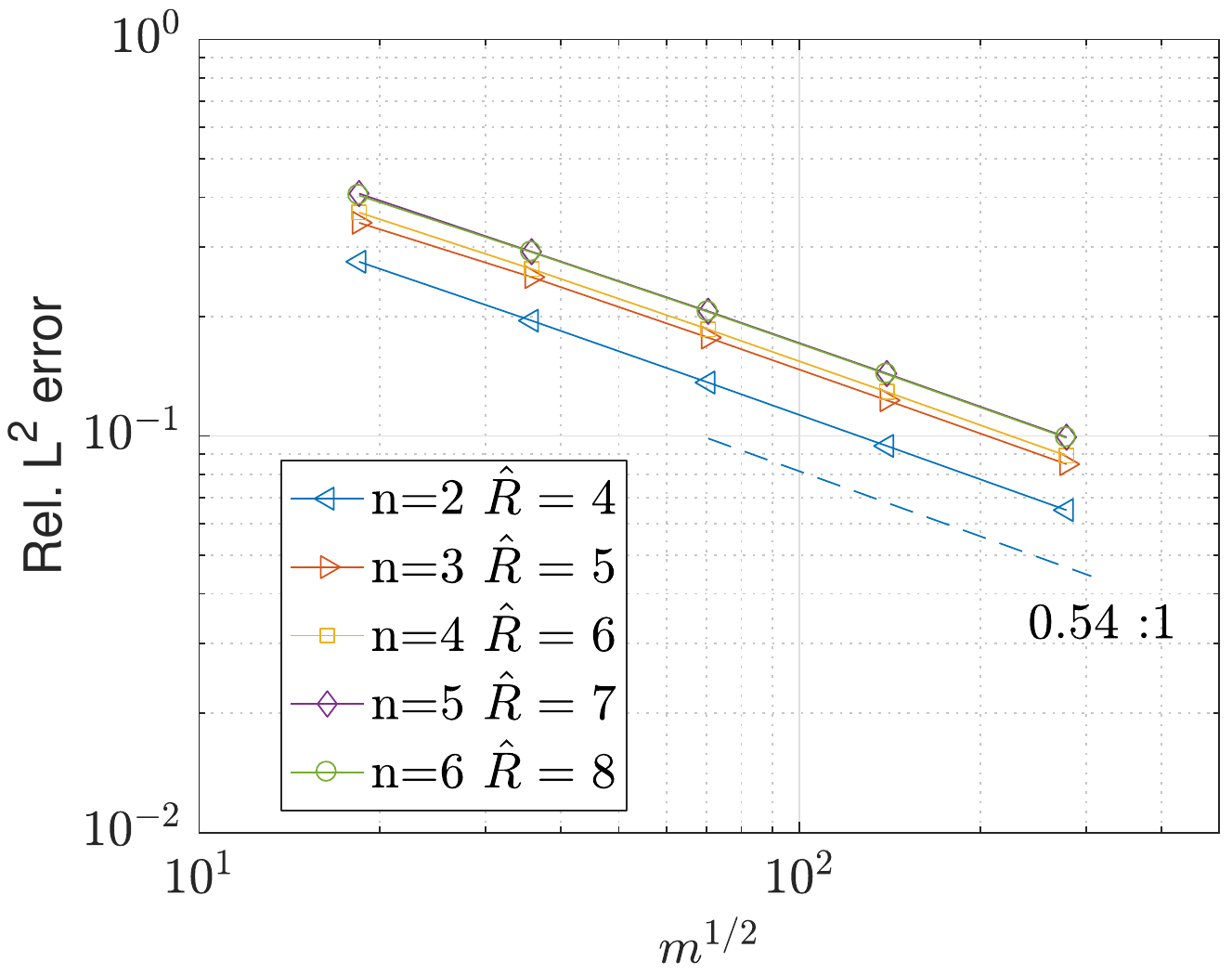}}
	\subfigure[Displacements ]
	{\includegraphics[width=0.49\textwidth,clip,keepaspectratio,angle=0]{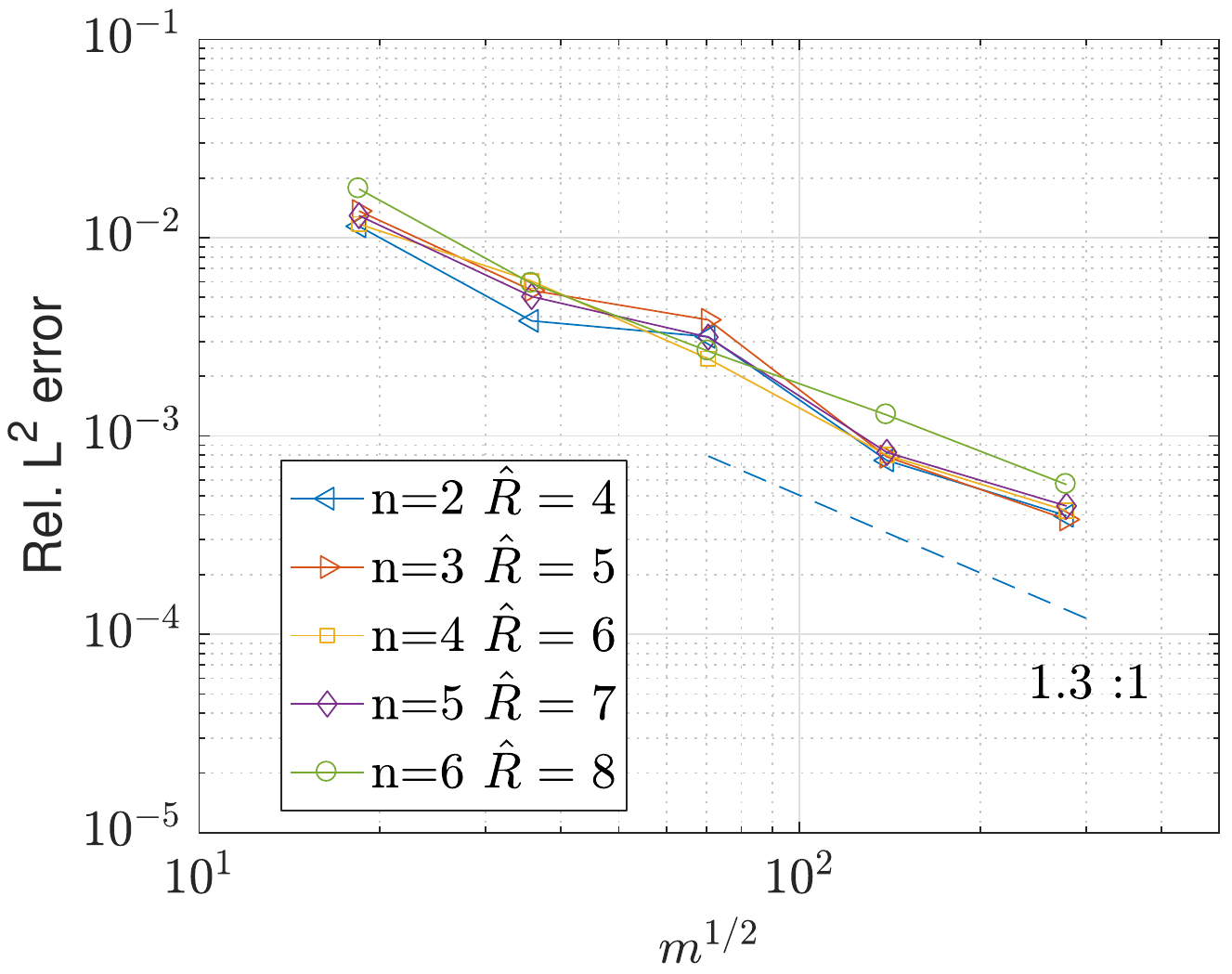}}
	\subfigure[Stress ]
	{\includegraphics[width=0.49\textwidth,clip,keepaspectratio,angle=0]{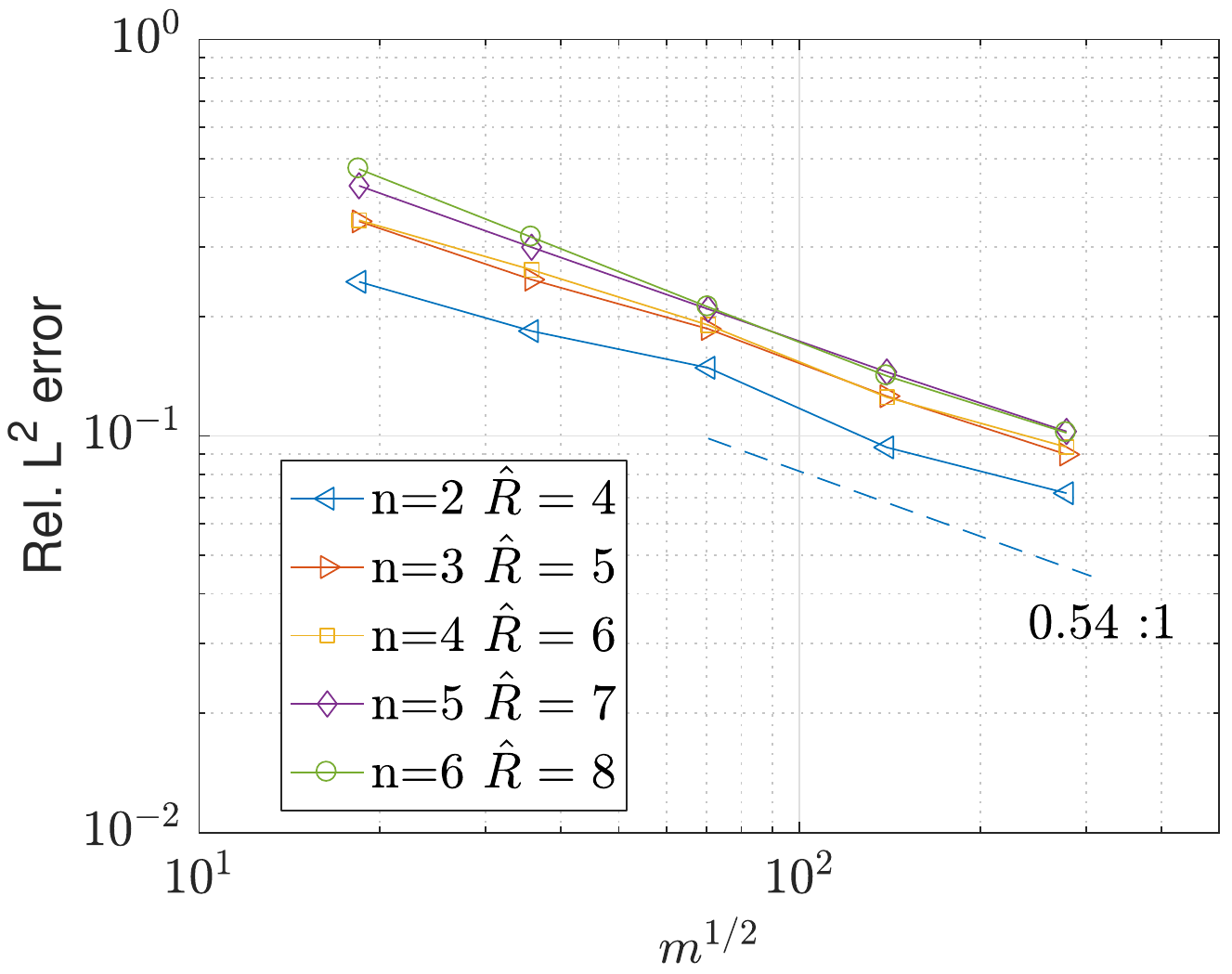}}
	\caption{
	Convergence of the relative $L^2$   error  of the displacements and stress fields for the L-shaped elastic plate with singularity problem, solved   for different values of $n$ and $p=2$.
	Regular nodal distributions are considered in (a) and (b), while a perturbation of $0.2h$ is considered in (c) and (d).
	} 
	\label{fig:conv_L}
\end{figure}

\end{MyColorPar}

\subsection{Poisson equation on a domain represented by a NURBS curve}
\label{section:laplace}

The numerical results presented in the previous sections showed the accuracy of HOLMES collocation in different applications. 
Both   the acoustic  and the infinite plate with hole problems are defined on circular domains (spherical in 3D) and the expected convergence rates are obtained for HOLMES of different orders.
For the acoustic Helmholtz problem with natural boundary conditions, this is not the case of Galerkin formulations integrated on the Delaunay triangulation of the nodes where, even if high-order basis functions are employed, the convergence rate is limited to the value of  $2:1$ given by the geometrical error
unless special integration rules are used \cite{Greco2017d}.
On the other hand, imposing the boundary conditions on the exact boundary description is enough for the collocation formulation to converge with the expected rates.

As discussed in the introduction, many efforts have been made over the last few years to seek a tighter integration between CAD and numerical analysis and, in this context, meshfree collocation can be a possible way to directly approximate PDEs on CAD geometric models \cite{Mirfatah2019a}.
This  concept is examined in this section with an example of a star-shaped domain    defined by a NURBS curve, which is the standard employed in CAD.
\red{The geometry is shown in Fig.~\ref{fig:star} and the coordinates of the control points are given in the Appendix.}

To implement the collocation procedure, the computational grid is generated as a proof of concept with the FEM mesher \textit{distmesh}, like in the other examples considered in this work. However, as discussed in \cite{Mirfatah2019a}, specific algorithms can be developed for the automatic grid generation on complex CAD models.

\begin{figure} 
	\centering
	\subfigure[ 74 nodes]
	{\includegraphics[width=0.40\textwidth,clip,keepaspectratio,angle=0]{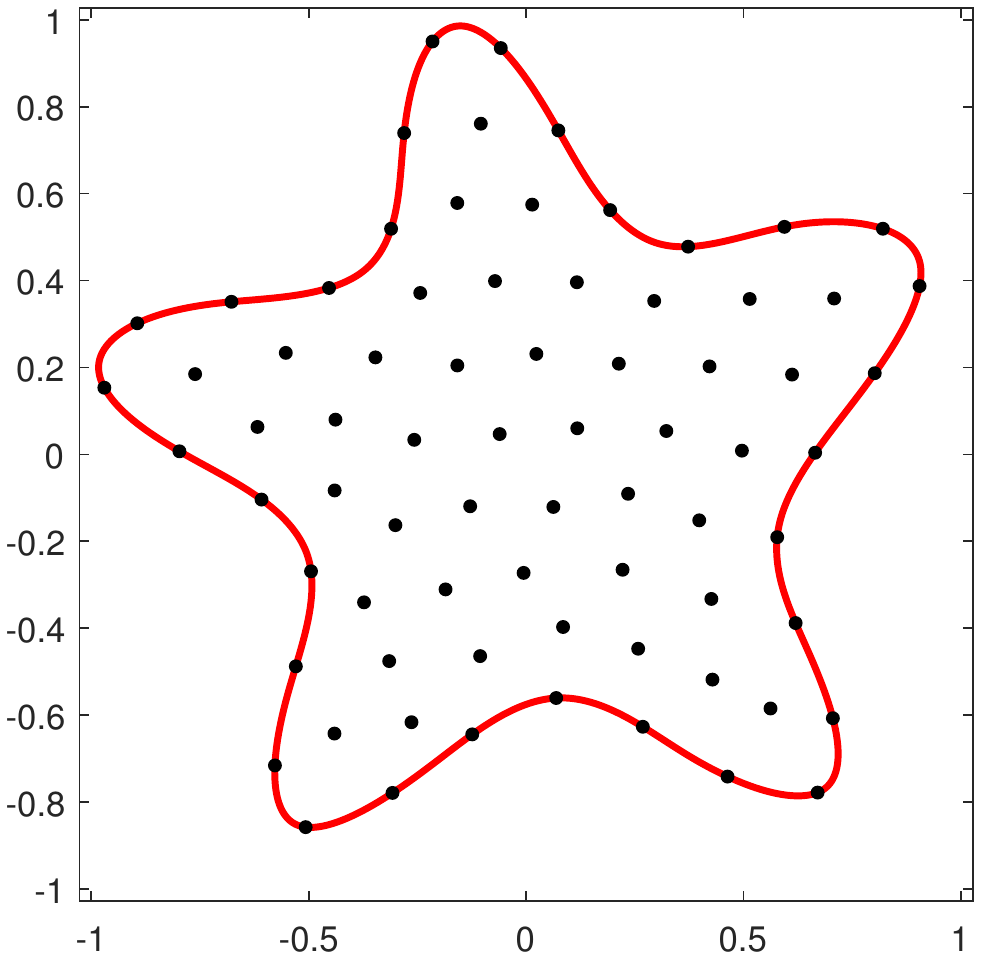}}\;\;\;
	\subfigure[ 244 nodes]
	{\includegraphics[width=0.40\textwidth,clip,keepaspectratio,angle=0]{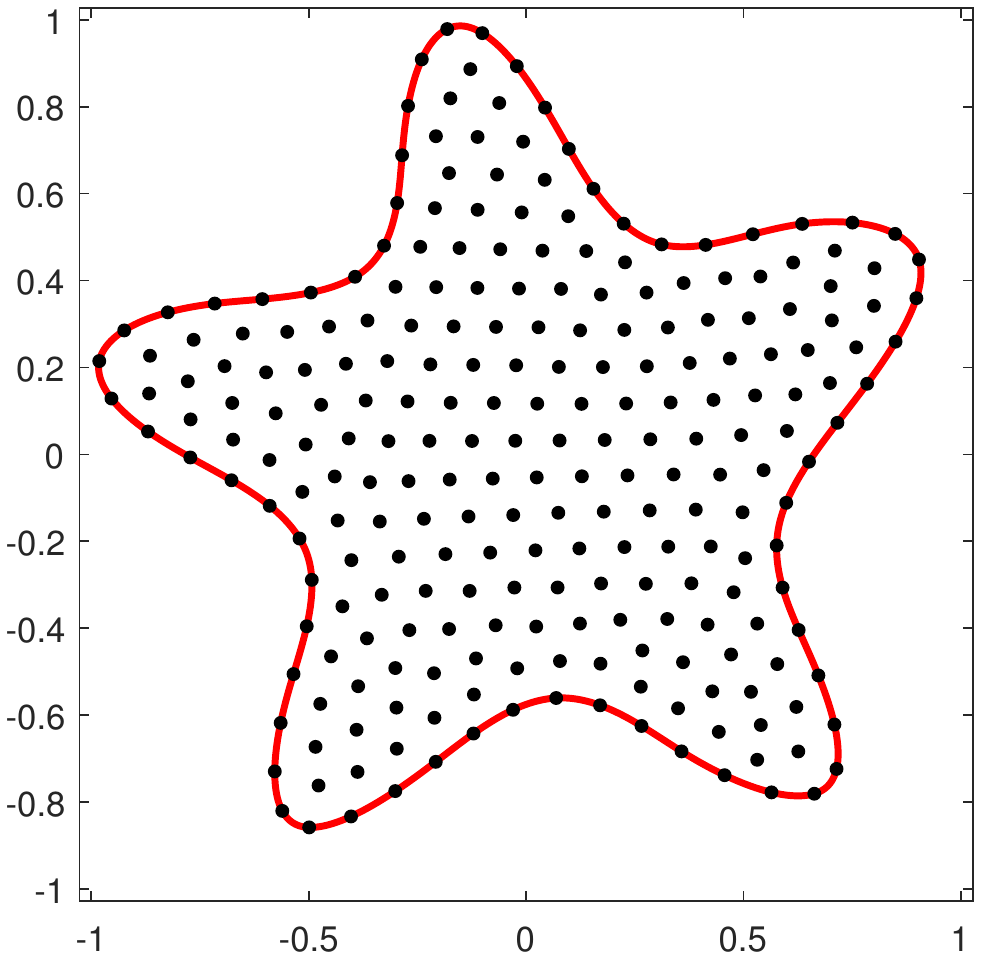}}
	\caption{A star-shaped domain defined by a NURBS curve and two discretizations used for the convergence study.
	} 
	\label{fig:star}
\end{figure}

The following Poisson problem is solved
\begin{equation}
\Delta u(\vx)  = b (\vx), \;\;\; x \in \Omega,
\label{eq:Poisson}
\end{equation}
and, as for the 1D case of Section \ref{section:1d},  source terms and essential boundary conditions are imposed to reproduce the following functions
\begin{equation}
\begin{aligned}
u_1(\vx)=&\sinh(x) \cos(y) \\
u_2(\vx)=&e^{x^2+y^2}+\sinh(x) \cos(2y).  
\end{aligned}
\end{equation}

\begin{figure} 
	\centering
	\subfigure[ $u_1(\vx)$]
		{\includegraphics[width=0.49\textwidth,clip,keepaspectratio,angle=0]{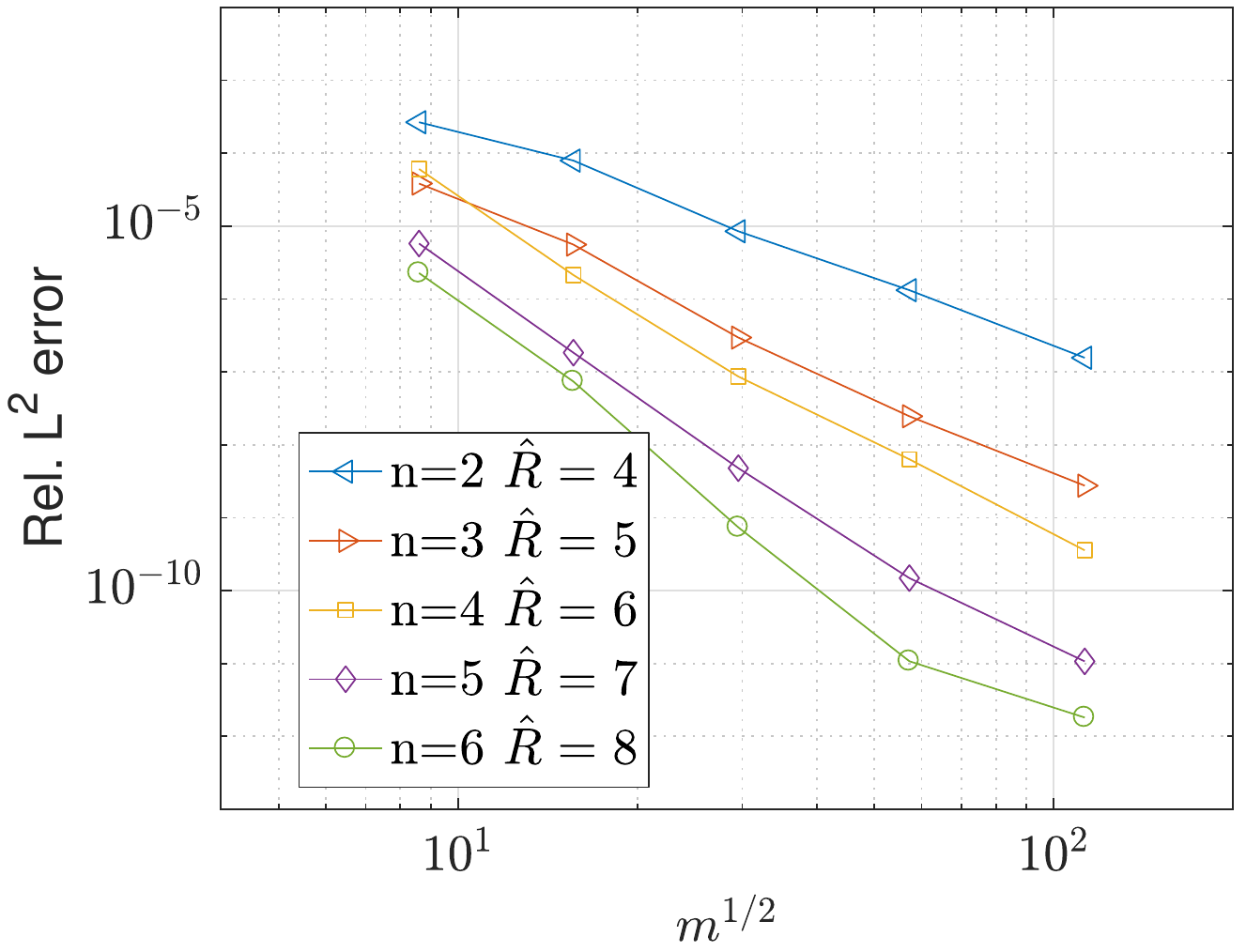}}
	\subfigure[$u_2(\vx)$ ]
	{\includegraphics[width=0.49\textwidth,clip,keepaspectratio,angle=0]{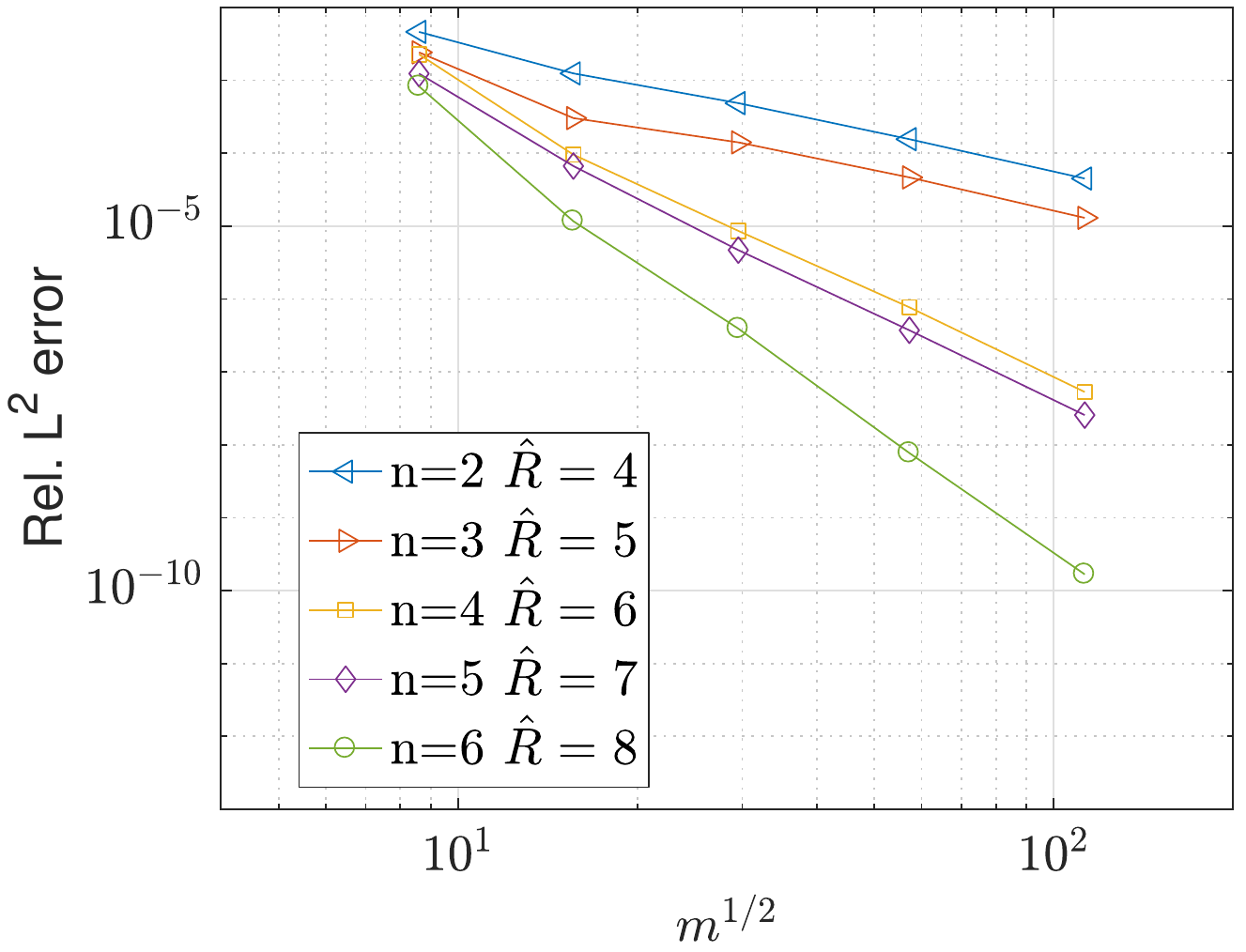}}
	\caption{
		Convergence of the relative $L^2$  error   for a Poisson problem solved on a star shaped domain defined by a NURBS curve  for different values of $n$, and $p=2$.
	} 
	\label{fig:conv_stel}
\end{figure}

The convergence curves of the relative  $L^2$ error are reported in Fig.~\ref{fig:conv_stel}, where the  expected rates  are found in all the cases, with a super-convergent behavior in most of the curves for $u_1(\vx)$.
These results confirm how the discretization with HOLMES approximants and the imposition of the boundary conditions directly on the points belonging to the NURBS boundary ensure the correct convergence behavior, allowing a direct integration between the CAD representation and the numerical analysis.

\section{Concluding remarks}
\label{section:conclusions}
In this work,  a collocation framework  based on HOLMES approximation  schemes  was applied to the resolution of PDEs.
This approach introduces some remarkable advantages with respect to the Galerkin formulation. 
A relevant one is that the computational times for assembling the stiffness matrix are reduced since the number of evaluation points of the basis functions and their derivatives is significantly lower and no memory bottlenecks for writing in the matrix occur. In addition, by avoiding a background integration grid, a truly meshfree character is obtained, which can be useful for many applications such as problems involving large deformations.   

The numerical examples   confirmed the validity and the accuracy of the proposed methodology in different applications.
They also suggested  that, in contrast to the Galerkin formulation, HOLMES collocation converges with a value of $p=2$ for the locality norm, independently of the order $n$. Furthermore, $p=2$ shows more regular convergence trends as compared to higher values of  $p$.
A theoretical study on the convergence of the method, to support these numerical findings, can be the topic of future research. However, a framework for the numerical analysis of collocation methods is not nearly as mature as for Galerkin methods \cite{Auricchio2010}. 
The character of the stiffness matrix, which is no longer symmetric and positive definite, is another drawback with respect to Galerkin methods.

This work also showed how the proposed HOLMES collocation procedure allows us to approximate PDEs on domains with a smooth geometric definition, given either by an explicit description or by a NURBS curve, which can be a very useful methodology to integrate CAD and numerical analysis.
Therefore, its application to complex 3D models with the development of an automated procedure for the grid generation, such as that used in \cite{Mirfatah2019a}, can also be an interesting topic of future research.

\section{Acknowledgments}
The research of F. Greco has received funding from the  European Union's Horizon 2020 research and innovation  programme, under the 
Marie Sklodowska-Curie grant agreement No 792028.

\bibliographystyle{ieeetr}

 \bibliography{/Users/Administrator/Documents/library}


\pagebreak

\begin{MyColorPar} {red}
	\section{Appendix}
	The NURBS curve representing the star domain of Section~\ref{section:laplace} is a periodic cubic curve obtained from an equi-spaced knot vector and 20 control points, whose coordinates and weights are given below.
	
	\begin{center}
		\begin{tabular}[t]{ |c| c| c| c|  }
\hline
Index i & $x_i$ & $y_i$ & $w_i$\\ \hline
1 & -0.315646 & -1.085310 & 1.515248\\ \hline
2 & 0.091382 & -0.554301 & 1.180104\\ \hline
3 & 0.471145 & -0.906028 & 1.275062\\ \hline
4 & 0.658987 & -0.724081 & 0.805530\\ \hline
5 & 0.717332 & -0.497103 & 1.085006\\ \hline
6 & 0.612211 & -0.207935 & 1.204812\\ \hline
7 & 1.225410 & 0.199055 & 1.602287\\ \hline
8 & 1.317630 & 0.653927 & 1.287513\\ \hline
9 & 0.781819 & 0.724652 & 1.279621\\ \hline
10 & 0.308763 & 0.447180 & 1.146973\\ \hline
11 & 0.104881 & 1.024695 & 1.371792\\ \hline
12 & -0.139101 & 0.885485 & 0.747637\\ \hline
13 & -0.328176 & 0.819813 & 1.104964\\ \hline
14 & -0.310237 & 0.389074 & 1.089473\\ \hline
15 & -1.013978 & 0.505792 & 1.398823\\ \hline
16 & -0.948944 & 0.187873 & 0.793557\\ \hline
17 & -1.055274 & -0.025958 & 1.393367\\ \hline
18 & -0.498210 & -0.237582 & 1.208597\\ \hline
19 & -0.769951 & -0.761041 & 1.349239\\ \hline
20 & -0.667820 & -1.073769 & 1.081645\\ \hline
\end{tabular}

	\end{center}
	
	%
	%
	%
	%
	%
	%
	%
	
\end{MyColorPar}

\end{document}